\documentclass[ aps,
reprint,
 prd,
 superscriptaddress,
 numerical,
 amsmath,amssymb,
nofootinbib,
longbibliography,
floatfix
]{revtex4-2}

\usepackage{bbold}
\usepackage{slashed}
\usepackage{combelow}
\usepackage{graphicx}
\usepackage[usenames,dvipsnames,svgnames,table]{xcolor}

\usepackage[colorlinks=True,linkcolor=blue,citecolor=blue,urlcolor=blue]{hyperref}

\definecolor{purple}{rgb}{0.8,0,0.6}

\let\oldciteauthor=\citeauthor
\def\citeauthor#1{\hypersetup{citecolor=black}\oldciteauthor{#1}}

\let\oldciten=\onlinecite
\def\onlinecite#1{\hypersetup{citecolor=blue}\oldciten{#1}}

\let\oldcite=\cite
\def\cite#1{\hypersetup{citecolor=blue}\oldcite{#1}}

\newcommand{\e}{{\mathrm{e}}}

\allowdisplaybreaks

\begin{document}

\title{Linear sigma model with quarks and Polyakov loop in rotation: phase diagrams, Tolman-Ehrenfest law and mechanical properties}

\author{Pracheta Singha}
\affiliation{Department of Physics, West University of Timi\cb{s}oara, Vasile P\^arvan Ave. 4, Timi\cb{s}oara 300223, Romania}

\author{Sergiu Busuioc}
\affiliation{Department of Physics, West University of Timi\cb{s}oara, Vasile P\^arvan Ave. 4, Timi\cb{s}oara 300223, Romania}

\author{Victor E. Ambru\cb{s}}
\affiliation{Department of Physics, West University of Timi\cb{s}oara, Vasile P\^arvan Ave. 4, Timi\cb{s}oara 300223, Romania}
\email{victor.ambrus@e-uvt.ro}
\thanks{corresponding author.}

\author{Maxim N. Chernodub}
\affiliation{Institut Denis Poisson, CNRS UMR 7013, Universit\'e de Tours, Universit\'e d'Orl\'eans, 
Parc de Grandmont, Tours, 37200, France}
\affiliation{Department of Physics, West University of Timi\cb{s}oara, Vasile P\^arvan Ave. 4, Timi\cb{s}oara 300223, Romania}

\begin{abstract}
We study the effect of rotation on the confining and chiral properties of QCD using the Polyakov-enhanced linear sigma model coupled to quarks. Working in the homogeneous approximation, we obtain the phase diagram at finite temperature, baryon density and angular frequency, taking into account the causality constraint enforced by the spectral boundary conditions at a cylindrical surface. We explicitly address various limits with respect to system size $R$, angular frequency $\Omega$ and chemical potential $\mu$. We demonstrate that, in this model, the critical temperatures of both the chiral restoration and the deconfinement transitions diminish in response to the increasing rotation, being in contradiction with the first-principle lattice results. We demonstrate that consistency between the thermodynamics of the model and the Tolman-Ehrenfest law is achieved in the limit of large volume. We also compute the mechanical characteristics of the rotating plasma, such as the moment of inertia and the $K_n$ shape coefficients describing the response of the thermodynamic potential with respect to the increase of angular velocity $\Omega$.
\end{abstract}

\maketitle

\section{Introduction}\label{sec:intro}

The experimental observation of highly vortical quark-gluon plasmas by the RHIC collaboration has ignited a strong interest in the theoretical community to explore and understand the properties of this most vortical fluid ever created~\cite{STAR:2007ccu, STAR:2017ckg}. The vortical motion of plasmas has been addressed from the point of view of hydrodynamic and transport-based models, which were reviewed, along with the relevant experimental results, in Refs.~\cite{Huang:2020dtn, Becattini:2021lfq}. On general grounds, one can expect that the presence of vorticity in the system can also substantially affect the thermodynamic properties and the phase structure of QCD, which will be the main topic of our paper.

Due to the strong non-perturbative nature of quark-gluon plasma, its dynamics cannot be reliably explored theoretically using conventional, perturbative methods. This fact leaves us with two options: we can obtain information from the first-principle approaches based on numerical Monte Carlo techniques (i.e., lattice QCD)~\cite{Yamamoto:2013zwa, Braguta:2020biu, Braguta:2021jgn, Braguta:2022str, Braguta:2023kwl, Braguta:2023yjn, Yang:2023vsw,  Braguta:2023iyx, Braguta:2024zpi}, or we can use effective theoretical models that describe the infrared properties of QCD. Unfortunately, numerical methods often cannot provide us with the relevant information about the mechanisms underlying the non-perturbative effects. However, numerical data can serve as a valuable source of knowledge that can be used to constrain effective theoretical models, consequently allowing us to acquire information about the effects that lie behind the physical phenomena in question. So far, most of these theoretical models fail to describe the lattice data under rotation.

The rotation of the quark-gluon plasma (QGP) implies several puzzles.
The earlier numerical simulations that focused on the global behavior of rotating gluon matter have revealed that the critical temperature of the deconfining phase transition is a rising function of the angular velocity (vorticity)~\cite{Braguta:2020biu, Braguta:2021jgn}. This result challenges the naive prediction, based on the Tolman-Ehrenfest (TE) law \cite{Tolman:1930ona, Tolman:1930zza}, that rotation makes the outer layers of the rotating system hotter, and thus rotation should decrease the critical temperature of the phase transition. Analytical predictions coming from the standard effective field theories, such as the Nambu-Jona-Lasinio (NJL) model~\cite{Nambu:1961fr, Nambu:1961tp}, the linear sigma model coupled to quarks (LSM$_q$)~\cite{GellMann1960} and other effective models~\cite{Fujimoto:2021xix, Chen:2022mhf} appear to confirm the aforementioned TE effect~\cite{Jiang:2024zsw, Morales-Tejera:2025qvh}, being, thus, in contradiction with the lattice data~\cite{Braguta:2020biu, Braguta:2021jgn}.

One possible explanation for the ``inverse Tolman-Ehrenfest effect'' revealed by lattice data is related to the melting of the gluonic condensate that alters dramatically the moment of inertia of the gluon plasma~\cite{Braguta:2023yjn}. Such an effect cannot be explained by standard effective models of QCD, however qualitative agreement with the lattice data can still be captured in certain analytical approaches in which the model couplings are specially selected functions of the rotational velocity to match the lattice data~\cite{Jiang:2021izj, Sun:2024anu, Jiang:2024zsw, Nunes:2024hzy} (see also Ref.~\cite{Chen:2024tkr} for evidence from perturbative QCD).

In this paper, we aim to study the phase diagram of strongly interacting matter under rotation. Without attempting to resolve the discrepancy between model calculations and lattice QCD results, we perform this study by considering a rigidly rotating plasma in global thermal equilibrium while rigorously enforcing the causality constraint by imposing boundary conditions such that the local speed of corotating observers does not exceed the speed of light. While the plasma produced in heavy-ion collisions has both space- and time-dependent vorticity~\cite{Huang:2020dtn}, rigid rotation corresponds to a state of global thermodynamic equilibrium, thus serving as a useful probe to study the response of the QGP to vorticity, also in numerical (lattice) approaches~\cite{Yamamoto:2013zwa, Braguta:2020biu, Braguta:2021jgn, Braguta:2022str, Braguta:2023kwl, Braguta:2023yjn, Yang:2023vsw,  Braguta:2023iyx}.

We employ the Polyakov-enhanced~\cite{Schaefer:2007pw} linear sigma model~\cite{GellMann1960} coupled to quarks (PLSM$_q$) as an effective model that captures both the deconfinement and the chiral phase transitions (see Ref.~\cite{Fukushima:2017csk} for a review). In this model, the potential for the mesonic self-interactions includes a small chiral symmetry breaking term that leads to a finite constitutive quark mass in the vacuum state. The Polyakov sector, modeled as a temperature-dependent effective potential, is related to a non-vanishing vacuum expectation value of the time component of the gluon four-potential. One drawback of this model is that neither the meson nor the Polyakov potentials depend directly on either the kinematic state of the system or on the presence of the boundary. Instead, the rotation and the boundary have an indirect effect on the meson and Polyakov loop expectation values through their coupling to fermionic loops. Although the inclusion of rotational effects either through the meson self-interactions \cite{Siri:2024cjw} or at the level of the gluonic sector \cite{Chen:2024tkr} may improve agreement with lattice data, we do not pursue this avenue in this work.

While it is known that rotation produces radial inhomogeneities in the QGP~\cite{Chernodub:2020qah, Braguta:2023iyx, Braguta:2024zpi}, we will use a simplified approach, treating both the chiral condensate $\sigma$ and the expectation values $L$ and $L^*$ of the Polyakov loop and its conjugate as homogeneous quantities. This approach has been adopted in various investigations of the rotating plasma (see, however, the study that incorporates inhomogeneities of rotating fields~\cite{Chernodub:2020qah, Chernodub:2022veq};
as well as a detailed comparison of homogeneous and inhomogeneous approaches in a complementary setup in Ref.~\cite{Morales-Tejera:2025qvh}).
On the other hand, lattice data reported in Ref.~\cite{Braguta:2024zpi} suggests that rotation has a sub-percent effect on the QCD plasma on the rotation axis, while drastically affecting the system phase away from the rotation axis.
Despite the fact that the effective model approach does not agree with the latest lattice results, we believe that it is essential to understand, down to the fine details, how these models respond to both boundary and rotational effects.

The particular flavor of PLSM$_q$ model that we consider involves a logarithmic potential for the Polyakov loop, as introduced in Ref.~\cite{Roessner:2006xn} (see also Ref.~\cite{Fukushima:2017csk} for a review). With respect to the polynomial ansatz \cite{Ratti:2005jh}, the logarithmic potential has the advantage of automatically limiting the values of the Polyakov loop and its conjugate between the physically accessible values $0 \le L, L^* < 1$. Moreover, we take, for simplicity, the deconfinement temperature $T_0$ at the value $0.27$ GeV, corresponding to the phase transition temperature in the ${\rm SU}(3)$ pure Yang-Mills theory. In principle, this temperature should be adjusted to some $T_0^f<T_0$ based on the number $N_f$ of fermionic flavours and in the presence of finite quark chemical potential $\mu$  (see Ref.~\cite{Schaefer:2007pw} for details). However, our numerical experiments indicate that, for the logarithmic potential with $T_0 < 0.27$ GeV, deconfinement occurs before chiral symmetry restoration, which contradicts lattice data. On the other hand, the polynomial potential \cite{Ratti:2005jh} with any implementation of $T_0$ leads to an unbounded growth of the expectation values of the Polyakov loop $L$ and its conjugate, $L^*$, which is exacerbated in the presence of rotation. Therefore, we discuss these extensions only briefly, in the appendix to this work (where we also benchmark our code against previously reported results \cite{Schaefer:2007pw}), and instead focus predominantly on the logarithmic model with fixed deconfinement temperature $T_0 = 0.27$ GeV.

As already mentioned, we investigate the effects of rotation on the phase diagram by considering a state of rigid rotation with constant angular velocity $\Omega$. To enforce the causality constraint of the resulting state, we enclose the system in a cylindrical boundary of radius $R$, such that $\Omega R \le 1$. On the boundary of the system, we follow Refs.~\cite{Ebihara:2016fwa,Zhang:2020hha} and employ the spectral boundary conditions \cite{Hortacsu:1980kv} (see Ref.~\cite{Ambrus:2015lfr} for the implementation within a cylindrical surface) to constrain the allowed fermionic states. This leads to a quantization of the transverse momentum by means of the roots of the Bessel functions, thus requiring numerical methods for the computation of all fermionic expectation values. The spectral boundary conditions have the advantage of preserving chiral symmetry \cite{Zhang:2020hha}, unlike the MIT bag model \cite{Chodos:1974je,Chernodub:2016kxh,Chernodub:2017ref,Chernodub:2017mvp}, which explicitly breaks this symmetry. 
The resulting geometry is still infinite, due to the assumed homogeneity along the $z$ direction. In order to consider the properties of strongly interacting matter inside a finite volume, one should impose suitable boundary conditions on the top and bottom lids of the cylinder \cite{Shanker:1983yd}.

We performed a thorough investigation of the phase structure of strongly interacting matter extensively over the parameter space of the model. While allowing the angular velocity $\Omega$ to vary between $0$ (static system) and $\Omega R = 1$ (causal limit), we considered system sizes from small (where the fermionic contributions are suppressed) to very large (thermodynamic limit) and studied the impact of both rotation and boundary effects on the phase diagram. In the limit of small volume, where fermionic contributions are suppressed, the Polyakov sector is dominant and the deconfinement transition proceeds as for the unbounded, non-rotating, pure-glue system. For large system sizes, it is expected that the thermodynamics of the system should be compatible with the classical TE law.

We formulate a quantitative implementation of the famous TE law by considering the thermodynamic limit of large volume (large cylinder radius $R$). To retain the causality constraint, we keep $|\Omega R| \le 1$ fixed, thus leading to decreasing $\Omega$ as $R$ is increased. Therefore, while the boundary effects become negligible, so do quantum effects, which are known to be proportional to $\Omega^2$ \cite{Becattini:2015nva,Ambrus:2017opa}. Using the local thermal equilibrium approximation for the density operator, introduced in Refs.~\cite{Becattini:2014yxa,Becattini:2015nva,Buzzegoli:2018wpy}, we consider the angular velocity dependence of the pseudo\-critical chiral restoration temperature $T_{pc}$ at vanishing chemical potential, as well as the critical chemical potential $\mu_c$ at vanishing temperature.

In the unbounded, non-rotating system, the deconfinement and chiral restoration temperatures are approximately the same both in QCD and in the PLSM$_q$ model that we consider. Recently, Ref.~\cite{Sun:2023kuu} reported that the chiral and deconfining transitions split under the influence of rotation. On the other hand, the existing numerical calculations imply that the window for this effect in the phase diagram is rather narrow~\cite{Yang:2023vsw}. Our recent effective model calculation found that the splitting is promoted by the finite-volume effects only and is absent for large enough volumes~\cite{Singha:2024tpo}. We now revisit this intriguing phenomenon by considering both the limit of small volume (discussed above) and that of fast rotation. In the latter case, the TE analogy provides unexpected evidence for a rotation-induced split occurring in the thermodynamic limit $R \to \infty$, as the causal limit is approached ($\Omega R \to 1$). Notice that, in the thermodynamic limit, the fastest possible rotation is necessarily slow in terms of its physical amplitude, as $\Omega \to 0$ when $R \to \infty$ in order to keep the product $\Omega R$ finite.

Another intriguing feature of the rotating QGP is related to its mechanical properties, namely to the moment of inertia $I$. Lattice calculations \cite{Braguta:2023yjn,Yang:2023vsw} indicate that $I$ becomes negative in a so-called ``supervortical'' range of temperatures. We therefore compute this quantity within the PLSM$_q$ model and compare it with the predictions of the TE law. Furthermore, we consider the shape coefficients $K_{2n}$, in particular, $K_2$ -- related to the moment of inertia at vanishing angular velocity -- and $K_4$ -- corresponding to shape deformations induced by rotation (see Refs.~\cite{Braguta:2023yjn,Ambrus:2023bid}). In the limit of both $T \to \infty$ (when the system becomes asymptotically free) and $R \to \infty$ (when boundary effects become negligible), these coefficients approach the values of $2!$ and $4!$, respectively, as expected \cite{Ambrus:2023bid,Patuleanu:2025zbn} for a free (non-interacting) conformal gas. At intermediate temperatures, we study the behavior of these coefficients in the context of the TE law.

The main results of this work are a rigorous numerical investigation of the bounded and rotating strongly interacting system within the PLSM$_q$ model, together with the predictions derived analytically, based on the Tolman-Ehrenfest law. We are able to explain quantitatively the dependence of the chiral restoration transition temperature on the angular frequency $\Omega$, as well as the mechanical properties (moment of inertia $I$ and shape coefficients $K_2$ and $K_4$), in the large-volume limit, when boundary effects become negligible.

The structure of the paper is as follows. In Section~\ref{sec:PLSM}, we review the PLSM${}_q$ in the absence of rotation, in the thermodynamic limit, thereby introducing our notation. We describe the meson, quark and Polyakov loop sectors of the model, the construction of the grand canonical potential, the mean-field approximation and the system thermodynamics. We also discuss the analytical and numerical strategy for determining the first-order or crossover transitions corresponding to deconfinement or chiral restoration. The vanishing-temperature limit is particularly discussed, as in this limit, PLSM$_q$ reduces to the LSM$_q$ without the Polyakov loop variable. As this section is mostly monographic, readers who are familiar with the PLSM$_q$ phenomenology are invited to skip ahead to Section~\ref{sec:rot}, where we construct the bounded PLSM$_q$ system under rotation, satisfying spectral boundary conditions on the cylindrical boundary.
Section~\ref{sec:res} starts with the phase diagram of the bounded, non-rotating PLSM$_q$ model in Subsec.~\ref{sec:res:rot0}, where we treat the limits of small, intermediate and large system sizes. Subsection~\ref{sec:res:rot} discusses the effect of rotation on the phase diagram of the system.
Section~\ref{sec:TE} presents the analysis of the rotating system based on the Tolman-Ehrenfest law. The general setup is presented in Subsec.~\ref{sec:TE:kinetic}. We consider the angular velocity dependence of the pseudocritical temperature at vanishing chemical potential in both the LSM$_q$ (Subsec.~\ref{sec:TE:LSM}) and the PLSM$_q$ (Subsec.~\ref{sec:TE:PLSM}) models, as well as of the critical chemical potential at vanishing temperature (Subsec.~\ref{sec:TE:T0}). Our analytical results are validated against the numerical ones in Subsec.~\ref{sec:TE:comparison}.
The mechanical properties (moment of inertia, shape coefficients) of the rotating plasma are studied in Section~\ref{sec:res:I}, where we also discuss the comparison with the TE prediction.
Our conclusions are presented in Sec.~\ref{sec:conc}. The paper also includes Appendix~\ref{app:poly}, which discusses subtleties of the choice of the potential for the Polyakov loop, including the shape of the potential (logarithmic vs. polynomial), as well as the dependence of the deconfinement temperature on both the number of quark flavors and on the quark chemical potential.

\section{Polyakov-loop extended linear sigma model }
\label{sec:PLSM}

In this section, we review the basic features of the Polyakov-loop-extended linear sigma model coupled to quarks for a two flavor system $(u,d)$ in a uniform, non-rotating background. The Lagrangian of the model reads as follows:
\begin{align}
    \mathcal{L} = \mathcal{L}_\mathcal{M} + \mathcal{L}_L + \mathcal{L}_q\,.
    \label{eq_L_PQSM}
\end{align}
We address the meson Lagrangian, $\mathcal{L}_\mathcal{M}$, in Subsec.~\ref{sec:PLSM:meson}. The Polyakov Lagrangian $\mathcal{L}_L$ is described in Subsec.~\ref{sec:PLSM:Polyakov}. Finally, the quark Lagrangian $\mathcal{L}_q$, including the quark-meson and quark-gluon interaction terms, is discussed in Subsec.~\ref{sec:PLSM:Dirac}. Subsection~\ref{sec:PLSM:F} presents the computation of the free energy. The saddle-point equations giving the mean-field values of the meson fields and of the color-traced Polyakov loops $L$ and $L^*$ are discussed in Subsec.~\ref{sec:PLSM:saddle}. Subsection~\ref{sec:PLSM:phase} discusses the thermodynamics of the model, while its vanishing temperature limit is presented in Subsec.~\ref{sec:PLSM:T0}.

\subsection{Meson Lagrangian}\label{sec:PLSM:meson}

The mesonic Lagrangian $\mathcal{L}_{\mathcal{M}} = T_{\mathcal{M}} - V_{\mathcal{M}}$ is composed of a kinetic and a potential part,
\begin{gather}
 T_{\mathcal{M}} = \frac{1}{2} \left(\partial_\mu \sigma\partial^\mu \sigma + \partial_\mu \vec\pi \cdot \partial^\mu \vec\pi\right), \nonumber\\
 V_{\mathcal{M}} = \frac{\lambda}{4} \left(\sigma^2+\Vec{\pi}^2-v^2\right)^2 - h\sigma\,,
 \label{eq.LagMeson}
\end{gather}
where $\sigma$ and ${\vec \pi} = (\pi^1,\pi^2,\pi^3)$ are the scalar isosinglet and pseudoscalar isotriplet meson fields, respectively.

In this paper, we work in the mean-field approximation, by which the meson fields are replaced by their (classical) expectation values, thus ignoring their fluctuating (quantum) contributions. The mean-field values $\sigma$ and $\vec{\pi}$ minimize the grand potential, thereby ensuring the thermodynamic stability of the solution. In this paper, we also ignore space-time gradients of the meson fields, such that the kinetic part $T_{\mathcal{M}}$ becomes thermodynamically irrelevant. The effects of the spatial variations of the sigma field, albeit in a simpler model without the Polyakov loop, have been explored in the recent study of Ref.~\cite{Morales-Tejera:2025qvh}.
However, for simplicity, we restrict ourselves to the homogeneous approximation.

Denoting by $\sigma_0$ and $\vec{\pi}_0$ the mean-field vacuum expectation values of the mesons,  the minimization of $V_{\mathcal{M}}$ gives the following equations:
\begin{equation}
 \lambda \vec{\pi}_0 (\sigma_0^2 + \vec{\pi}^2_0 - v^2) = 0, \qquad
 \lambda \sigma_0(\sigma_0^2 + \vec{\pi}^2_0 - v^2) - h = 0\,.
 \label{eq_classical_pi}
\end{equation}
The symmetry-breaking term $h$ in the second equation leads to a vanishing pion condensate, $\vec{\pi}_0 = 0$, while the scalar condensate $\sigma_0$ is obtained as a solution of the following cubic equation:
\begin{equation}
 \lambda \sigma_0 (\sigma_0^2 - v^2) - h = 0.
\end{equation}
In principle, Eqs.~\eqref{eq_classical_pi} receive corrections from fermion loops, which renormalize the coupling constants. Such corrections can be treated self-consistently beyond the mean-field approximation~\cite{Herbst:2010rf}, leading to the same vacuum structure. For brevity, we do not discuss these details in the present paper.

\begin{table*}
\centering
\begin{tabular}{|c|c|c|c|}
\hline
\multicolumn{4}{|l|}{{\bf Input parameters}} \\\hline
$\sigma_0 = f_\pi$ &
$m_\pi = \sqrt{\lambda(\sigma_0^2 - v^2)}$ &
$m_\sigma = \sqrt{\lambda(3\sigma_0^2 - v^2)}$ &
$m_q = g \sigma_0$ \\\hline
$0.093$ GeV & $0.138$ GeV & $0.600$ GeV & $0.307$ GeV \\\hline\hline
\multicolumn{4}{|l|}{\bf Model parameters} \\\hline
$h = f_\pi m_\pi^2$ & $\lambda = (m_\sigma^2 - m_\pi^2) / (2f_\pi^2)$ &
$v^2 = f_\pi^2 - m_\pi^2 / \lambda$ &
$g = m_q / \sigma_0$ \\\hline
$0.00177$ (GeV)$^3$ &
$19.7$ & $0.00768$ (GeV)$^2$ &
$3.3$\\
\hline\hline
\end{tabular}
\label{table:3}
\caption{Parameters of the meson Lagrangian $\mathcal{L}_{\mathcal{M}}$ in Eq.~\eqref{eq.LagMeson}.
}
\label{tbl:mesons}
\end{table*}

The model parameters $\lambda$, $v$ and $h$ are fixed by the vacuum properties of the model, namely the expectation value $\sigma_0$ of the sigma meson and the masses of the $\sigma$ and $\vec{\pi}$ mesons:
\begin{gather}
    \sigma_0 = f_\pi\,, \qquad m_\sigma^2 =\frac{\partial^2 V_\mathcal{M}}{\partial \sigma^2} = \lambda(3\sigma^2_0 - v^2)\,,
    \nonumber\\
    m_\pi^2 =\frac{\partial^2 V_\mathcal{M}}{\partial \pi^2} = \lambda(\sigma^2_0 - v^2)\,,
    \label{eq_f_pi}
\end{gather}
where $f_\pi = 0.093$ GeV is the pion decay constant, while $m_\sigma = 0.600$ GeV is the mass of the sigma meson~\cite{ParticleDataGroup:2024cfk} and $m_\pi = 0.138$ GeV is the pion mass \cite{Daum:2019dav}. The latter two relations in Eq.~\eqref{eq_f_pi} fix the parameters $\lambda$ and $v$, while the explicit symmetry breaking term $h$ is determined by Eq.~\eqref{eq_classical_pi}, as follows:
\begin{equation}
 \lambda = \frac{m_\sigma^2 - m_\pi^2}{2f_\pi^2}, \quad
 v^2 = f_\pi^2 - \frac{m_\pi^2}{\lambda}\,, \quad
 h = f_\pi m_\pi^2.
\end{equation}
The values of the input and model parameters are summarized in Table~\ref{tbl:mesons}~.

\subsection{Polyakov loop Lagrangian}\label{sec:PLSM:Polyakov}

The Polyakov Lagrangian $\mathcal{L}_L$ models the (non-linear) thermal properties of the gluon sector of QCD and contains only the potential term,
\begin{align}
 \frac{\mathcal{L}_L}{T^4} &= -\frac{V_L}{T^4} = \frac{1}{2} a(T) L^* L-b(T)\ln \mathcal{H}, \nonumber\\
 \mathcal{H} &= 1-6L^* L+4\left(L^{* 3} +L^3\right)-3(L^* L)^2,
 \label{eq_L_Polyakov}
\end{align}
where $\mathcal{H}$ is the ${\rm SU}(3)$ Haar measure associated with the gluonic path integral, while $a(T)$ and $b(T)$ are temperature-dependent functions, parametrized as
\begin{align}
    a(T) =a_0+a_1\left(\frac{T_0}{T}\right)+a_2\left(\frac{T_0}{T}\right)^2,
    \,\,\ b(T) =b_3\left(\frac{T_0}{T}\right)^3.
    \label{eq_Polya_para}
\end{align}
The values of the constant parameters are 
\begin{equation}
 a_0 = 3.51, \,\,\,\,\,
 a_1 = -2.47, \,\,\,\,\,
 a_2 = 15.2, \,\,\,\,\,
 b_3 = -1.75.
 \label{eq:Polyakov}
\end{equation}
For simplicity, we set $T_0 = 0.27$ GeV, corresponding to the transition temperature of the pure-glue ${\rm SU}(3)$ theory. In Eq.~\eqref{eq_L_Polyakov}, we considered the logarithmic model for the Polyakov potential, which will be used in the main text of this work. We dedicate Appendix~\ref{app:poly} to a discussion of other models for the Polyakov potential, as well as of different values and models for $T_0$.

The degrees of freedom appearing in $\mathcal{L}_L$ are the color-traced Polyakov loops, $L$ and $L^*$ (for brevity we refer to $L$ and $L^*$ as Polyakov loops in the rest of the paper), defined as:
\begin{equation}
 L = \frac{1}{3} {\rm Tr}_c\, \Phi(x), \quad
 \Phi(x) = \mathcal{P} \exp\biggl[-i\,\int_0^{\beta} d\tau \mathcal{A}_4({\boldsymbol{x}},\tau)\biggr],
\end{equation}
where $\mathcal{A}_4 \equiv \mathcal{A}_\tau = i \mathcal{A}_0$ is the temporal component of the gluon field, expressed with respect to the imaginary time. In the mean-field approximation, $\mathcal{A}_4$ is a constant, classical field. In the Polyakov gauge \cite{tHooft:1981bkw}, $\mathcal{A}_4 = \frac{1}{2} \mathcal{A}_4^{(3)} \lambda_3 + \frac{1}{2} \mathcal{A}_4^{(8)} \lambda_8$ is diagonal, with $\lambda_3 = {\rm diag}(1,-1,0)$ and $\lambda_8 = \frac{1}{\sqrt{3}}{\rm diag}(1, 1, -2)$ being the diagonal Gell-Mann matrices representing the Abelian (Cartan) subalgebra of the ${\rm SU}(3)$ Lie algebra. The diagonal entries of $\mathcal{A}_4$, corresponding to its eigenvalues, can be taken as
\begin{equation}
 \mathcal{A}_4 = {\rm diag}(\phi, \phi', -\phi - \phi'),
 \label{eq:A4_diag}
\end{equation}
where the scalar parameters $\phi$ and $\phi'$ are related to the vector components of the gluonic fields $\mathcal{A}^{(3)}_4$ and $\mathcal{A}^{(8)}_4$ via
\begin{gather}
 \phi = \frac{1}{2} \mathcal{A}^{(3)}_4 + \frac{1}{2\sqrt{3}} \mathcal{A}^{(8)}_4, \quad
 \phi' = -\frac{1}{2} \mathcal{A}^{(3)}_4 + \frac{1}{2\sqrt{3}} \mathcal{A}^{(8)}_4, \nonumber \\
 -\phi-\phi' = -\frac{1}{\sqrt{3}} \mathcal{A}^{(8)}_4.
\end{gather}

With the above notations, the Polyakov loops $L$ and $L^*$ are
\begin{gather}
  L = \frac{1}{3} (e^{-i\beta \phi} + e^{-i\beta \phi'} + e^{i\beta(\phi + \phi')}), \nonumber\\
  L^* = \frac{1}{3} (e^{i\beta \phi} + e^{i\beta \phi'} + e^{-i\beta(\phi + \phi')}).
 \label{eq_Polyakov_loop}
\end{gather}
In the effective model description, $L$ and $L^*$ are treated as two independent, real fields, chosen by imposing the minimization of the grand potential. In the absence of the quark degrees of freedom, $L$ and $L^*$ satisfy the following equations:
\begin{subequations}\label{eq_Polyakov_minimization}
\begin{align}
 \frac{1}{T^4} \frac{\partial V_L}{\partial L} &= -\frac{1}{2} a L^* - \frac{6b}{\mathcal{H}} (L^* - 2L^2 + L L^*{}^2) = 0, \\
 \frac{1}{T^4} \frac{\partial V_L}{\partial L^*} &= -\frac{1}{2} a L - \frac{6b}{\mathcal{H}} (L - 2L^*{}^2 + L^* L^2) = 0.
\end{align}
\end{subequations}
Imposing $L = L^*$, the above equations allow for three solutions: $L = 0$ and $L = L^{(\pm)}$, with
\begin{equation}
 L^{(\pm)} = \frac{1}{3} \pm \frac{2}{3} \sqrt{1 + \frac{9 b(T)}{a(T)}}.
 \label{eq:logarithmic_Lpm}
\end{equation}
In general, $L = 0$ and $L = L^{(+)}$ correspond to the minima of the effective potential, while $L = L^{(-)}$ gives a local maximum. At small $T$, the global minimum is attained for $L = 0$, corresponding to the confined phase, while $L^{(\pm)}$ are complex-valued quantities, corresponding to the deconfinement phase. When $T \to \infty$, the system becomes deconfined and $L,L^*$ asymptotically approach the value $1$, as follows:
\begin{equation}
 L, L^* \to 1 + \frac{3b_3}{a_0} \left(\frac{T_0}{T}\right)^3 + O(T^{-4})\ \qquad [{\rm as}\ T \to \infty].
\end{equation}
Imposing the Stefan-Boltzmann limit for the gluon pressure, $P_g = -V_L = \mathcal{L}_L$, fixes the $a_0$ parameter:
\begin{equation}
 \lim_{T \rightarrow \infty} P_g = \frac{1}{2} a_0 T^4 = \nu_g \frac{\pi^2 T^4}{90},
\end{equation}
where $\nu_g = 2(N_c^2 - 1) = 16$ is the gluon degeneracy factor.

The model also consistently imposes in the pure-glue ${\rm SU}(3)$ gauge theory a first-order phase transition when $T = T_0$. This property is achieved by demanding that $V_L = 0$ when $T = T_0$ and $L = L^{(+)}$, such that
\begin{gather}
 \ln[(1 - L^{(+)})^3(1+3L^{(+)})] = \frac{a_c (L^{(+)})^2}{2b_3}, \nonumber\\
 a_c = a(T_0) = a_0 + a_1 + a_2.
\end{gather}
Using the relation $L^{(+)} = \frac{1}{3} + \frac{2}{3} \sqrt{1 + 9b_3 / a_c}$, we get
$b_3 = -0.107817 (a_0 + a_1 + a_2)$. This fixes $b_3$ once $a_1$ and $a_2$ are known. These latter two parameters are found by fitting $P_g$ to the lattice data. At the first-order phase transition, the Polyakov loop jumps to the value $L^{(+)}_c \simeq 0.448$.

\subsection{Quark Lagrangian} \label{sec:PLSM:Dirac}

The quark Lagrangian $\mathcal{L}_q$ reads
\begin{align}
    \mathcal{L}_q =
    \overline{\psi}\left[i\slashed{D} + \mu \gamma^0 - g(\sigma +i\gamma_5 \Vec{\tau}\cdot \Vec{\pi})\right]\psi,
    \label{eq_L_quark}
\end{align}
where the Feynman slash denotes contraction with the Dirac matrices, $\slashed{D} = \gamma^\mu D_\mu$, $D_\mu = \partial_\mu - i \mathcal{A}_\mu$ is the covariant derivative, $\mathcal{A}_\mu = \frac{1}{2} A_\mu^a \lambda_a$ is the gluon potential and $\mu$ is the quark chemical potential. The strength $g$ of the coupling to the meson fields is determined by fixing the effective mass of the fermions in the vacuum to $M_q = 0.307$ GeV:
\begin{equation}
 g = \frac{M_q}{\sigma_0} = \frac{M_q}{f_\pi} \simeq 3.3,
\end{equation}
as shown in Table~\ref{tbl:mesons}.

This work considers the $N_f = 2$ flavor model, such that $\psi = (u, d)^T$. Each of the components of the field $\psi$ is a spinor that also carries an ${\rm SU}(3)$ color index $c = 1,2,3$ (not explicitly shown).
The gamma matrices $\gamma^\mu = (\gamma^0, \boldsymbol{\gamma})$ are taken in the Dirac-Pauli representation,
\begin{align}
    \gamma^0 = \begin{pmatrix}
        I_2 & 0 \\ 0 & -I_2
    \end{pmatrix}, \qquad
    \boldsymbol{\gamma} = \begin{pmatrix}
        0 & \boldsymbol{\sigma} \\
        -\boldsymbol{\sigma} & 0
    \end{pmatrix},
    \label{eq:gamma}
\end{align}
where $\boldsymbol{\sigma}$ are the Pauli matrices,
\begin{equation}
 \sigma^1 = \begin{pmatrix}
  0 & 1 \\ 1 & 0
 \end{pmatrix}, \quad
 \sigma^2 = \begin{pmatrix}
  0 & -i \\ i & 0
 \end{pmatrix}, \quad
 \sigma^3 = \begin{pmatrix}
  1 & 0 \\ 0 & -1
 \end{pmatrix}.
 \label{eq:sigma}
\end{equation}

\subsection{Grand canonical potential} \label{sec:PLSM:F}

The partition function of PLSM${}_q$ reads
\begin{subequations}
\label{eq.part}
\begin{equation}
    \mathcal{Z} =
    e^{-\beta V (V_{\mathcal{M}} + V_L)}\mathcal{Z}_q,\,\,\,\,\,
    \mathcal{Z}_q = \int [i d \psi^\dagger] [d\psi] e^{-S_E}, 
\end{equation}
where the Euclidean action $S_E = \int_0^\beta d\tau \int_V d^3x \,\mathcal{L}_E$ is written in terms of the Euclidean Lagrangian, $\mathcal{L}_E(\tau, \mathbf{x}) = -\mathcal{L}_q(it, \mathbf{x})$:
\begin{equation}
    \mathcal{L}_E = -\overline{\psi} \left[\gamma^0(-\partial_\tau - i \mathcal{A}_4 + \mu) + i \boldsymbol{\gamma} \cdot \boldsymbol{\nabla} - g \sigma \right] \psi\,,    
\end{equation}
\end{subequations}
where we omitted an irrelevant overall normalization factor. The kinetic parts of the meson and gluon sectors were neglected, $V$ is the spatial volume, while the potentials $V_{\mathcal{M}}$ and $V_L$ are given in Eqs.~\eqref{eq.LagMeson} and \eqref{eq_L_Polyakov},
respectively. The volume-averaged grand canonical potential of the model, $F = -(T \ln \mathcal{Z}) / V$, is given by
\begin{align}
    F = V_\mathcal{M} + V_L + F_q\,, \qquad
    F_q = -\frac{T}{V} \ln \mathcal{Z}_q.
    \label{eq_F}
\end{align}
Note that $F$ has units of [energy]/[length]$^3$.

The fermionic part $\mathcal{Z}_q$ of the partition function can be evaluated by considering a decomposition $\psi(X) = \sum_j \hat{b}_j \psi_j(X)$ of the Dirac spinor with respect to the plane-wave basis $\psi_j(X)$, where $X_\mu = (\tau, \mathbf{x})$ denotes the space-time coordinates on the Euclidean manifold. The basis modes $\psi_j(X)$ satisfy the following eigenvalue equations:
\begin{equation}
 (-i \partial_\tau, \sigma^3_f, \mathcal{A}_4, h, \mathbf{P}) \psi_j = (w_j, f_j, A_{4;j}, \lambda_j, \mathbf{p}_j) \psi_j,
 \label{eq:static_eigen}
\end{equation}
with $w_j = 2\pi T (n_j + \frac{1}{2})$ the Matsubara frequencies; the flavor $f_j = \pm 1$ for the $u$/$d$ quarks; $A_{4;j} \equiv A_{4;c_j}$ the eigenvalues of the gluon field shown in Eq.~\eqref{eq:A4_diag}, namely $(A_{4;1}, A_{4;2}, A_{4;3}) = (\phi, \phi', -\phi - \phi')$; $\lambda_j = \pm 1/2$ the eigenvalues of the helicity operator $h = \mathbf{S} \cdot \mathbf{P} / |\mathbf{P}|$; and $\mathbf{P} = -i \boldsymbol{\nabla}$ the momentum operator.
In addition, we require $\psi_j$ to be eigenmodes of the Hamiltonian operator, $H = i \mathcal{A}_4 - \mu + 2 \gamma^5 \mathbf{S} \cdot \mathbf{P} + g \sigma \gamma^0$:
\begin{equation}
 H \psi_j = \varsigma_j \mathcal{E}_j \psi_j, \quad
 \mathcal{E}_j = E_j - \varsigma_j(\mu - i A_{4;j}),
 \label{eq:static_H}
\end{equation}
where $\varsigma_j = \pm 1$ distinguishes between the particle and anti-particle branches and $E_j = \sqrt{\mathbf{p}_j^2 + g^2 \sigma^2}$ is the relativistic particle energy.
In the above discussions, we used $j \equiv (\varsigma_j, w_j, f_j, c_j, \lambda_j, \mathbf{p}_j)$ to collectively denote the eigenvalues characterizing the mode $\psi_j$.

The modes described above have the following form:
\begin{multline}
 \psi_j(X) = \frac{e^{i w_j \tau + i \mathbf{p}_j \cdot \mathbf{x}}}{(2\pi)^{3/2}} \begin{pmatrix}
  \frac{1}{2}(1 + f_j) \\[2mm]
  \frac{1}{2}(1 - f_j)
 \end{pmatrix}\otimes
 \begin{pmatrix}
  \delta_{c_j, 1} \\
  \delta_{c_j, 2} \\
  \delta_{c_j, 3}
 \end{pmatrix}\\
 \otimes \frac{1}{\sqrt{2}}
 \begin{pmatrix}
  \mathfrak{E}^{\varsigma_j}_j \\
  2\varsigma_j \lambda_j \mathfrak{E}^{-\varsigma_j}_j
 \end{pmatrix} \otimes \xi_j,
 \label{eq:static_modes}
\end{multline}
where $\mathfrak{E}^\pm_j = \sqrt{1 \pm g \sigma / E_j}$ and the two-spinors $\xi_j \equiv \xi_{\lambda_j}(\mathbf{p}_j)$ satisfying $\boldsymbol{\sigma} \cdot \mathbf{p} \xi_\lambda(\mathbf{p}) = 2\lambda p \xi_\lambda(\mathbf{p})$ given by
\begin{align}
 \xi_{1/2}(\mathbf{p}) &= \begin{pmatrix}
  e^{-\frac{i}{2} \varphi_p} \cos\frac{\theta_p}{2} \\
  e^{\frac{i}{2} \varphi_p} \sin\frac{\theta_p}{2} \\
 \end{pmatrix}, \nonumber\\
 \xi_{-1/2}(\mathbf{p}) &= \begin{pmatrix}
 -e^{-\frac{i}{2} \varphi_p} \sin\frac{\theta_p}{2} \\
  e^{\frac{i}{2} \varphi_p} \cos\frac{\theta_p}{2}
 \end{pmatrix},
 \label{eq:static_xi}
\end{align}
with $(\theta_p, \varphi_p)$ being the angular coordinates of the momentum vector $\mathbf{p}$.
It can be easily checked that the modes are orthonormal:
\begin{equation}
 \int_0^\beta d\tau \int d^3x \, \psi_j^\dagger \psi_{j'} = \beta \delta(j,j'), 
 \label{eq:static_ortho}
\end{equation}
where $\delta(j,j') = \delta_{\varsigma_j \varsigma_{j'}} \delta_{n_j n_{j'}} \delta_{f_j f_{j'}} \delta_{c_j c_{j'}} \delta_{\lambda_j \lambda_{j'}} \delta^3(\mathbf{p}_j - \mathbf{p}_{j'})$.

Considering now that the spatial volume $V$ is a fictitious rectangular box and the momentum modes are discretized via $p_i = 2 \pi n_i / L_i$, with $n_i$ being integer numbers and $L_i$ being the length of side $i$, the fermionic partition function can be written as
\begin{align}
 \mathcal{Z}_q & = \int \prod_j \left(id\hat{b}^\dagger_j d\hat{b}_j \right) e^{-S_E}, 
 \nonumber \\
 S_E & = -\sum_{j,j'} i \hat{b}^\dagger_j D(j, j') \hat{b}_{j'},
 \label{eq:static_Zq_aux}
\end{align}
with $D(j,j')$ being a diagonal matrix:
\begin{equation}
 D(j,j') = \beta (- w_j + i \varsigma_j \mathcal{E}_j) \delta(j,j').
 \label{eq:static_D}
\end{equation}
Using the integration formula $\int \prod_j [id\hat{b}^\dagger_j d\hat{b}_j]\, e^{i\hat{b}^\dagger \cdot D \cdot \hat{b}} = \det D$ over the Grassmann variables $i\hat{b}^\dagger_j$ and $\hat{b}_j$, the partition function becomes simply
\begin{align}
 \mathcal{Z}_q & \, = \prod_j \left[\int i d\hat{b}^\dagger_j d\hat{b}_j e^{i \hat{b}^\dagger_j [\beta(i\varsigma_j \mathcal{E}_j - w_j)] \hat{b}_j}\right] \nonumber\\
 & = \prod_j [\beta(i \varsigma_j \mathcal{E}_j - w_j)].
 \label{eq:static_Zq}
\end{align}

The fermion contribution to the volume-averaged grand potential, $F_q = -\frac{T}{V} \ln \mathcal{Z}_q$, becomes
\begin{equation}
 F_q = -2 N_f T \sum_{n = -\infty}^\infty \sum_{c = 1}^3 \sum_{\varsigma = \pm 1} \int\frac{d^3p}{(2\pi)^3} \ln[\beta(i \varsigma \mathcal{E}^c_\varsigma - w_n)],
 \label{eq:static_Fq_aux}
\end{equation}
where we introduced the notation
\begin{equation}
 \mathcal{E}^c_\varsigma = \mathcal{E}_\varsigma + i \varsigma A_{4;c}, \,\,\,
 \mathcal{E}_\varsigma = E - \varsigma \mu, \,\,\,
 E = \sqrt{\mathbf{p}^2 + g^2 \sigma^2}.
 \label{eq:static_Ecs}
\end{equation}
Averaging over positive and negative $n$ leads to
\begin{equation}
 F_q = -N_f T \hspace{-5pt} \sum_{n = -\infty}^\infty \sum_{c = 1}^3 \sum_{\varsigma = \pm 1} \int\frac{d^3p}{(2\pi)^3} \ln[\beta^2 w_n^2 + (\beta \mathcal{E}^c_\varsigma)^2].
 \label{eq:static_Fq_aux2}
\end{equation}
The summation over $n$ can be performed using the method described in Ref.~\cite{kapusta_gale_2006}, leading to:
\begin{multline}
\frac{1}{2} \sum_{n = -\infty}^\infty \ln [(2n+1)^2 \pi^2+ (\beta \mathcal{E}^c_\varsigma)^2] = \frac{1}{2} \beta \mathcal{E}^c_\varsigma + \ln(1 + e^{-\beta \mathcal{E}^c_\varsigma})\\ 
+ \mathcal{N},
\end{multline}
where $\mathcal{N} = \frac{1}{2} - \ln(1 + e) + \frac{1}{2} \sum_{n = -\infty}^\infty 1$ is an (infinite) constant which is independent of the thermodynamic parameters of the state $(T,\mu,V)$ and can therefore be discarded. The two terms above give rise to the zero-point (vacuum) energy and the thermal contribution, such that $F_q = F^{\rm vac}_q + F^{\beta}_q$, with \cite{Scavenius:2000qd, Ratti:2007jf}:
\begin{align}
 F^{\rm vac}_q &= -2N_f N_c \int \frac{d^3p}{(2\pi)^3} E,\nonumber
 \\
 F^\beta_q &= -2N_fT \sum_{\varsigma = \pm 1} \int \frac{d^3p}{(2\pi)^3} \sum_c \ln \left(1+\e^{-\beta \mathcal{E}^c_\varsigma}\right).
 \label{eq:static_F}
\end{align}
In the vacuum term, we took into account that $\sum_{\varsigma, c} \mathcal{E}^c_\varsigma = 2N_c E$, with $N_c=3$ being the number of colors, since $\sum_c A_{4;c} = \phi + \phi' - (\phi + \phi') = 0$. In the linear sigma model, the (regularized) vacuum contribution can be employed to renormalize the parameters of the meson Lagrangian and can therefore be safely neglected. In this work, we follow Ref.~\cite{Scavenius:2000qd} and ignore such vacuum contributions, therefore tacitly assuming that $F^{\rm vac}_q = 0$. Since now $F_q = F^{\beta}_q$, we will omit the $\beta$ superscript and instead identify the fermionic contribution to the volume-averaged grand potential with its thermal contribution.

The sum over the colour index can be performed as follows:
\begin{align}
F_\varsigma &\equiv \frac{1}{3} \sum_{c = 1}^3 \ln(1 + e^{-\beta \mathcal{E}^c_\varsigma}) \nonumber\\
&= 
\frac{1}{3} \ln (1 + 3L_\varsigma e^{-\beta \mathcal{E}_\varsigma} + 3L_{-\varsigma} e^{-2\beta \mathcal{E}_\varsigma} + e^{-3\beta\mathcal{E}_\varsigma}),
 \label{eq:static_Fs}
\end{align}
where we introduced the notation $L_\pm = \frac{1}{3} \sum_c e^{\mp i \beta A_{4;c}} = \frac{1}{3}(e^{\mp i \beta \phi} + e^{\mp i \beta \phi'} + e^{\pm i \beta(\phi + \phi')})$, such that $L_+ \equiv L$ and $L_- \equiv L^*$ [see Eq.~\eqref{eq_Polyakov_loop}]. Finally, the fermion free energy becomes
\begin{align}
 F_q &=-2 N_f N_c T \sum_{\varsigma = \pm 1} \int \frac{d^3p}{(2\pi)^3} F_\varsigma.
 \label{eq:static_Fq}
\end{align}

\subsection{saddle-point approximation}\label{sec:PLSM:saddle}

The thermodynamic and phase space structure of the model are guided by the mean-field values: $\Phi=\Phi_{\rm mf}$, $\sigma=\sigma_{\rm mf}$ and $\vec{\pi}=\vec{\pi}_{\rm mf} = 0$. We find these mean-field values by imposing the minimization of the total free energy at thermodynamic equilibrium by solving the following equations:
\begin{align}
    \frac{\partial F}{\partial \sigma} =
    \frac{\partial F}{\partial L} = \frac{\partial F}{\partial L^*} =0,
    \label{eq_saddle}
\end{align}
where we remember that within the scope of the Polyakov loop-enhanced models, the two conjugate fields $L$ and $L^*$ are treated as real but independent numbers.

The derivatives appearing in Eq.~\eqref{eq_saddle} can be expressed with the help of several distribution functions: 
$f_\pm \equiv -\partial F_\pm / \partial (\beta \mathcal{E}_\pm)$, 
$f_{\pm,\pm} \equiv \partial F_\pm / \partial L_\pm$, and 
$f_{\pm,\mp} \equiv \partial F_\pm / \partial L_\mp$. Explicitly,
\begin{subequations}\label{eq_fall}
\begin{align}
 f_\pm &= \frac{L_\pm e^{-\beta \mathcal{E}_\pm} + 2 L_\mp e^{-2\beta\mathcal{E}_\pm} + e^{-3\beta \mathcal{E}_\pm}}{1 + 3 L_\pm e^{-\beta \mathcal{E}_\pm} + 3L_\mp e^{-2 \beta \mathcal{E}_\pm} + e^{-3 \beta \mathcal{E}_\pm}}, \label{eq_f}\\
 f_{\pm,\pm} &= \frac{e^{-\beta \mathcal{E}_\pm}}{1 + 3 L_\pm e^{-\beta \mathcal{E}_\pm} + 3L_\mp e^{-2 \beta \mathcal{E}_\pm} + e^{-3 \beta \mathcal{E}_\pm}}, \label{eq_fL}\\
 f_{\pm,\mp} &= \frac{e^{-2\beta \mathcal{E}_\pm}}{1 + 3 L_\pm e^{-\beta \mathcal{E}_\pm} + 3L_\mp e^{-2 \beta \mathcal{E}_\pm} + e^{-3 \beta \mathcal{E}_\pm}},
 \label{eq_fLb}
\end{align}
\end{subequations}
where the upper (lower) sign corresponds to the (anti)particle sector.
With the above definitions in hand, we obtain
\begin{align}
 \frac{\partial F}{\partial \sigma} &= \frac{\partial V_\mathcal{M}}{\partial \sigma} + g \langle \Bar{\psi} \psi \rangle\,, \nonumber\\
 \frac{\partial F}{\partial L_\pm} &= \frac{\partial V_L}{\partial L_\pm} - 2 N_f N_c T \sum_{\varsigma = \pm 1} \int \frac{d^3p}{(2\pi)^3} f_{\varsigma, \pm}\,,
 \label{eq_dFdp}
\end{align}
where we identified the fermion condensate in the first equation,
\begin{equation}
 \langle \Bar{\psi} \psi \rangle = 2 N_f N_c g \sigma \sum_{\varsigma = \pm 1} \int \frac{d^3p}{(2\pi)^3 E} f_\varsigma\,,
 \label{eq_ppsi}
\end{equation}
while the derivatives of the mesonic and of the Polyakov loop potentials are given by
\begin{subequations}\label{eq_dVdp}
\begin{align}
 \frac{\partial V_\mathcal{M}}{\partial \sigma} &= \lambda(\sigma^2 - v^2) \sigma - h, \label{eq_dVds}\\
 \frac{\partial V_L}{\partial L_\pm} &= - \frac{6 T^4 b(T)}{\mathcal{H}} (L_\mp - 2 L_\pm^2 + L_\mp^2 L_\pm) - \frac{T^4}{2} a(T) L_\mp. \label{eq_dVdL}
\end{align}
\end{subequations}
where $\mathcal{H} = 1 - 6 L^* L + 4 (L^*{}^3 + L^3) - 3(L^* L)^2$ was introduced in Eq.~\eqref{eq_L_Polyakov}.

As the terms coming from Eqs.~\eqref{eq_dVdp} involve cubic powers of the order parameters, the saddle-point equations \eqref{eq_saddle} will typically give multiple solutions. As a rule of thumb, the physically acceptable solutions must satisfy the constraints:
\begin{equation}
 0 < \sigma \le f_\pi, \qquad\
 0 \le L, L^* < 1\,.
 \label{eq_constraints}
\end{equation}
The first constraint arises on physical grounds, as we expect that both the thermal fluctuations and the finite-density environment restore the chiral symmetry ($\sigma \to 0$) instead of enhancing its breaking ($\sigma \to f_\pi$). Similarly, in the vacuum, the Polyakov loop expectation value vanishes. As the medium becomes more energetic (denser, at high $\mu$; or hotter, at large $T$), the $\sigma$ condensate melts and the system becomes deconfined, i.e.~$L, L^* \rightarrow 1$. Note that $L$ and $L^*$ cannot exceed $1$ as a result of the mathematical definition of the Polyakov loop~\eqref{eq_Polyakov_loop}. Values outside the intervals indicated in Eq.~\eqref{eq_constraints} cannot represent physically meaningful systems.

When Eqs.~\eqref{eq_saddle} allow for multiple solutions satisfying Eqs.~\eqref{eq_constraints}, one must distinguish between unstable and (meta)stable solutions. This selection can be done at the level of the volume-averaged grand potential, $F = V_{\mathcal{M}} + V_L + F_q$, where the fermionic contribution $F_q = -P_q$ can be written in terms of the pressure $P_q = -\frac{i}{6} \Bar{\psi} \boldsymbol{\gamma} \cdot \overleftrightarrow{\boldsymbol{\nabla}} \psi$,
\begin{align}
 P_q &= 2 N_f N_c \sum_{\varsigma = \pm 1} \int \frac{d^3p}{(2\pi)^3} \frac{p^2}{3E} f_\varsigma, \nonumber\\
 \epsilon_q &= 2 N_f N_c \sum_{\varsigma = \pm 1} \int \frac{d^3p}{(2\pi)^3} E f_\varsigma,
 \label{eq_P}
\end{align}
where we also listed the expression for the energy density, for later convenience.
The unstable solutions correspond to points of local maxima or saddle points. Points of local (but not lowest) minima correspond to metastable states. The global (lowest-free-energy) minimum represents the true stable ground state of the system. Under quantum and thermal fluctuations, it is reasonable to expect that the quantum system will always find its way from any of the metastable states to the true ground state.
Typical examples of the dependence of the total grand potential $F$ on $\sigma$ can be found in Fig.~1 of Ref.~\cite{Singha:2024tpo}.

In order to find the solutions of the saddle-point equations, we write Eqs.~\eqref{eq_saddle} in a compact form as
\begin{equation}
 \frac{\partial F}{\partial p_i} = 0\,, \qquad
 p_i \in \{\sigma, L, L^*\}.
 \label{eq_saddle_p}
\end{equation}
Equation~\eqref{eq_saddle_p} is satisfied when the control parameters take their mean-field values, $p_i \rightarrow p_i^{\rm mf}$. Considering now some values $\tilde{p}_i$ of the control parameters, which are in a small vicinity of their mean-field values, we can set up the Newton-Raphson algorithm by expanding $(\partial F / \partial p_i)$ in a Taylor series around the mean-field set of values $p_i^{\rm mf}$:
\begin{equation}
 \frac{\partial F}{\partial p_i} {\biggl|}_{p = p_{\rm mf}} =
 \frac{\partial F}{\partial p_i}{\biggl|}_{p = \tilde{p}} +
 \frac{\partial^2 F}{\partial p_i \partial p_j}{\biggl|}_{p = \tilde{p}} (p_j^{\rm mf} - \tilde{p}_j) + \dots\,.
\end{equation}
We can find $p_i^{\rm mf}$ by solving
\begin{equation}
 \mathcal{M}_{ij}(\tilde{p})(p_j^{\rm mf} - \tilde{p}_j) = -\frac{\partial F}{\partial \tilde{p}_i}, \,\,
 \mathcal{M}_{ij}(\tilde{p}) = \frac{\partial^2 F}{\partial p_i \partial p_j}{\biggl|}_{p = \tilde{p}}\,,
 \label{eq_NR}
\end{equation}
where $\mathcal{M}_{ij}$ is the $3 \times 3$ Jacobian (or Hessian) matrix, also discussed in Ref.~\cite{Haensch:2023sig}. We split this matrix into the potential and fermionic parts, $\mathcal{M}_{ij} = \mathcal{M}^V_{ij} + \mathcal{M}^{q}_{ij}$, where
\begin{gather}
 \mathcal{M}^V_{ij} = \frac{\partial^2(V_\mathcal{M} + V_L)}{\partial p_i \partial p_j}\,, \quad 
 \mathcal{M}^q_{ij} = \frac{\partial^2 F_{q}}{\partial p_i \partial p_j}\,.
\end{gather}
The contributions $\mathcal{M}^V_{\sigma\sigma}$ and $\mathcal{M}^V_{\sigma\pm}$ are simply
\begin{equation}
 \mathcal{M}^{V}_{\sigma\sigma} = \lambda(3\sigma^2 - v^2)\,, \quad
 \mathcal{M}^{V}_{\sigma \pm} = \mathcal{M}^{V}_{\pm \sigma} = 0\,.
 \label{eq:unb_MV}
\end{equation}
The double derivatives with respect to $L$ and $L^*$ are:
\begin{align} \label{eq.d2vdp2}
 \mathcal{M}^{V}_{\pm \pm} &= \frac{6 T^4 b(T)}{\mathcal{H}} (4L_\pm - L_\mp^2) \nonumber\\
 & -\frac{1}{T^4 b(T)}  \left(\frac{\partial V_L}{\partial L_\pm} + \frac{T^4}{2} a(T) L_\mp\right)^2\,,\nonumber\\
 \mathcal{M}^{V}_{\pm \mp} &= -\frac{T^4}{2} a(T) - \frac{6 T^4 b(T)}{\mathcal{H}^2} 
 [1 + 2 L_+ L_- + 21 L_+^2 L_-^2 \nonumber\\
 &- 4(2 + L_+ L_-) (L_+^3 + L_-^3)]\,.
\end{align}
The elements of the fermionic contribution $\mathcal{M}^q_{ij}$ read
\begin{align}
 \mathcal{M}^q_{\sigma\sigma} &= 2 N_f N_c g^2 \sum_{\varsigma = \pm 1} \int \frac{d^3p}{(2\pi)^3 E} \left(\frac{\mathbf{p}^2}{E^2} f_\varsigma +
 \frac{g^2 \sigma^2}{E T} f'_\varsigma\right)\,,\nonumber\\
 \mathcal{M}^q_{\sigma \pm} &= 2 N_f N_c g^2 \sigma \sum_{\varsigma = \pm 1} \int \frac{d^3p}{(2\pi)^3 E} f_{\varsigma,\pm} (1 + \delta_{\varsigma,\mp} - 3 f_\varsigma)\,, \nonumber\\
 \mathcal{M}^q_{\pm\pm} &= 2 N_f N_c^2 T \sum_{\varsigma = \pm 1} \int \frac{d^3p}{(2\pi)^3} f^2_{\varsigma,\pm}\,, \nonumber\\
 \mathcal{M}^q_{+-} &= 2 N_f N_c^2 T \sum_{\varsigma = \pm 1} \int \frac{d^3p}{(2\pi)^3} f_{\varsigma,+} f_{\varsigma,-}\,,
 \label{eq:unb_Mij}
\end{align}
where $\delta_{\varsigma,\mp}$ is the Kronecker symbol and
\begin{equation}
 f'_\varsigma \equiv \frac{\partial f_\varsigma}{\partial (\beta \mathcal{E}_\varsigma)} = -3f_\varsigma(1 - f_\varsigma) + 2 L_+ f_{\varsigma,+} + 2 L_{-} f_{\varsigma,-}\,.
 \label{eq_df}
\end{equation}
The expressions in Eqs.~\eqref{eq.d2vdp2} and \eqref{eq:unb_Mij} are in agreement with similar ones in Ref.~\cite{Haensch:2023sig}.

While the solution of Eq.~\eqref{eq_NR} can be obtained by a straightforward inversion of the $3 \times 3$ matrix $\mathcal{M}_{ij}$, using the standard Newton iteration $p_i^{(n+1)} = p_i^{(n)} - [\mathcal{M}^{(n)}]^{-1}_{ij} \frac{\partial F}{\partial p^{(n)}_j}$, the stability of this method is guaranteed only when the initial guess is (sufficiently) close to the convergence, mean-field values $p_i^{\rm mf}$. In our numerical code, we instead employ the Broyden method \cite{Broyden:1965,NumericalRecipes}, which adjusts the numerical step implied by Eq.~\eqref{eq_NR} to increase the overall stability. Another advantage of this routine is that it provides a numerical approximation to the Jacobian matrix $\mathcal{M}_{ij}$ for the cases where obtaining the Jacobian analytically becomes cumbersome.

In our numerical analysis, we employed the Jacobian in a two-fold way. Firstly, we used the analytically calculated Jacobian to obtain the control parameters as functions of temperature and chemical potential (and also as functions of angular velocity for the bounded case considered in later sections). Secondly, we used the numerical Jacobian to obtain the phase diagrams.

\subsection{Thermodynamics of the model}
\label{sec:PLSM:phase}

In the absence of a thermodynamic discontinuity, the system goes continuously from one phase to the other, undergoing a so-called crossover phase transition. Pinpointing an exact position of the pseudo-critical parameters of the crossover transition has a certain degree of arbitrariness, which depends on the choice of the observables used to identify the crossover. In this paper, we employ the phase-space gradients of the control parameters, defined as
\begin{equation}
 dp_{\rm mf} = \left(\frac{\partial p}{\partial T}\right)_{\rm mf} dT + \left(\frac{\partial p}{\partial \mu}\right)_{\rm mf} d\mu,
 \label{eq.dpmf2var}
\end{equation}
where $p \in \{\sigma, L, L^*\}$ is one of the control parameters and the ``${\rm mf}$'' subscript reminds that the variations are taken along the mean-field, thermodynamic trajectories of the control parameters imposed by the saddle-point equations~\eqref{eq_saddle_p}. We compute these derivatives following Ref.~\cite{Chernodub:2020yaf} by noting that due to the constraint in Eq.~\eqref{eq_saddle_p}, the free energy $F$ becomes an implicit function only of the thermodynamic parameters $T$ and $\mu$ since the control parameters $p_i \rightarrow p_i^{\rm mf} \equiv p_i^{\rm mf}(T, \mu)$ are replaced by their mean-field values. Considering now that $F(T, \mu) \rightarrow F(T, \mu; p_i)$ is an explicit function of the thermodynamic parameters $(T,\mu)$ and of the control parameters $p_i$, we can take the derivative of Eq.~\eqref{eq_saddle_p} with respect to a thermodynamic parameter $G_a$ (say, $G_1 = T$ and $G_2 = \mu$), leading to
\begin{equation}
 \mathcal{M}_{ij} \frac{\partial p_j}{\partial G_a} + \frac{\partial^2 F}{\partial G_a \partial p_i} = 0,
\end{equation}
where the Jacobian matrix $\mathcal{M}_{ij}$ was introduced in Eq.~\eqref{eq_NR}.
Therefore, the derivatives of the control parameters $p_i$ with respect to the thermodynamic parameters $G_a$ can be obtained by inverting the $\mathcal{M}_{ij}$ matrix,
\begin{equation}
 \frac{\partial p_i}{\partial G_a} = -\mathcal{M}_{ij}^{-1} \frac{\partial^2 F}{\partial G_a \partial p_j}.
 \label{eq:dpdA}
\end{equation}
The terms appearing on the right-hand side of the above equation can be computed explicitly starting from Eq.~\eqref{eq_dFdp}:
\begin{align}\label{eq:d2FdAdp}
 \frac{\partial^2 F}{\partial T \partial \sigma} &= -\frac{2 N_f N_c g^2 \sigma}{T^2} \sum_{\varsigma = \pm 1} \int \frac{d^3p}{(2\pi)^3 E} \mathcal{E}_\varsigma f'_\varsigma \,,\nonumber\\
 \frac{\partial^2 F}{\partial \mu \partial \sigma} &= -\frac{2 N_f N_c g^2 \sigma}{T} \sum_{\varsigma = \pm 1} \varsigma \int \frac{d^3p}{(2\pi)^3 E} f_\varsigma\,,\nonumber\\
 \frac{\partial^2 F}{\partial T \partial L_\pm} &= \frac{\partial^2 V_L}{\partial T \partial L_\pm} - 2 N_f N_c \sum_{\varsigma = \pm 1} \int \frac{d^3p}{(2\pi)^3} f_{\varsigma,\pm} \nonumber\\
 & \qquad \times [1 + \beta \mathcal{E}_\varsigma (1 + \delta_{\varsigma,\mp} - 3 f_\varsigma)]\,,\nonumber\\
 \frac{\partial^2 F}{\partial \mu \partial L_\pm} &= -2 N_f N_c \sum_{\varsigma = \pm 1} \varsigma \int \frac{d^3p}{(2\pi)^3} f_{\varsigma,\pm} \nonumber\\ 
 & \times (1 + \delta_{\varsigma,\mp} - 3 f_\varsigma),
\end{align}
where the second derivatives of the Polyakov potential read:
\begin{align}
 \frac{\partial^2 V_L}{\partial T \partial L_\pm} &= - \frac{6}{\mathcal{H}} \frac{\partial(T^4 b(T))}{\partial T} (L_\mp - 2 L_\pm^2 + L_\mp^2 L_\pm) \nonumber\\
 &- \frac{L_\mp}{2} \frac{\partial(T^4 a(T))}{\partial T},\nonumber\\
 \frac{\partial(T^4 a(T))}{\partial T} &= 4T^3 a_0 + 3 T_0 T^2 a_1 + 2 T_0^2 T a_2,
\end{align}
while $\partial[T^4 b(T)] / \partial T = b_3 T_0^3$.

\begin{figure*}
    \centering
\begin{tabular}{cc}
\includegraphics[width=0.46\linewidth]{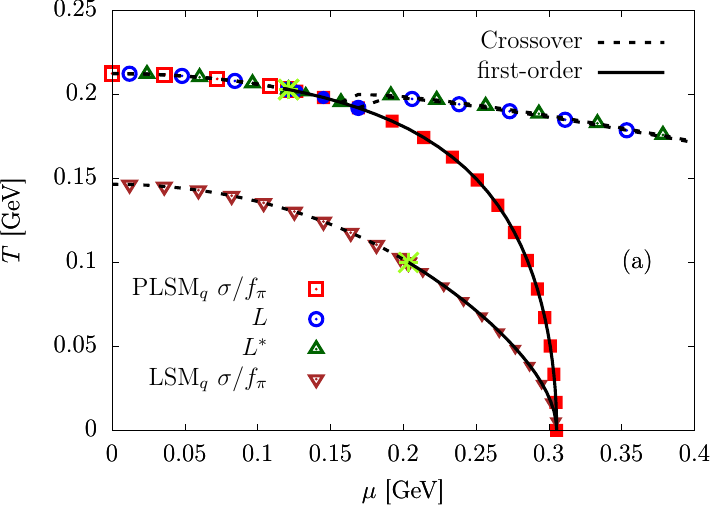} & \includegraphics[width=0.46\linewidth]{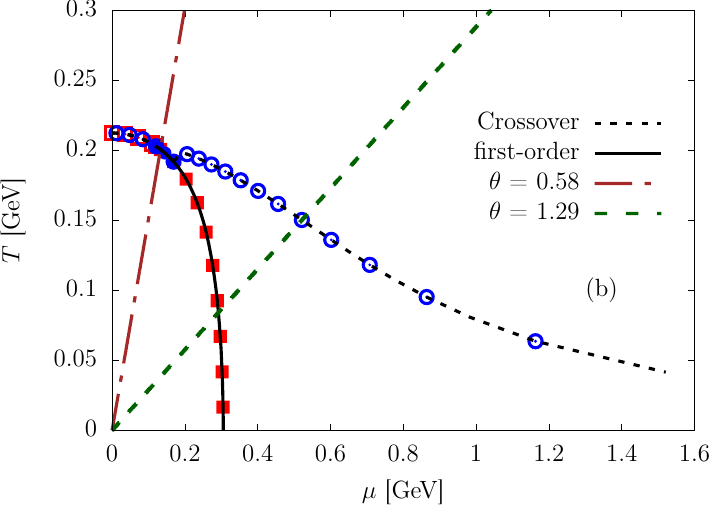}\\
\includegraphics[width=0.46\linewidth]{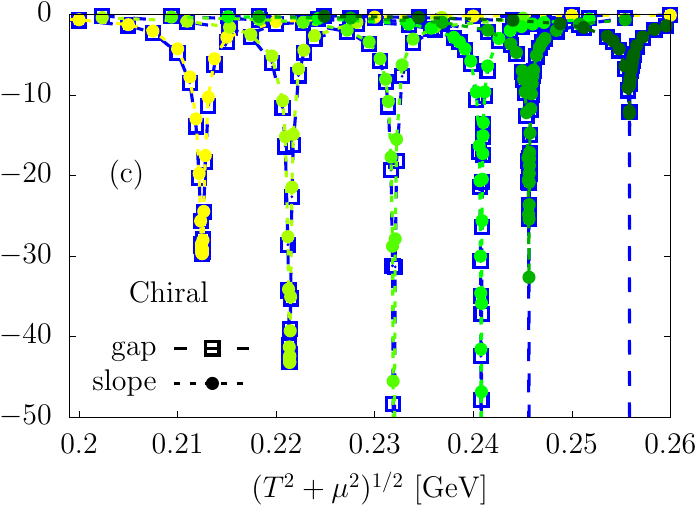} & \includegraphics[width=0.46\linewidth]{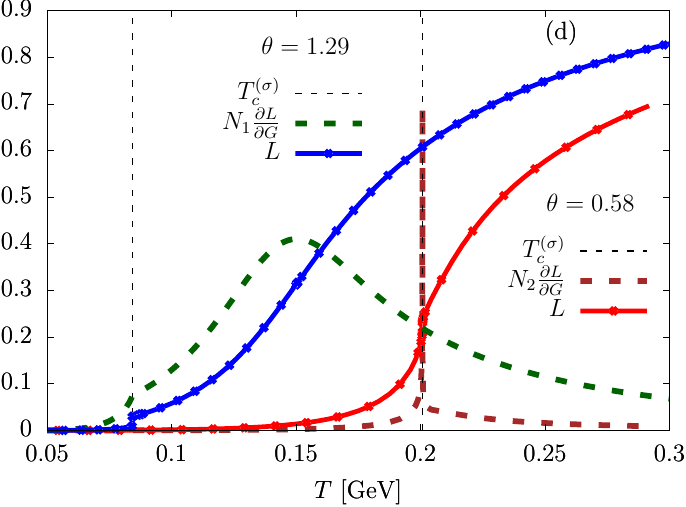}
\end{tabular}
\caption{(a) Chiral and deconfinement phase diagram for the unbounded, nonrotating LSM$_q$ and PLSM$_q$ models.
The critical points are denoted by the star ``*''. (b) The splitting between chiral and deconfinement phase diagram in $T{-}\mu$ plane. (c) A comparison between the numerically-calculated gap (shown with dashed blue lines and open squares) and the analytically-calculated slope (shown with colored dotted lines and filled circles) for the chiral phase diagram, as a function of $G = (T^2+\mu^2)^{1/2}$, with increasing values of $\theta = \tan^{-1}(\mu/T)$ denoted by the color gradient from yellow to dark-green colors. (d) The Polyakov loop and its slope $(\partial L/\partial G)$ as a function of temperature $T=G\cos{\theta}$ for two different values $\theta_1 = 0.58$ rad and $\theta_2 = 1.29$ rad, marked in the phase diagram of (b). The slopes are scaled by the constant factors $N_1=0.2$ and $N_2=0.005$, respectively, for presentation purposes. The corresponding chiral transition temperatures are marked by the black dashed lines. See text for more details.
\label{fig:PD}
}
\end{figure*}

We now study the phase diagram of the (P)LSM$_{q}$ models discussed so far. To help identify the transition line, we run a scan over the $(T,\mu)$ plane along diagonal lines passing through the origin, $T = \mu = 0$, which makes an angle $\theta$ to the vertical axis. Introducing $G = \sqrt{T^2 + \mu^2}$, the temperature and chemical potential along this diagonal line search have the expressions $(T,\mu) = G (\cos\theta, \sin\theta)$, while the slope is computed as
\begin{equation}
 \frac{\partial{p_i}}{\partial G} = \cos\theta \frac{\partial p_i}{\partial T} + \sin\theta \frac{\partial p_i}{\partial \mu},
 \label{eq:diagonal_slope}
\end{equation}
with $p_i \in \{\sigma, L, L^*\}$.

In Fig.~\ref{fig:PD}(a), we present the phase diagram obtained by such a diagonal search for our 3 order parameters: $\sigma/f_\pi$, $L$ and
$L^*$. For all three cases, the open symbols signify the pseudocritical point of a crossover transition, while the solid symbols
mark the first-order phase transition. The chiral phase diagram for the LSM$_{q}$ model is also presented. Comparing the PLSM$_q$ and LSM$_q$ phase diagrams shows that, due to the coupling with the background Polyakov loop, the chiral critical point moves towards
higher temperature and lower chemical potential [denoted by the green star ``*'' in Fig.~\ref{fig:PD}(a)].

Let us discuss the criteria we used to mark a first-order phase transition in our numerical analysis.
We have already mentioned that, to pinpoint the crossover temperature for different order parameters, one can calculate the slope along diagonal lines of search, as given in  Eq.~\eqref{eq:diagonal_slope}, using Eq.\eqref{eq:dpdA}. Along with this, one can define a numerical derivative, using the second-ordered centered scheme, for the control parameters ($\sigma,
L, L^*$) with respect to $T$ and $\mu$ (and $\Omega$, for the rotating case). We refer to this
numerical derivative as ``gap'' to distinguish it from the ``slope'' calculated via Eq.~\eqref{eq:diagonal_slope}.
In the crossover region, the two evaluations agree with each other up to $\mathcal{O}(\delta G^2)$ and the ratio between ``gap'' and ``slope'' will approach $1$ as the step $\delta G$ is decreased.
However, in the first-order region, the slope evaluated on either side of the discontinuity is finite, but the ``gap'' goes to infinity as we refine around the region of discontinuity. Thus, the ``gap'' to ``slope'' ratio tends to $\infty$ when the transition is of first order. Around the critical point, both the slope and the gap tend to infinity. In our analysis, we have used the heuristic criterion that when ``gap'' $/$ ``slope'' $ > 1.3$, the transition is marked as first order. We marked the critical point on the curve where the ratio between the gap and the slope first exceeds the value $1.3$.
We remark that the first-order phase transition occurs when the saddle-point Eqs.~\eqref{eq_saddle_p} allow for multiple solutions. When the criterion for the first-order phase transition is met, indicating a jump from one solution branch to another, we perform a detailed comparison of the two solution branches to ensure that the thermodynamic potential is continuous at the transition point, thereby ensuring its global minimization.

Our calculation results for the slope (the dashed lines with the filled circles and the color gradient) and gap (the solid blue lines and the empty squares) for the order parameter $\sigma / f_\pi$, computed in the PLSM$_q$ model, are shown in Fig.~\ref{fig:PD}(c). Here, we considered 6 values of $\theta(=\tan^{-1}(\mu/T)$) and we represented our results with respect to the parameter $G = \sqrt{T^2+\mu^2}$ of our diagonal line search. The color gradient in the slope from yellow ($\theta = 0$) to dark-green ($\theta = 0.722$ rad) corresponds to increasing values of $\theta$. For each value of $\theta$, the comparison between the gap and the slope shows that they both have a distinct peak structure at the same value of $G$ and the magnitudes of the two peaks agree for smaller values of $\theta$. We mark these points as points of crossover transition in the $T{-}\mu$ chiral phase diagram shown in Figs.~\ref{fig:PD}(a) and \ref{fig:PD}(b). As $\theta$ increases, it can be seen that, although the peak positions of the gap and slope still agree, their magnitudes differ. While the peak of the slope has a finite value, the value of the gap tends to infinity. Such a feature corresponds to a first-order phase transition.

Note that for the chiral phase transition, in Fig.~\ref{fig:PD}(c) the peak position for the gap and slope always agree. This situation changes as we focus on the deconfinement sector.
In Fig.~\ref{fig:PD}(d), we plotted the expectation value of the traced Polyakov loop ($L$) and the scaled slope, $N \partial L / \partial G$,  where $N$ is a constant scaling factor, as a function of temperature for two particular values of $\theta$: a smaller one, $\theta_1 = 0.58$ rad, where the chiral and deconfinement phase transitions are simultaneous; and a larger one, $\theta_2 = 1.29$ rad, where we see a split between these transitions. These particular values of $\theta$ are highlighted on the phase diagram in Fig.~\ref{fig:PD}(b). The slopes are appropriately scaled, $N_1 = 0.2$ and $N_2 = 0.005$, for presentation purposes.

In Fig.~\ref{fig:PD}(d), we also showed the chiral first-order phase transition temperatures for these two values of $\theta$. Since the fields are coupled to each other, the gap in all control parameters ($\sigma$, $L$ and $L^*$) will simultaneously exhibit a peak at the point of discontinuity in the system, i.e. when $T = T_c^{(\sigma)}$. For $\theta=0.58$ rad, the slope of $L$ also shows a peak there and the value of $L$ changes significantly over the chiral phase transition. In this instance, we see that the chiral and deconfinement phase transitions are simultaneous. However, for $\theta=1.29$ rad, one can see that, although there is a small jump in $L$ around $T = T_c^{(\sigma)} \simeq 0.08$ GeV, the value of $L$ after the jump is significantly smaller, $L \approx 0.03$.
Moreover, the slope only shows a small ridge at $T = T_c^{(\sigma)}$, having a much more prominent peak at a larger temperature, $T = T_c^{(L)} \simeq 0.15$ GeV. After the small jump at $T_c^{(\sigma)}$, we can observe a smooth increment in $L$, and around the peak of the slope (at $T = T_c^{(L)}$), $L \approx 0.3$. We argue that the value of $L$ as small as $0.03$ cannot be associated with the deconfinement phase transition. Rather, a confinement-deconfinement crossover can be marked at the position of the slope peak, shown as the intersection point between the green dashed line and blue open circles in Fig.~\ref{fig:PD}(b), where $L \approx 0.3$.
\begin{figure}
    \centering
    \includegraphics[width=0.92\linewidth]{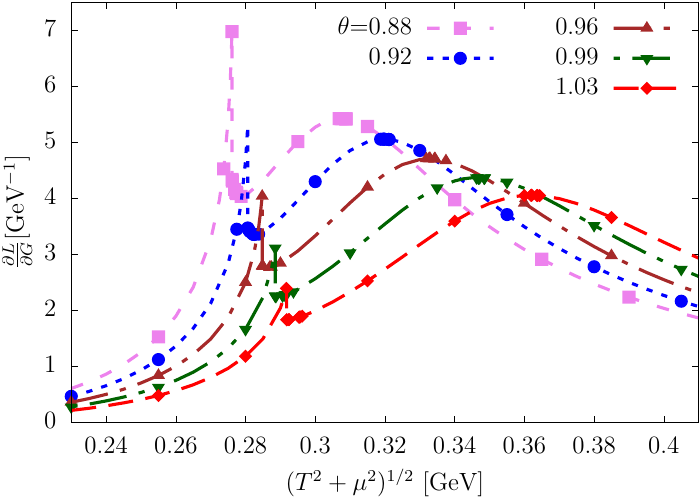}
    \caption{Double-peak structure of the slope $\partial L / \partial G$ of the Polyakov loop with respect to $G = \sqrt{T^2 + \mu^2}$, computed on diagonal trajectories under various angles $\theta$. The deconfinement transition point is taken where the slope has the highest peak: the left peak for $\theta = 0.88$ and $0.92$ rad; the right peak for $\theta = 0.96$, $0.99$ and $1.03$ rad.}
    \label{fig:Lslope}
\end{figure}

There is a splitting between the chiral and deconfinement phase diagram for higher values of the chemical potential and, while the chiral transition shows a first-order behaviour, the deconfinement transition becomes of crossover type. This phenomenon, resembling a first-order phase transition, can be understood by looking at the slope $\partial L / \partial G$, introduced in Eq.~\eqref{eq:diagonal_slope}. The pseudocritical parameters for $L$ are defined as those parameters $(T_{\rm pc}, \mu_{\rm pc})$ for which $\partial L / \partial G$, computed along the line with $(T, \mu) = G (\cos\theta, \sin\theta)$ at fixed $\theta$, attains a maximum value. As shown in Fig.~\ref{fig:Lslope}, $\partial L / \partial G$ has a double peak structure around the point where the deconfinement and chiral restoration transition temperatures start to disagree, $T_{\rm pc}(L) \neq T_{\rm pc}(\sigma)$. For small values of the chemical potential $\mu$, the first peak, developed at a lower temperature at the point of the chiral restoration, dominates over the second peak. As $\mu$ increases, the magnitude of the second peak also increases, eventually becoming dominant, leading to a split between the deconfinement and chiral restoration transitions.

\subsection{Vanishing temperature limit}
\label{sec:PLSM:T0}

At vanishing temperature, the distribution function $f_\pm$  becomes a step function:
\begin{equation}
 \left. f_\pm \right|_{T \rightarrow 0} = \theta(\pm \mu - E)\,,
\end{equation}
while $\left. f_{\pm,+} \right|_{T\rightarrow 0} = \left.f_{\pm,-} \right|_{T\rightarrow 0} = 0$.
Since both functions $f_{\varsigma, L}$ and $f_{\varsigma, L^*}$ vanish, the equations $\partial F / \partial L = 0$ and $\partial F / \partial L^* = 0$ are automatically satisfied for any value of $L$ and $L^*$. We therefore focus on the $\sigma$ sector, where we expect a first-order phase transition. As a consequence, the Polyakov loop variable does not play any role at vanishing temperature and the model reduces to the linear sigma model coupled to quarks only (LSM${}_q$).

We must therefore evaluate the fermion condensate [cf.~Eq.~\eqref{eq_ppsi}] to enforce $\partial F / \partial \sigma = 0$, and calculate the pressure [cf.~Eq.~\eqref{eq_P}], to ensure the minimization of the thermodynamic potential. Both of these quantities can be assessed analytically:
\begin{align} \label{eq_unb_T0}
 \langle \Bar{\psi} \psi \rangle &= \frac{N_f N_c g\sigma}{\pi^2} \theta(p_f^2)\int_0^{p_f} \frac{dp\,p^2}{E} 
 = \frac{N_f N_c g \sigma}{2\pi^2} \theta(|\mu| - g\sigma) \nonumber\\
 & \times \left[p_f |\mu| - (g\sigma)^2 {\rm artanh}\frac{p_f}{|\mu|}\right], \nonumber\\
 P_q &= \frac{N_fN_c}{3\pi^2} \theta(p_f^2)\int_0^{p_f} \frac{dp\, p^4}{E} =  \frac{N_f N_c}{24\pi^2} \theta(|\mu| - g\sigma) \nonumber\\
 & \times \left[|\mu| p_f (2\mu^2 - 5g^2 \sigma^2) + 3 (g \sigma)^4 {\rm artanh} \frac{p_f}{|\mu|} \right],
\end{align}
where $p_f = \sqrt{\mu^2 - g^2 \sigma^2}$ is the Fermi momentum, and we employed the integration of the phase space $d^3p = p^2 dp d\Omega_2$ with the $d\Omega_2$ angular integral trivially giving $4\pi$.
The above results are valid only when the Fermi energy exceeds the effective mass, i.e., $|\mu| > g \sigma$, as implied by the step function appearing on the right-hand sides of~\eqref{eq_unb_T0}.

\section{Rotating PLSM\texorpdfstring{${}_q$}{q} model}
\label{sec:rot}

To facilitate the analysis of the rotating PLSM${}_q$ model, we will employ corotating coordinates, defined by the angular velocity $\Omega$ entering the grand canonical ensemble. To avoid superluminal corotating velocities, the system must be enclosed in a cylindrical box, where suitable boundary conditions must be enforced.
In Subsec.~\ref{sec:rot:L}, we briefly present the model Lagrangian in corotating coordinates. In Subsec.~\ref{sec:rot:F}, we derive the grand potential for the bounded, rotating system, obeying spectral boundary conditions. In Subsec.~\ref{sec:rot:saddle}, we discuss the saddle-point equations giving the mean-field values for the mesons and Polyakov loops. Finally, Subsections~\ref{sec:rot:LSM} and \ref{sec:rot:T0} discuss the LSM$_q$ limit and the vanishing temperature limit, respectively.

\subsection{Corotating frame}\label{sec:rot:L}

We now switch to corotating coordinates, defined with respect to the static cylindrical coordinates (denoted with the subscript $M$) by
\begin{equation}
 t = t_M, \qquad \varphi = \varphi_M - \Omega t_M,
\end{equation}
where $\Omega$ represents the angular velocity of the corotating observer.
The line element $ds^2 = g_{\mu\nu} dx^\mu dx^\nu$ expressed with respect to the corotating coordinates reads
\begin{equation}
 ds^2 = (1 - \rho^2 \Omega^2) dt^2 - 2\rho^2 \Omega dt d\varphi - d\rho^2 - \rho^2 d\varphi^2 - dz^2.
\end{equation}
The metric can be brought to Minkowski form by introducing the non-holonomic, orthonormal tetrad (vielbein) vector field, $e_{\hat{\alpha}} = e_{\hat{\alpha}}^\mu \partial_\mu$, and the associated one-forms $\omega^{\hat{\alpha}} = \omega^{\hat{\alpha}}_\mu dx^\mu$, defined by
\begin{align}
 e_{\hat{t}} &= \partial_t + \Omega(y \partial_x - x \partial_y), &
 e_{\hat{x}} &= \partial_x, &\nonumber\\
 e_{\hat{y}} &= \partial_y, &
 e_{\hat{z}} &= \partial_z, \nonumber\\
 \omega^{\hat{t}} &= dt, &
 \omega^{\hat{x}} &= dx - \Omega y dt, &\nonumber\\
 \omega^{\hat{y}} &= dy + \Omega x dt, &
 \omega^{\hat{z}} &= dz,
\end{align}
such that $g_{\mu\nu} = \eta_{\hat{\alpha}\hat{\beta}} \omega^{\hat{\alpha}}_\mu \omega^{\hat{\beta}}_\nu$ and $\omega^{\hat{\alpha}}_\mu e^\mu_{\hat{\beta}} = \delta^{\hat{\alpha}}_{\hat{\beta}}$.

The Cartan coefficients associated with the tetrad are defined by the commutation relation $[e_{\hat{\alpha}}, e_{\hat{\rho}}] = c_{\hat{\alpha}\hat{\rho}}{}^{\hat{\sigma}} e_{\hat{\sigma}}$.
Taking into account the commutation relations $[e_{\hat{t}}, e_{\hat{x}}] = \Omega e_{\hat{y}}$ and $[e_{\hat{t}}, e_{\hat{y}}] = -\Omega e_{\hat{x}}$, the non-vanishing Cartan coefficients are  $c_{\hat{t}\hat{x}}{}^{\hat{y}} = c_{\hat{y}\hat{t}}{}^{\hat{x}} = \Omega$.
The connection coefficients $\Gamma_{\hat{\rho}\hat{\sigma} \hat{\alpha}} = \frac{1}{2} (c_{\hat{\rho}\hat{\sigma} \hat{\alpha}} + c_{\hat{\rho} \hat{\alpha} \hat{\sigma}} - c_{\hat{\sigma} \hat{\alpha} \hat{\rho}})$ vanish identically, except for
\begin{equation}
 \Gamma_{\hat{x}\hat{y}\hat{t}} = \Omega.
\end{equation}

The quark Lagrangian \eqref{eq_L_quark} can be written in curvilinear Minkowski coordinates as
\begin{equation}
 \mathcal{L}_q = \overline{\psi} \left(i \slashed{D} + i \slashed{\Gamma} + \mu \gamma^0  - g \sigma \right) \psi,
 \label{eq_L_quark_rot}
\end{equation}
where $\Gamma_\mu = -\frac{i}{2} \omega_\mu^{\hat{\alpha}} \Gamma_{\hat{\beta}\hat{\gamma}\hat{\alpha}} S^{\hat{\beta}\hat{\gamma}}$ is the spin connection coefficient and $S^{\hat{\beta}\hat{\gamma}} = \frac{i}{4} [\gamma^{\hat{\beta}}, \gamma^{\hat{\gamma}}]$ represent the spin part of the generators of the Lorentz transformations.
We take the gamma matrices $\gamma^{\hat{t}}$ and $\gamma^{\hat{\imath}}$ with respect to the tetrad vector fields in the Dirac representation, shown in Eq.~\eqref{eq:gamma}. With the above convention, $\Gamma_{\hat{\imath}} = 0$ and $\Gamma_{\hat{t}} = -i \Omega S^z$ \cite{Ambrus:2015lfr} and Eq.~\eqref{eq_L_quark_rot} becomes
\begin{equation}
 \mathcal{L}_q = \bar{\psi}
 [\gamma^{\hat{0}} (i \partial_t + A_0 + \mu + \Omega J^z) + i \boldsymbol{\gamma} \cdot \boldsymbol{\nabla} - g\sigma] \psi,
 \label{eq:rot_Dirac_eq}
\end{equation}
where $J^z = -i \partial_\varphi + S^z$ and $S^z = \frac{i}{2} \gamma^{\hat{1}} \gamma^{\hat{2}}$. Switching now to imaginary time, $\tau = i t$, the Euclidean Lagrangian $\mathcal{L}_E(\tau, \mathbf{x}) = -\mathcal{L}_q(it, \mathbf{x})$ reads
\begin{equation}
 \mathcal{L}_E = -\overline{\psi} \left[\gamma^{\hat{0}} (-\partial_\tau + \Omega J^z - iA_4 + \mu) + i \boldsymbol{\gamma} \cdot \boldsymbol{\nabla} - g \sigma\right] \psi.
 \label{eq:LE}
\end{equation}

\subsection{Grand canonical potential}\label{sec:rot:F}

Let us now compute the partition function $\mathcal{Z}_q$ and the associated thermodynamic potential, $F_q = -\frac{T}{V} \ln \mathcal{Z}_q$. As outlined in Sec.~\ref{sec:PLSM:F}, we seek a mode decomposition of the spinor $\psi$ that diagonalizes the Euclidean Lagrangian. Since the momentum operator $\mathbf{P}$ does not commute with the angular momentum operator $J^z$ appearing in Eq.~\eqref{eq:LE}, we must employ cylindrical modes that are instead eigenfunctions of $J^z$ and $P^z = -i \partial_z$. Thus, Eqs.~\eqref{eq:static_eigen} and \eqref{eq:static_H} are replaced by
\begin{multline}
 (H, -i \partial_\tau, \sigma^3_f, \mathcal{A}_4, h, J^z, P^z) \psi_j = \\ 
 (\varsigma_j \widetilde{\mathcal{E}}_j, w_j, f_j, A_{4;j}, \lambda_j, m_j, k_j) \psi_j,
\end{multline}
where the Hamiltonian $H = i \mathcal{A}_4 - \mu - \Omega J^z + 2 \gamma^5 \mathbf{S} \cdot \mathbf{P} + g \sigma \gamma^0$, has eigenvalues satisfying
\begin{equation}
 \widetilde{\mathcal{E}}_j = \mathcal{E}_j - \Omega m_j,
\end{equation}
where $\mathcal{E}_j$ is given in Eq.~\eqref{eq:static_H}.
The modes described above have the following form \cite{Ambrus:2019cvr}:
\begin{multline}
 \psi_j(X) = \mathcal{C}_j \frac{e^{i w_j \tau + i k_j z}}{2\pi}
 \begin{pmatrix}
  \frac{1}{2}(1 + f_j) \\[1mm]
  \frac{1}{2}(1 - f_j)
 \end{pmatrix}\otimes
 \begin{pmatrix}
  \delta_{c_j, 1} \\
  \delta_{c_j, 2} \\
  \delta_{c_j, 3}
 \end{pmatrix} \\
 \otimes
 \frac{1}{\sqrt{2}}
 \begin{pmatrix}
  \mathfrak{E}^{\varsigma_j}_j \\
  2\varsigma_j \lambda_j \mathfrak{E}^{-\varsigma_j}_j
 \end{pmatrix} \otimes \phi_j(\rho,\varphi),
 \label{eq:cyl_modes}
\end{multline}
where $\mathcal{C}_j$ is a normalization constant, $\mathfrak{E}^\pm_j = \sqrt{1 \pm g \sigma / E_j}$ was introduced in Eq.~\eqref{eq:static_modes} and $\phi_j$ satisfying $(-i \partial_\varphi + S^z) \phi_j = m_j \phi_j$ and $-i \boldsymbol{\sigma} \cdot \boldsymbol{\nabla} \phi_j = 2\lambda_j p_j \phi_j$ is given by
\begin{equation}
 \phi_j(\rho, \varphi) = \frac{1}{\sqrt{2}}
\begin{pmatrix}
 \mathsf{p}^+_j e^{i (m_j -\frac{1}{2}) \varphi} J_{m_j - \frac{1}{2}}(q_j \rho) \\
 2i \lambda_j \mathsf{p}^-_j e^{i (m_j +\frac{1}{2}) \varphi} J_{m_j + \frac{1}{2}}(q_j \rho)
\end{pmatrix},
\label{eq:cyl_modes_phi}
\end{equation}
with $\mathsf{p}^\pm_j = \sqrt{1 \pm 2\lambda_j k_j / p_j}$.

We now wish to impose orthogonality of the modes $\psi_j$ with respect to the integration over a cylinder of radius $R$. As discussed in Ref.~\cite{Ambrus:2015lfr}, this can be ensured in a number of ways. In the spectral boundary conditions model, the transverse momentum $q_j$ is quantized according to
\begin{equation}
 q_j R = \begin{cases}
  \xi_{m_j - \frac{1}{2}, l_j}, & m_j > 0,\\
  \xi_{-m_j - \frac{1}{2}, l_j}, & m_j < 0,
 \end{cases}
 \label{eq:cyl_spectral}
\end{equation}
with $\xi_{nl}$ being the $l$'th nonzero root of the Bessel function $J_n$, i.e. $J_n(\xi_{nl}) = 0$ ($n$ is a non-negative integer).
We thus impose the normalization $\int_0^\beta d\tau \int_V d^3x \, \psi_j^\dagger \psi_{j'} = \beta\delta(j,j')$, following Eq.~\eqref{eq:static_ortho},
which gives the normalization constant $\mathcal{C}_j$ as
\begin{equation}
 \mathcal{C}_j = \frac{\sqrt{2}}{R|J_{|m_j| + \frac{1}{2}}(\xi_{|m_j| - \frac{1}{2}, l_j})|}.
\end{equation}
The delta function reads
\begin{multline}
  \delta(j,j') = \delta_{\varsigma_j \varsigma_{j'}} \delta_{n_j n_{j'}} \delta_{f_j f_{j'}} \delta_{c_j c_{j'}} \delta_{\lambda_j \lambda_{j'}} \delta_{m_j m_{j'}} \delta_{l_j l_{j'}} \\\times 
  \delta(k_j - k_{j'}).
\end{multline}

Taking again the decomposition $\psi(X) = \sum_j \hat{b}_j \psi_j(X)$ using the cylindrical modes introduced in Eq.~\eqref{eq:cyl_modes}, the partition function looks formally identical to the one in the static case, with $\mathcal{E}_j$ replaced by $\widetilde{\mathcal{E}}_j$ in Eqs.~\eqref{eq:static_Zq_aux}--\eqref{eq:static_Zq}. Repeating for definiteness Eq.~\eqref{eq:static_Zq}, we have $\widetilde{\mathcal{Z}}_q = \prod_j [\beta(i \varsigma_j \widetilde{\mathcal{E}}_j - w_j)]$. Upon taking the logarithm to express the grand potential, $\widetilde{F}_q = -\frac{T}{V} \ln \widetilde{\mathcal{Z}}_q$, the continuous momentum integration in Eq.~\eqref{eq:static_Fq_aux} is replaced using the following rule:
\begin{equation}
 \int \frac{d^3p}{(2\pi)^3} \rightarrow \frac{1}{\pi R^2} \sum_{\mathrm{b}} \int \frac{dk}{2\pi},
 \qquad
 \sum_{\mathrm{b}} = \sum_{m = -\infty}^\infty \sum_{l = 1}^\infty\,,
 \label{eq:rot_int_measure}
\end{equation}
keeping in mind that $m = \pm \frac{1}{2}, \pm \frac{3}{2}, \dots$ runs over all odd half-integer values. This leads to the following modification of Eq.~\eqref{eq:static_F}:
\begin{equation}
 \widetilde{F}_q = -\frac{2N_f N_c T}{\pi R^2} \sum_{\varsigma = \pm 1} \sum_{\mathrm{b}} \int_{-\infty}^\infty \frac{dk}{2\pi} \widetilde{F}_\varsigma,
 \label{eq:rot_Fq}
\end{equation}
where $\widetilde{F}_\varsigma$ is $F_\varsigma$ from Eq.~\eqref{eq:static_Fs}, with $\mathcal{E}_\varsigma$ replaced by $\widetilde{\mathcal{E}}_\varsigma$:
\begin{equation}
 \widetilde{F}_\varsigma \equiv \frac{1}{3} \ln (1 + 3L_\varsigma e^{-\beta \widetilde{\mathcal{E}}_\varsigma} + 3L_{-\varsigma} e^{-2\beta \widetilde{\mathcal{E}}_\varsigma} + e^{-3\beta\widetilde{\mathcal{E}}_\varsigma}).
\end{equation}

Note that for the numerical analysis, one can further simplify the expression given in Eq.~\eqref{eq:rot_int_measure} by replacing the sum over the odd half-integers $m$ with a sum over non-negative integers, $\sum_{m=-\infty}^{\infty} f_m = \sum_{n=0}^\infty [f_{n + \frac{1}{2}} + f_{-n-\frac{1}{2}}]$, with the convention $n = |m| - \frac{1}{2}$. The corotating effective energies are modified to
\begin{equation}
 \widetilde{\mathcal{E}}^\varsigma_{m > 0,l} \rightarrow \widetilde{E}_{n + \frac{1}{2},l} - \varsigma \mu, \qquad
 \widetilde{\mathcal{E}}^\varsigma_{m < 0,l} \rightarrow \overline{E}_{n + \frac{1}{2},l} - \varsigma \mu,
 \label{eq:rot_Em_refl}
\end{equation}
where we introduced
\begin{align}
 \widetilde{E}_{n+\frac{1}{2},l} &= E_{n+\frac{1}{2},l} - \Omega\left(n + \frac{1}{2}\right),
 \nonumber\\
 \overline{E}_{n+\frac{1}{2},l} &= E_{n+\frac{1}{2},l} + \Omega\left(n + \frac{1}{2}\right),
\end{align}
while $E_{n + \frac{1}{2}, l} = \sqrt{k^2 + R^{-2} \xi_{nl}^2 + g^2 \sigma^2}$. In arriving at Eq.~\eqref{eq:rot_Em_refl}, we took into account the fact that $E_{-ml} = E_{ml}$, due to the property \eqref{eq:cyl_spectral}.

\subsection{saddle-point approximation}\label{sec:rot:saddle}

As in the unbounded case discussed in Sec.~\ref{sec:PLSM:saddle}, the control parameters of the model are $p \in \{\sigma, L, L^*\}$. Their expectation values at finite $T$, $\mu$ and $\Omega$ can be obtained by imposing the minimization of the free energy via Eqs.~\eqref{eq_saddle}, with $F$ replaced by $\widetilde{F} = V_{\mathcal{M}} + V_L + \widetilde{F}_q$, where the meson ($V_{\mathcal{M}}$) and Polyakov ($V_L$) contributions remain unchanged. The derivatives $\partial \widetilde{F} / \partial p_i$ remain as given in Eqs.~\eqref{eq_dFdp}--\eqref{eq_dVdp}, with the replacement of the integration measure as shown in Eq.~\eqref{eq:rot_int_measure}, as well as the replacement of the distributions $f_\varsigma$ and $f_{\varsigma;\pm}$ by their rotating counterparts:
\begin{subequations}\label{eq_rot_fall}
\begin{align}
 \tilde{f}_\pm  &= \frac{L_\pm e^{-\beta \widetilde{\mathcal{E}}_\pm} + 2 L_\mp e^{-2\beta\widetilde{\mathcal{E}}_\pm} + e^{-3\beta \widetilde{\mathcal{E}}_\pm}}{1 + 3 L_\pm e^{-\beta \widetilde{\mathcal{E}}_\pm} + 3L_\mp e^{-2 \beta \widetilde{\mathcal{E}}_\pm} + e^{-3 \beta \widetilde{\mathcal{E}}_\pm}}, \label{eq_rot_f}\\
 \tilde{f}_{\pm,\pm} &= \frac{e^{-\beta \widetilde{\mathcal{E}}_\pm}}{1 + 3 L_\pm e^{-\beta \widetilde{\mathcal{E}}_\pm} + 3L_\mp e^{-2 \beta \widetilde{\mathcal{E}}_\pm} + e^{-3 \beta \widetilde{\mathcal{E}}_\pm}}, \label{eq_rot_fL}\\
 \tilde{f}_{\pm,\mp}  &= \frac{e^{-2\beta \widetilde{\mathcal{E}}_\pm}}{1 + 3 L_\pm e^{-\beta \widetilde{\mathcal{E}}_\pm} + 3L_\mp e^{-2 \beta \widetilde{\mathcal{E}}_\pm} + e^{-3 \beta \widetilde{\mathcal{E}}_\pm}}, \label{eq_rot_fLb}\\
 \tilde{f}'_\varsigma  &= -3\tilde{f}_\varsigma(1 - \tilde{f}_\varsigma) + 2 L_+ \tilde{f}_{\varsigma,+} + 2 L_{-} \tilde{f}_{\varsigma,-}\,. \label{eq_rot_df}
\end{align}
\end{subequations}

As in the unbounded case, we solve the saddle-point equations using Eq.~\eqref{eq_NR}, with the matrix $\mathcal{M}_{ij}$ replaced by $\widetilde{\mathcal{M}}_{ij}$, constructed as discussed above: the meson and Polyakov parts remain the same, as given in Eqs.~\eqref{eq:unb_MV} and \eqref{eq.d2vdp2}, while the fermion contributions in  Eq.~\eqref{eq:unb_Mij} are modified by replacing the momentum integration via Eq.~\eqref{eq:rot_int_measure} and by replacing $(f_\varsigma, f_{\varsigma,\pm}, f'_\varsigma)$ by $(\tilde{f}_\varsigma, \tilde{f}_{\varsigma,\pm}, \tilde{f}'_\varsigma)$.

When discussing the pseudocritical parameters corresponding to a crossover transition, we define the gradient of the control parameters $p \in \{\sigma, L, L^*\}$ by extending Eq.~\eqref{eq.dpmf2var} to include the effect of a finite angular velocity $\Omega$:
\begin{equation}
 dp_{\rm mf} = \left(\frac{\partial p}{\partial T}\right)_{\rm mf} dT + \left(\frac{\partial p}{\partial \mu}\right)_{\rm mf} d\mu +  \left(\frac{\partial p}{\partial \Omega}\right)_{\rm mf} d\Omega.
 \label{eq.dpmf3var}
\end{equation}
The evaluation of the derivatives $\partial p_i / \partial G_a$ of the control parameters with respect to the thermodynamic parameters $G_a \in \{T, \mu, \Omega\}$ can be performed as shown in Eq.~\eqref{eq:dpdA}, where the derivatives of the free energy appearing on the right-hand side of this equation with respect to $T$ and $\mu$ are identical to those given in Eqs.~\eqref{eq:d2FdAdp}, with $(F, f_\varsigma, f_{\varsigma,\pm}, \mathcal{E}_\varsigma)$ replaced by $(\widetilde{F}, \tilde{f}_\varsigma, \tilde{f}_{\varsigma,\pm}, \widetilde{\mathcal{E}}_\varsigma)$ and the integration measure modified according to Eq.~\eqref{eq:rot_int_measure}.
In addition, we compute the derivatives involving the rotation angular velocity:
\begin{align} \label{eq:rot_d2FdOdL}
\frac{\partial^2 \widetilde{F}}{\partial \Omega \partial \sigma} &= -\frac{N_f N_c g^2 \sigma}{\pi^2 R^2 T} \sum_{\mathrm{b},\varsigma = \pm 1} m \int_{-\infty}^\infty \frac{dk}{ E} f'_\varsigma,\\
 \frac{\partial^2 \widetilde{F}}{\partial \Omega \partial L_\pm} &= -\frac{N_f N_c}{\pi^2 R^2} \sum_{\mathrm{b}, \varsigma = \pm 1} m \int_{-\infty}^\infty dk \tilde{f}_{\varsigma,\pm} (1 + \delta_{\varsigma,\mp} - 3 \tilde{f}_\varsigma),\nonumber
\end{align}
where we remind the reader that $m = \pm\frac{1}{2}, \pm \frac{3}{2}, \dots$ takes odd half-integer values, while $\delta_{\varsigma,\varsigma'}$ is the Kronecker symbol.

\subsection{Rotating LSM\texorpdfstring{${}_q$}{q} model with boundary}\label{sec:rot:LSM}

We now discuss the limit $L, L^* \rightarrow 1$, when the system is deconfined and the PLSM${}_q$ model reduces to the $\mathrm{LSM}_q$ model. Here, the only control parameter is $\sigma$, and the corresponding volume-averaged grand potential for the quark sector can be obtained by putting  $L=L^*=1$ in Eq.~\eqref{eq:rot_Fq}:
\begin{equation}
    \widetilde{F}_q = -\frac{N_f N_c T}{\pi^2 R^2}\sum_{\mathrm{b},\varsigma=\pm 1}\int_{-\infty}^\infty dk\, \widetilde{F}_\varsigma,
    \label{eq:rot_LSM_Fq}
\end{equation}
with $\widetilde{F}_\varsigma = \ln(1 + e^{-\beta\widetilde{\mathcal{E}}_\varsigma})$. 
The expectation value $\sigma_{\rm mf}$ of the $\sigma$ meson can be found by imposing the minimization of the grand potential,
\begin{align}
\frac{\partial \widetilde{F}}{\partial \sigma} &= \frac{\partial {V}_\mathcal{M} }{\partial \sigma} + \frac{\partial \widetilde{F}_q}{ \partial \sigma} = 0, \nonumber&
\\
\frac{\partial \widetilde{F}_q}{\partial \sigma} &= \frac{N_f N_c g^2 \sigma}{\pi^2 R^2} \sum_{\mathrm{b},\varsigma=\pm 1}\int_{-\infty}^\infty\frac{d k}{E}  \tilde{f}_\varsigma~,
\label{eq:rot_LSM_sad}
\end{align}
where $\partial V_{\mathcal{M}} / \partial \sigma$ is given in Eq.~\eqref{eq_dVds} and
\begin{align}
    \tilde{f}_\varsigma = - \frac{\partial \widetilde{F}_\varsigma}{\partial (\beta \widetilde{\mathcal{E}}_\varsigma)}  =
    \frac{1}{e^{\beta \widetilde{\mathcal{E}}_\varsigma} + 1}. \label{eq.rot_LSM_sad}
\end{align}
Equation~\eqref{eq:rot_LSM_sad} is solved numerically using Eq.~\eqref{eq_NR} and the following contribution from the fermionic part,
\begin{multline}
 \widetilde{\mathcal{M}}^q_{\sigma\sigma} =
 \frac{\partial^2 \widetilde{F}_q}{\partial \sigma^2} = \frac{N_f N_c g^2}{\pi^2 R^2} \sum_{\mathrm{b},\varsigma=\pm 1}\int_{-\infty}^\infty \frac{dk}{E} \\\times \left(\frac{\mathbf{p}^2}{E^2}\Tilde{f}_\varsigma + \frac{g^2 \sigma^2}{E T} \tilde{f}'_\varsigma\right),
 \label{eq.rot_LSM_dFds}
\end{multline}
with $\mathbf{p}^2 \equiv \mathbf{p}_{kml}^2 = k^2 + R^{-2} \xi_{|m| - \frac{1}{2},l}^2$ and $\tilde{f}'_\varsigma = -\tilde{f}_\varsigma(1-\tilde{f}_\varsigma)$. The mesonic contribution $\widetilde{\mathcal{M}}^V_{\sigma\sigma} = \mathcal{M}^V_{\sigma\sigma}$ is identical to that of the unbounded case, given in Eq.~\eqref{eq:unb_MV}. When searching for the pseudocritical parameters of a crossover phase transition, the slopes $\partial \sigma / \partial G_a$, with $G_a \in \{T, \mu, \Omega\}$, can be evaluated using:
\begin{multline}
 \left(\frac{\partial^2 \widetilde{F}}{\partial T \partial \sigma}, \frac{\partial^2 \widetilde{F}}{\partial \mu \partial \sigma}, \frac{\partial^2 \widetilde{F}}{\partial \Omega \partial \sigma}\right) = -\frac{N_f N_c g^2 \sigma}{\pi^2 R^2 T}\\
 \times \sum_{\mathrm{b},\varsigma = \pm 1} \int_{-\infty}^\infty \frac{dk}{ E} (\beta \tilde{\mathcal{E}}_\varsigma, \varsigma, m) \tilde{f}'_\varsigma.
 \label{eq:rot_LSM_d2FdAdp}
\end{multline}

\subsection{Vanishing temperature limit}\label{sec:rot:T0}

\begin{figure}
    \centering
    \begin{tabular}{c}
    \includegraphics[width=0.92\linewidth]{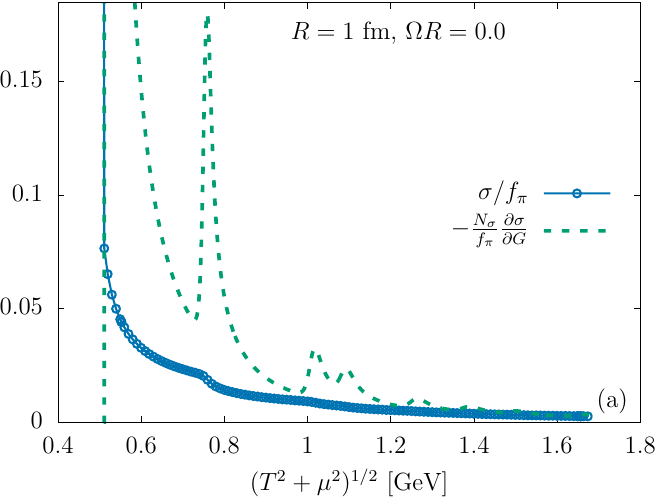}         \\
    \includegraphics[width=0.92\linewidth]{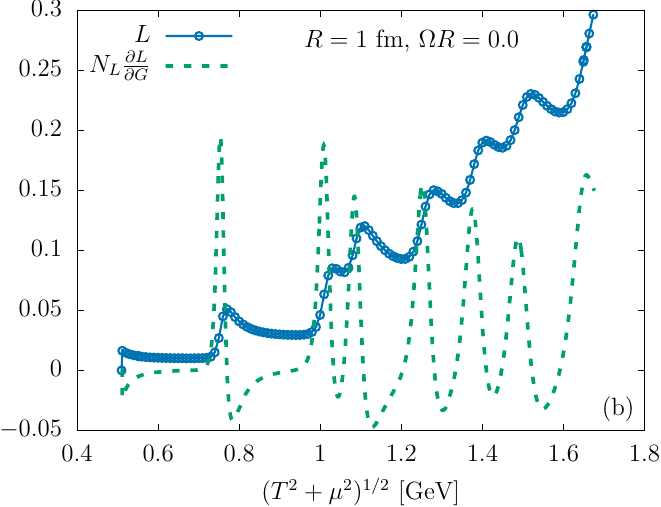}
    \end{tabular}
    \caption{The order parameters (a) $\sigma/f_\pi$  and (b) $L$, and their associated analytically calculated slope (shown with colored dashed lines) as functions of $G = (T^2+\mu^2)^{1/2}$, with
    $\tan \theta \simeq 54.5$ and $R=1$ fm, in the absence of rotation. The slopes are scaled by the constant factors (a) $N_\sigma=0.1$ and (b) $N_L=10$, respectively. 
    }
    \label{fig:lowT}
\end{figure}

In the $T\rightarrow 0$ limit, the distributions $\tilde{f}_\varsigma$ and $\tilde{f}_{\varsigma;\pm}$ reduce to
\begin{equation}
 \tilde{f}_\varsigma \rightarrow \theta(\varsigma \mu - \widetilde{\mathcal{E}}_\varsigma), \qquad
 \tilde{f}_{\varsigma,+}, \tilde{f}_{\varsigma,-} \rightarrow 0.
\end{equation}
The Polyakov sector completely decouples from the quark sector and the resulting model becomes equivalent to the LSM${}_q$ model without the Polyakov extension. The thermodynamic potential $\widetilde{F}_{T=0} = V_\mathcal{M}+\widetilde{F}_q^{T=0}$ receives no contribution from the Polyakov sector, while the fermionic contribution reads
\begin{multline} \label{eq.F0T}
 \widetilde{F}_q^{T= 0} = - \frac{2 N_f N_c}{\pi^2 R^2}\hspace{-5pt}\sum_{\mathrm{b}, \varsigma = \pm 1} \int_0^\infty \hspace{-5pt} \frac{dk\, k^2}{E_{ml}(k)} \theta[\varsigma \mu + \Omega m - E_{ml}(k)]
 \\
 = -\frac{N_f N_c}{\pi^2 R^2} \hspace{-5pt} \sum_{\mathrm{b}, \varsigma = \pm 1} \hspace{-5pt} \theta_{ml}\left(E^f_{ml} k^f_{ml}  - M_{ml}^2 {\rm artanh} \frac{k^f_{ml}}{E^f_{ml}}\right),
\end{multline}
where $E_{ml}(k) = \sqrt{k^2 + M_{ml}^2}$ is the Minkowski energy, $M_{ml} = \sqrt{q_{ml}^2 + g^2 \sigma^2}$ is the transverse mass, and $\theta_{ml} = \theta(\varsigma \mu + \Omega m - M_{ml})$. For notational brevity, we introduced the Fermi energy $E^f_{ml} = |\varsigma \mu + \Omega m|$ and the Fermi momentum, $k^f_{ml} = \sqrt{(E^f_{ml})^2 - M_{ml}^2}$. The saddle-point equation becomes
\begin{equation}
 \frac{\partial \widetilde{F}_q^{T= 0}}{\partial \sigma} =  \frac{2 g^2 \sigma N_f N_c}{\pi^2 R^2} \sum_{\mathrm{b}, \varsigma = \pm 1} \theta_{ml}
 \mathrm{artanh} \frac{k^f_{ml}}{E^f_{ml}}.
\end{equation}
For the evaluation of the slope, $\partial \sigma / \partial \mu$, we require:
\begin{align} \label{eq:d2F0T}
 \frac{\partial^2 \widetilde{F}^{T=0}_q}{\partial\mu \partial \sigma} &= \frac{2 g^2 \sigma N_f N_c}{\pi^2 R^2} \sum_{\mathrm{b}, \varsigma = \pm 1} \frac{\varsigma}{k^f_{ml}} \theta_{ml},\nonumber\\
 \frac{\partial^2 \widetilde{F}^{T=0}_q}{\partial \sigma^2} &= -\frac{2 g^2 N_f N_c}{\pi^2 R^2} \sum_{\mathrm{b}, \varsigma = \pm 1} \theta_{ml} \nonumber\\
 &\times \left(\frac{g^2 \sigma^2 E^f_{ml}}{M_{ml}^2 k^f_{ml}} - \mathrm{artanh} \frac{k^f_{ml}}{E^f_{ml}}\right). 
\end{align}

The presence of $k_{ml}^f$ in the denominators of some of the terms appearing in Eq.~\eqref{eq:d2F0T} indicates that, at small temperatures, the slope of $\sigma$ will diverge whenever modes for which $k_{ml}^f$ vanishes are supported by the system. In particular, the step function $\theta(\varsigma \mu + \Omega m - M_{ml})$ ``fires'' precisely when $k_{ml}^f = 0$, such that when $\mu$ is continuously increased, leading to the inclusion of a new mode in the $\mathrm{b}$ summation, the contribution of this mode will be arbitrarily large. In practice, this is unlikely to happen, as the mean-field value of $\sigma$ is constantly adjusted, eventually leading to a finite $k_{ml}^f$ at the point where the grand potential takes its minimum. Signatures of this effect are still visible, as the slope can still develop peaks whenever a new root is allowed within the Fermi level. These oscillations are analogues of the Shubnikov-de Haas effect arising from the periodic modulation of the density of states at the Fermi level in a conducting material subjected to a background external magnetic field~\cite{Schubnikow1930}. Similar effects have also been observed in other model calculations, for example, in Refs.~\cite{Almasi:2016zqf, Xu:2019gia, Kovacs:2023kbv}, where irregular oscillations of thermodynamic quantities are associated with finite-size effects.

In panel (a) of Fig.~\ref{fig:lowT}, we demonstrate this effect at small, but non-vanishing temperature, for a small, non-rotating system of $R = 1$\,fm, by considering the variation of $\sigma$ and $\partial \sigma / \partial G$ along a diagonal phase-space line with $(T,\mu) = G (\cos\theta, \sin\theta)$, at a fixed value of $\theta$ (we took $\tan \theta = 54.5$). While our analysis in this subsection is focused mainly on the behavior of $\sigma$ and its slope, panel (b) of Fig.~\ref{fig:lowT} indicates that the peak structure is also inherited in the slope $\partial L / \partial G$ of the Polyakov loop. Moreover, our results show that following the development of one peak, the slope can become negative, leading to a non-monotonic dependence of $L$ on the control parameter, $G$.

The above discussion, together with Fig.~\ref{fig:lowT}, illustrates the conceptual difficulties in pinpointing the parameters of the deconfinement phase transition. At low temperatures, the first-order chiral transition unambiguously occurs at some value of $G$. At this point, the Polyakov loop exhibits only a modest jump, such that the system remains in a confined state. As $G$ increases, the Polyakov loop exhibits a generally increasing trend, but its nonmonotonicity makes it difficult to identify the point of highest peak in its slope. In practice, our algorithm picks the parameters of the deconfinement transition by looking for the closest peak in the vicinity of $L_{\rm th} = 0.3$.

\section{Phase Diagram}\label{sec:res}

\begin{figure}
\centering
\begin{tabular}{c}
\includegraphics[width=0.92\linewidth]{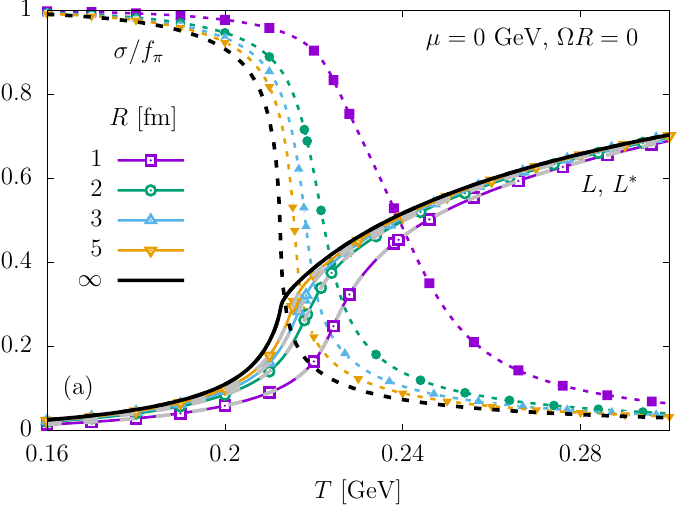} \\
\includegraphics[width=0.92\linewidth]{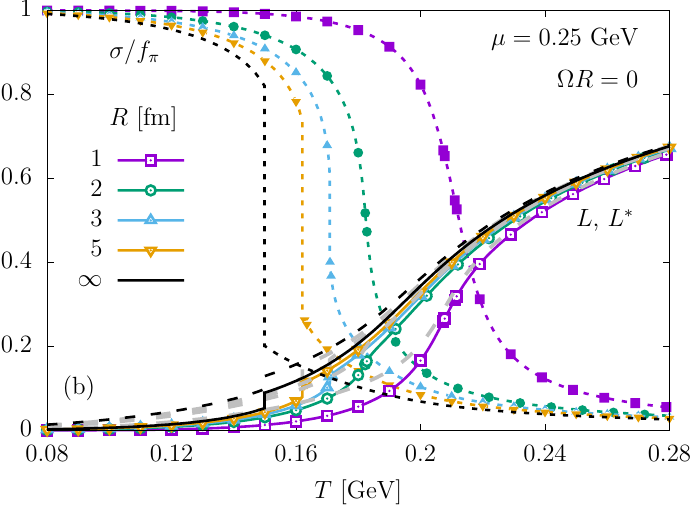}
\end{tabular}
\caption{Variation of $\sigma/f_\pi$ (dotted curves with filled symbols, starting at $1$), $L$ (solid curves with empty symbols) and $L^*$ (gray dashed lines) for a static ($\Omega R = 0$) cylindrical system of various radii for (a) $\mu = 0$ GeV and (b) $\mu = 0.25$ GeV.
}
\label{fig:rot0_sL}
\end{figure}

We now discuss the phase transition properties of strongly interacting matter enclosed in a cylindrical boundary, with or without rotation. For brevity, we will only consider the logarithmic potential~\eqref{eq_L_Polyakov} for the Polyakov loop in what follows.

We first remark that the effects of the boundary and/or of the rotation manifest themselves only through the fermionic contributions to the saddle-point equations. On one hand, we anticipate that a reduced volume (low radius $R$) will lead to a suppression of the fermionic contributions, due to the additional mass gap introduced by the transverse momentum quantization. Thus, the presence of the boundary produces an effect of delaying chiral restoration. Deconfinement also takes place at higher temperatures compared to the case without boundaries, up to the point where the temperature approaches that of the pure glue system, $T_0$. Since the Polyakov potential $V_L$ is not affected by either boundary or rotation, the deconfinement transition, in the absence of fermionic contributions, occurs as in the pure glue system described in Sec.~\ref{sec:PLSM:Polyakov}, namely as a first-order transition at $T = T_0$. When the radius increases, the gap due to the boundary becomes smaller and we can expect to recover the thermodynamics of the unbounded system.

On the other hand, rotation has the effect of amplifying the thermal fermionic contributions, as suggested by the TE law~\cite{Tolman:1930ona, Tolman:1930zza}.
The strength of the effect is characterized by the velocity of the system at the boundary, $v_R = \Omega R$, or, more precisely, by its Lorentz factor $\Gamma_R = (1 - \Omega^2 R^2)^{-1/2}$. Compared to the static system, the chiral phase transition in the rotating medium should occur at lower temperatures, as indicated in Eq.~\eqref{eq:TE_PLSM}, discussed in the following section of this paper. While Eq.~\eqref{eq:TE_PLSM} predicts that the chiral restoration temperature $T_\sigma^{\Omega R}$ vanishes when $\Omega R \rightarrow 1$, this is not the case at any finite radius $R < \infty$. The finite mass gap induced by the boundary is sufficient to render the system causal and all expectation values finite, even in the causality limit $\Omega R \rightarrow 1$. Thus, at finite $R$, quantum effects have a sizable influence
on the system thermodynamics, especially at low $R$. When the radius of the system is pushed to infinity, keeping $\Omega R$ fixed, we can expect these quantum effects to become subleading and to recover the Tolman-Ehrenfest predictions.

We address the non-rotating case in Subsec.~\ref{sec:res:rot0}, with an emphasis on the suppression effects of the boundary and the restoration of the unbounded system thermodynamics as $R \rightarrow \infty$. The rotating system is discussed in Subsec.~\ref{sec:res:rot}, where we explore the influence of the rigid rotation on the chiral and deconfinement phase transitions in cylindrical systems with radii of a few fermi that match the sizes of typical realistic systems.

\subsection{Static system: Effect of boundary}
\label{sec:res:rot0}

We begin the analysis of the effect of the boundary with a set of plots, shown in Fig.~\ref{fig:rot0_sL}, presenting the temperature dependence of $\sigma$, $L$ and $L^*$ for different values of the radius $R$ and two different values of the chemical potential: $\mu=0$ (panel a) and $\mu=0.25$ GeV (panel b). For vanishing chemical potential, all control parameters $\sigma$, $L$ and $L^*$ exhibit a smooth variation with respect to $T$. For $\mu = 0.25$ GeV, the first-order phase transition expected from the unbounded case (shown with black lines, marked as ``$R=\infty$'' and discussed in Sec.~\ref{sec:PLSM}) gradually softens into a crossover transition as the system size $R$ is decreased. The softening effect becomes observable for $R \lesssim 3$ fm. The transition point, be it of first-order or crossover type, moves to higher temperatures as $R$ is decreased.
This behavior agrees with the previously reported results with the bag boundary conditions in Ref.~\cite{Chernodub:2016kxh}.

\begin{figure}
\centering
\begin{tabular}{cc}
\includegraphics[width=0.92\linewidth]{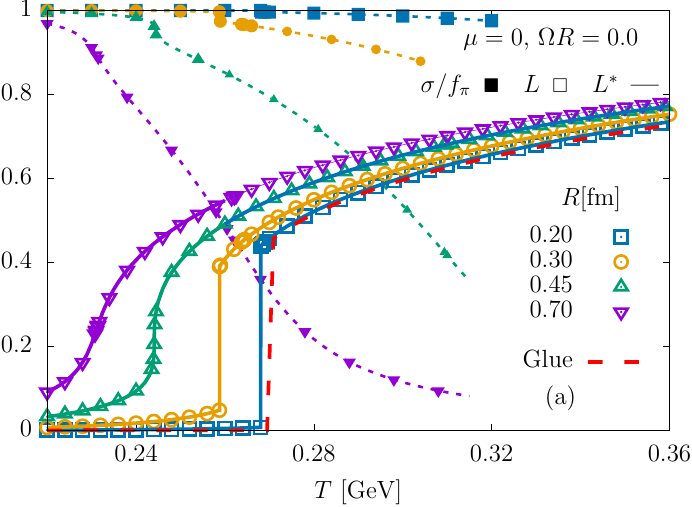} \\
\includegraphics[width=0.92\linewidth]{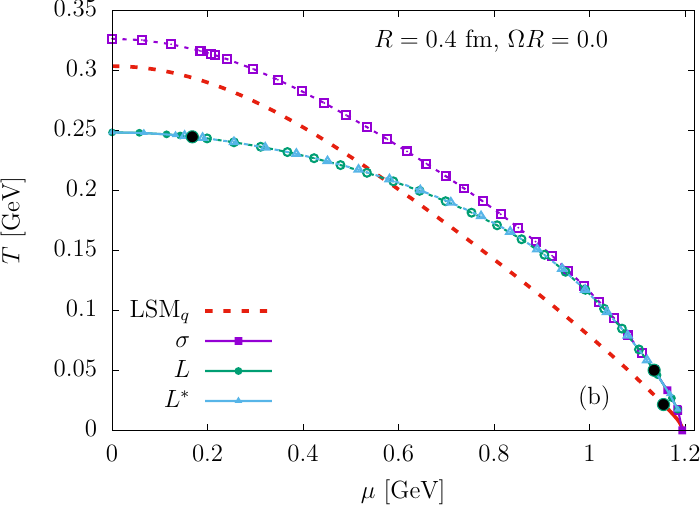}
\end{tabular}
\caption{
(a) Temperature dependence of $\sigma/f_\pi$ (the dotted lines with filled symbols), $L$ (the empty symbols) and $L^*$ (the solid lines) for a static system at vanishing chemical potential, for small system sizes. The dashed red line represents the Polyakov loop for the pure glue system in the absence of fermions, obtained by solving Eq.~\eqref{eq_Polyakov_minimization}.
(b) Phase diagram in the $T$-$\mu$ plane for the static system with $R = 0.4$ fm, showing the chiral (purple) and deconfinement (green and cyan for $L$ and $L^*$) transitions in the PLSM$_q$ model and the chiral transition (red) in the LSM$_q$ model at zero angular velocity. The dashed (solid) lines denote a crossover (first-order) transition. The black circles indicate critical points.
}
\label{fig:smallR}
\end{figure}

We now investigate the restoration of the first-order deconfinement phase transition in the limit where the fermionic contributions are suppressed due to the small system size. For simplicity, we focus on the case of vanishing chemical potential.
Figure~\ref{fig:smallR}(a) confirms that as $R < 1$\,fm, both the deconfinement and the chiral restoration transitions are inhibited. As $R \rightarrow 0$, the Polyakov sector decouples from the fermionic sector and the deconfinement transition takes place as predicted by the pure glue theory discussed in Sec.~\ref{sec:PLSM:Polyakov}, namely $L = L^* = 0$ for $T < T_0$, while for $T > T_0$, $L = L^* = L^{(+)}$ introduced in Eq.~\eqref{eq:logarithmic_Lpm}. This solution is shown with a red dashed line in Fig.~\ref{fig:smallR}(a) and it can be seen that it approximates very well the results obtained at $R = 0.2$ fm. As $R$ increases, the transition temperature decreases, but the deconfinement transition remains of first order, until a critical radius $R_c \simeq 0.45$ fm is reached. When the radius is further increased, the deconfinement transition becomes of crossover type. Remarkably, the chiral restoration transition takes place at significantly larger temperatures.

Panel (b) of Fig.~\ref{fig:smallR} shows the phase diagram in the $T$-$\mu$ plane for $R = 0.4$ fm. As $\mu$ increases, the deconfinement phase transition softens as the fermion contributions become more sizable, thereby developing a second endpoint. In the region of intermediate $\mu$, both the chiral and the deconfinement phase transitions are of crossover type. At large $\mu$ and small $T$, the two transitions merge into a single critical point, signaling the onset of the first-order regime for the chiral phase transition. We conclude that at small $R$, the chiral and deconfinement transitions not only split with respect to temperature, but they also possess different strengths (the chiral crossover vs. the first-order deconfinement transition). In Ref.~\cite{Singha:2024tpo}, we have shown that the boundary-induced splitting disappears when $R$ is increased. Thus, the emergence of the second, low-$\mu$ endpoint can be attributed to the effect of the small transverse size of the system.

However, on physical grounds, one could expect that since the confining QCD string has a transverse size of the order of $1$~fm, smaller sizes of the system can substantially affect confinement properties. The finite-size effect has indeed been observed in simulations in a pure SU(3) gluonic system~\cite{Chernodub:2018aix}. The motivation to study the small-radius region in our paper is to point out that our model does not incorporate such effects, since the Polyakov potential does not depend on the type of boundary conditions employed. 

\begin{figure}
\centering
\begin{tabular}{cc}
\includegraphics[width=0.92\linewidth]{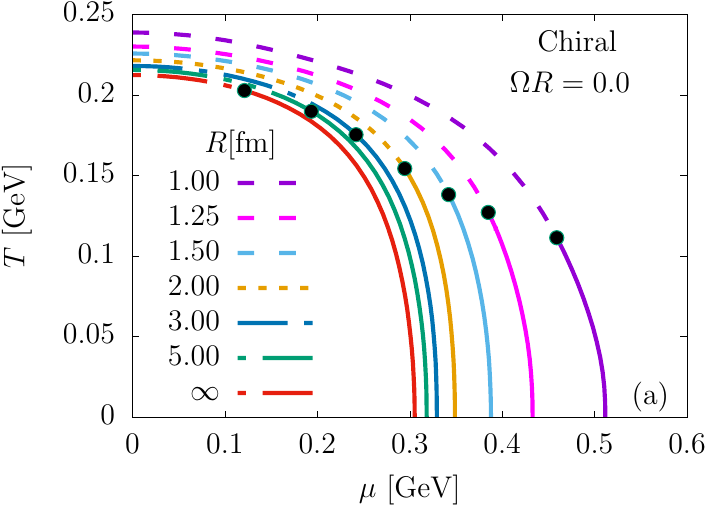} \\
\includegraphics[width=0.92\linewidth]{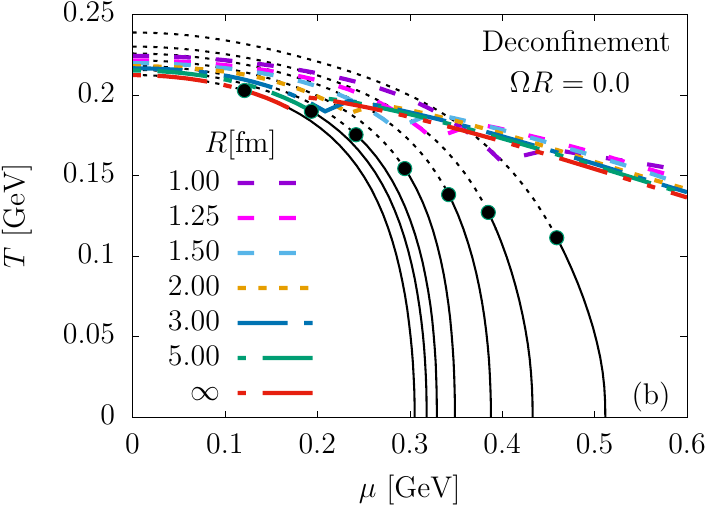}
\end{tabular}
\caption{ Phase diagram in the $T$-$\mu$ plane for (a) the chiral restoration and (b) deconfinement at zero angular velocity for different radii $R$. The solid and dashed line segments mark the positions of, respectively, the first-order phase transitions and crossovers separated by the endpoints (the black-filled circles). The black lines in panel (b) indicate the chiral transition lines.
}
\label{fig:rot0_PD}
\end{figure}

We now investigate the effect of the boundary on the phase diagram of the system. Figure~\ref{fig:rot0_PD} shows the phase diagram in the $T$-$\mu$ plane indicating separately (a) the chiral symmetry restoration and (b) the deconfinement transition lines. Starting from $R = 1$ fm,  the transition lines for the bounded system approach gradually the transition of the unbounded system (indicated as $R = \infty$) as the radius $R$ increases. Generally, the critical point separating the crossover and first-order phase transition regions travels towards higher temperatures and smaller chemical potentials as the system size is increased. The behavior shown in Fig.~\ref{fig:rot0_PD}(a) is qualitatively similar to that found in Fig.~3(b) of Ref.~\cite{Kovacs:2023kbv}, in the frame of the extended Polyakov quark-meson model with $N_f = 3$ flavors, with the boundary emulated by a low-momentum cutoff, in the case when the vacuum term is treated as in the unbounded case (as also done in our present work). Fig.~3(a) of Ref.~\cite{Kovacs:2023kbv} shows that including the low-momentum cutoff also in the vacuum term leads to an opposite behavior, in which the decrease of the system size leads to a decrease of the temperature and chemical potential characterizing the chiral transition, a situation that we have not explored in our work.

\subsection{System under rotation}\label{sec:res:rot}

\begin{figure*}
\centering
\begin{tabular}{cc}
\includegraphics[width=0.46\linewidth]{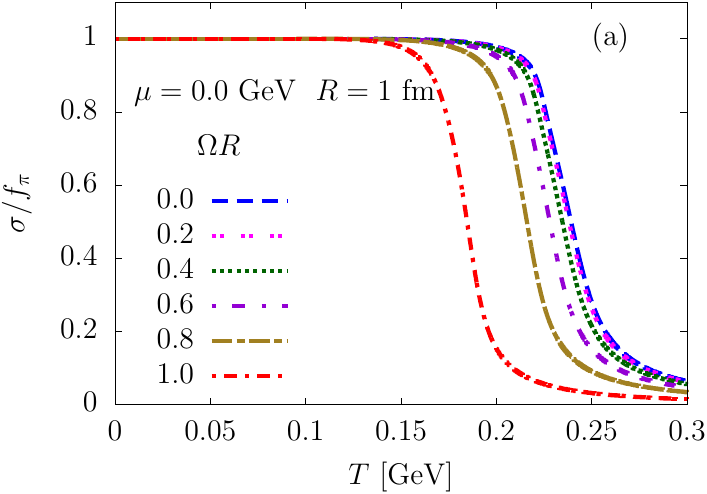}
\includegraphics[width=0.46\linewidth]{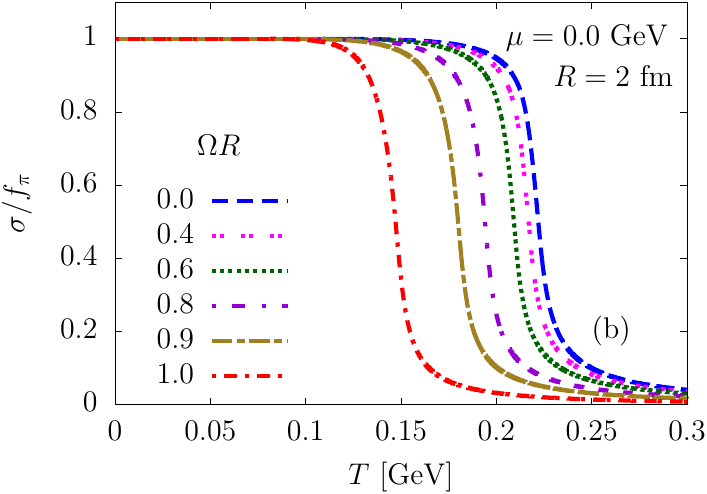}
\\
\includegraphics[width=0.46\linewidth]{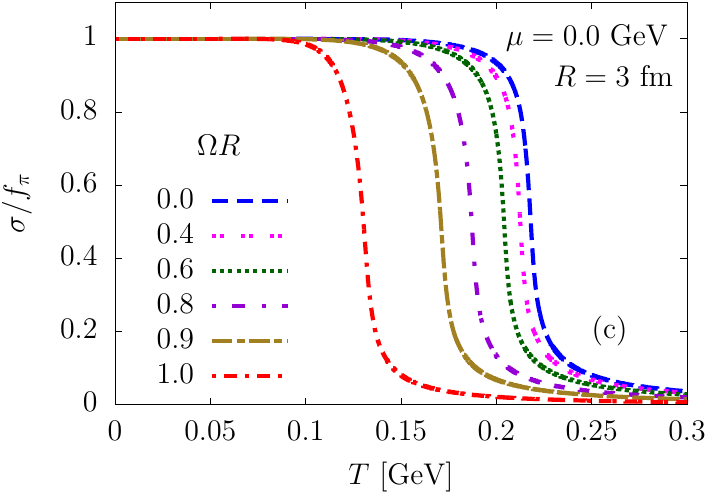}
\includegraphics[width=0.46\linewidth]{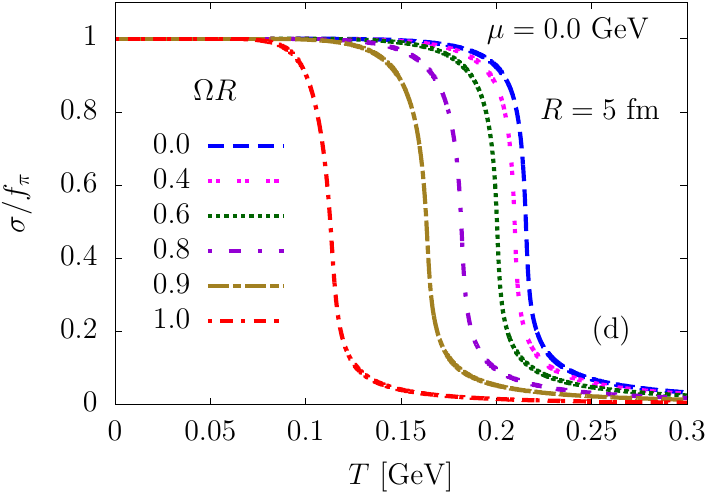}
\end{tabular}
\caption{Normalized scalar condensate $\sigma/f_
\pi$, at $\mu=0$, as a function of temperature $T$, for different angular velocities and various system sizes: (a) $R = 1$, (b) $2$, (c) $3$ and (d) $5$ fm.}
\label{fig:sigmaRotmu0}
\end{figure*}

\begin{figure*}
\centering
\begin{tabular}{cc}
\includegraphics[width=0.46\linewidth]{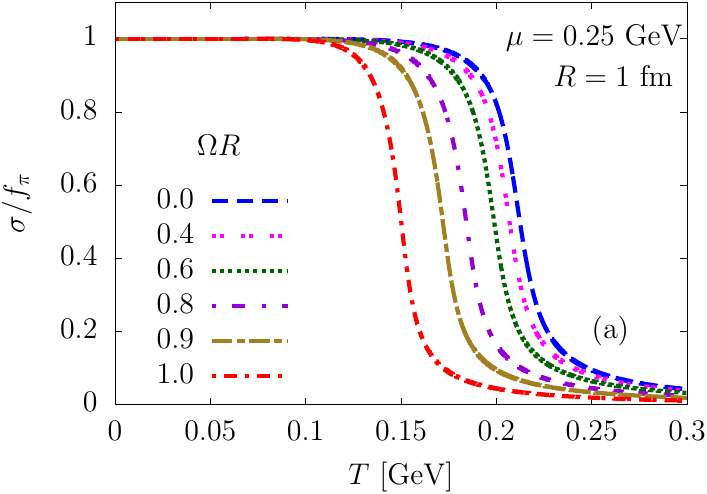}
\includegraphics[width=0.46\linewidth]{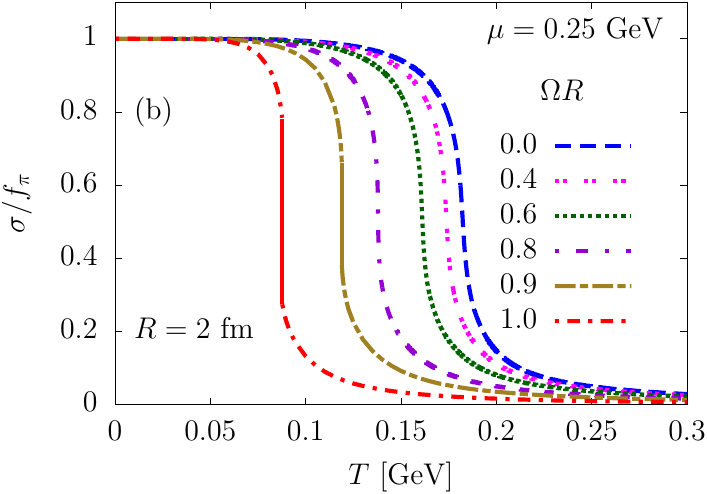}
\\
\includegraphics[width=0.46\linewidth]{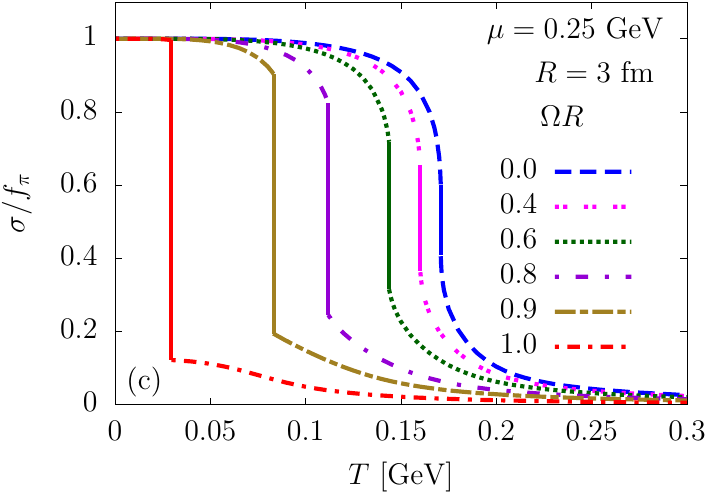}
\includegraphics[width=0.46\linewidth]{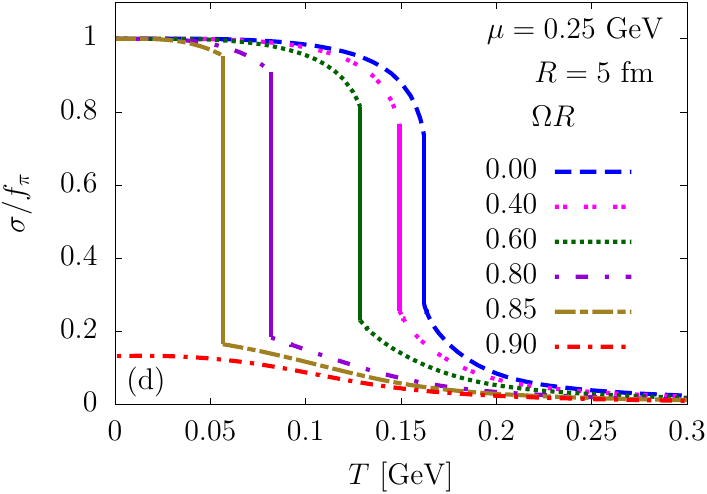}
\end{tabular}
\caption{Same as Fig.~\ref{fig:sigmaRotmu0} for non-vanishing chemical potential $\mu = 0.25$ GeV.}
\label{fig:sigmaRotmu0.25}
\end{figure*}

We now turn on rotation. Figures~\ref{fig:sigmaRotmu0} and \ref{fig:sigmaRotmu0.25} show the effect of rotation on the temperature dependence of the $\sigma$ expectation value for various system sizes ($R = 1$, $2$, $3$ and $5$ fm), at $\mu = 0$ and $\mu = 0.25$ GeV, respectively. It can be seen that for all these parameters, rotation promotes the chiral symmetry restoration, leading to a decrease of $\sigma$ as $\Omega R$ is increased at fixed $R$, $\mu$ and $T$. This result was reported in the recent literature~\cite{Sun:2023kuu}, albeit for a system without boundaries.
Furthermore, panels (b)--(d) of Fig.~\ref{fig:sigmaRotmu0.25} indicate that increasing $\Omega R$ pushes the system towards a first-order phase transition, similar to the effect of increasing the radius $R$. In the case of a system of radius $R = 5$ fm at $\mu = 0.25$ GeV, chiral symmetry is restored at any temperature when $\Omega R \gtrsim 0.9$.
Similarly, we can expect that increasing $\Omega R$ leads to an increase of the expectations of the traced Polyakov loop $L$ and its conjugate, $L^*$, leading to a decrease of the deconfinement transition temperature.

\begin{figure}
\centering
\begin{tabular}{c}
\includegraphics[width=0.92\linewidth]{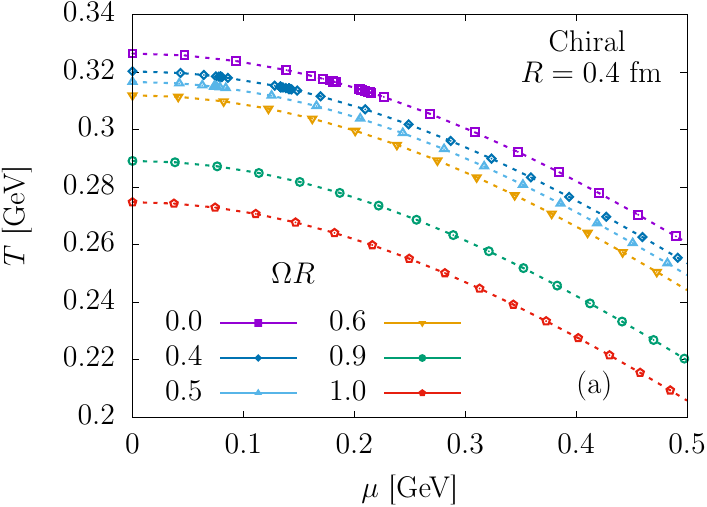} \\
\includegraphics[width=0.92\linewidth]{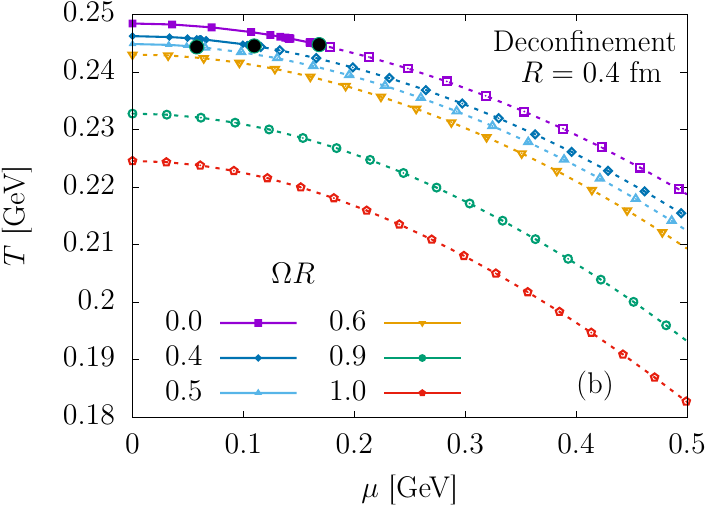}
\end{tabular}
\caption{Phase diagrams for the chiral (a) and deconfinement (b) transitions at $R = 0.4$ fm, for various angular velocities.
The dashed (solid) lines denote crossover (first-order) transitions. The black circles indicate critical points. Both panels show a small range of chemical potential, focusing on the region where the two transitions split. The low-$\mu$ critical point for the deconfinement transition touches the vertical axis for $\Omega R \simeq 0.53$, disappearing at higher angular velocities.
}
\label{fig:PD_smallR}
\end{figure}

\begin{figure*}
\centering
\begin{tabular}{cc}
\includegraphics[width=0.46\linewidth]{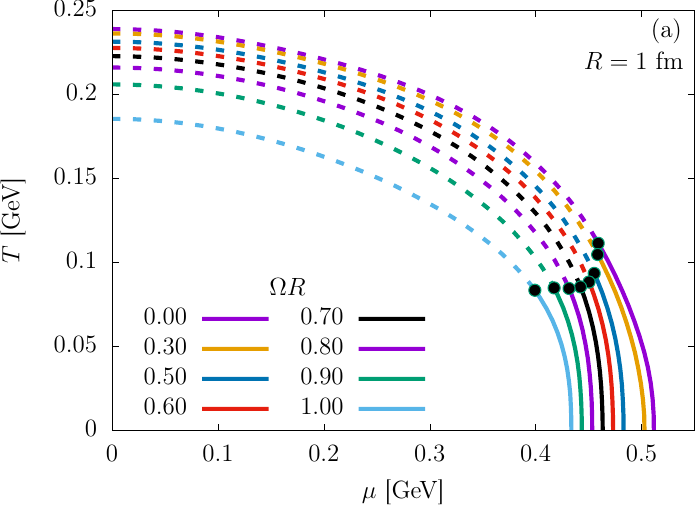} &
\includegraphics[width=0.46\linewidth]{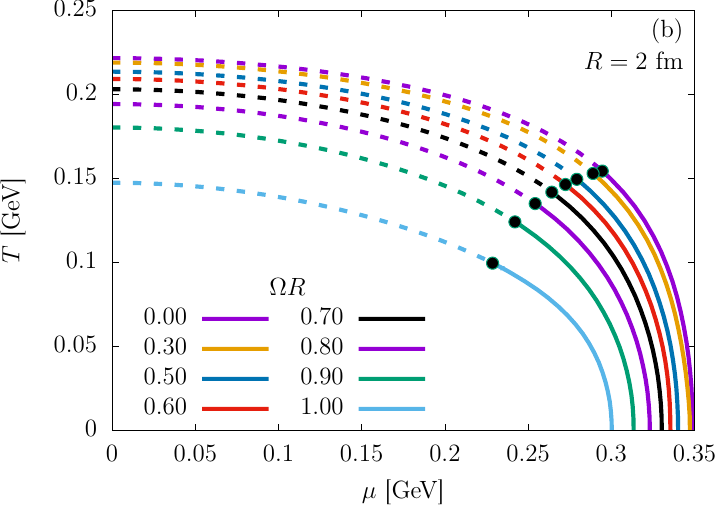}\\
\includegraphics[width=0.46\linewidth]{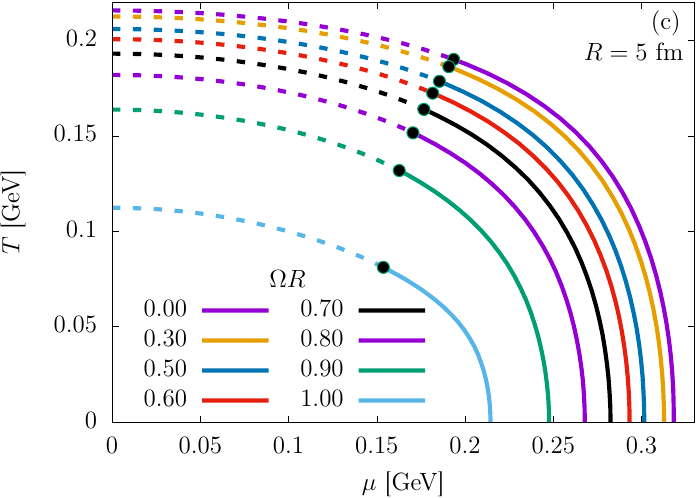}&
\includegraphics[width=0.46\linewidth]{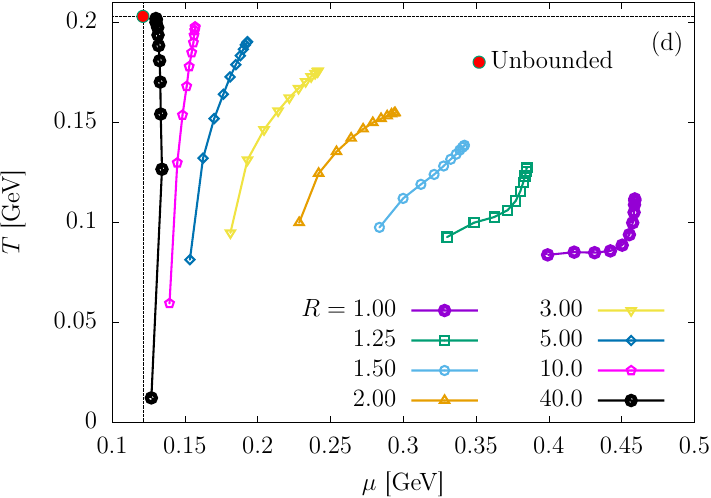}
\end{tabular}
\caption{Phase diagram for the chiral transition in the $T$-$\mu$ plane, at fixed angular velocities and various sizes: (a) $R=1$; (b) $2$; and (c) $5$ fm. (d) The trajectory of the critical point in the $T$-$\mu$ plane as a function of $\Omega R$, for various radii $R$, expressed in fm. The critical point corresponding to the static, unbounded system is indicated by a red dot at the top left of panel (d).
}
\label{fig:Phase_diagramRot}
\end{figure*}

\begin{figure}
\centering
\begin{tabular}{c}
\includegraphics[width=0.92\linewidth]{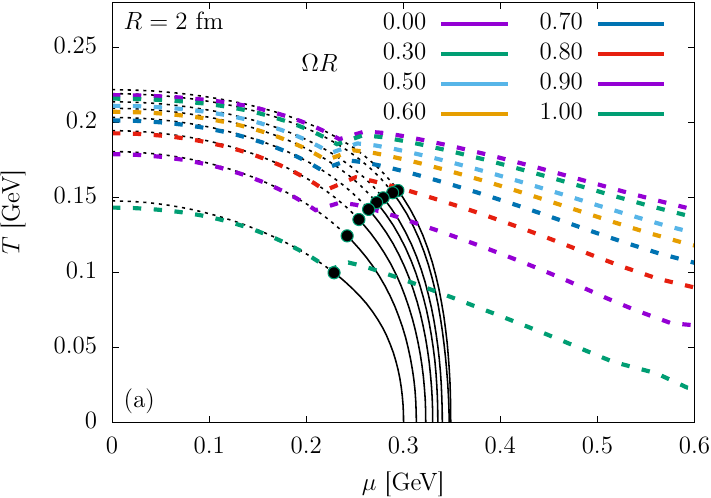} \\
\includegraphics[width=0.92\linewidth]{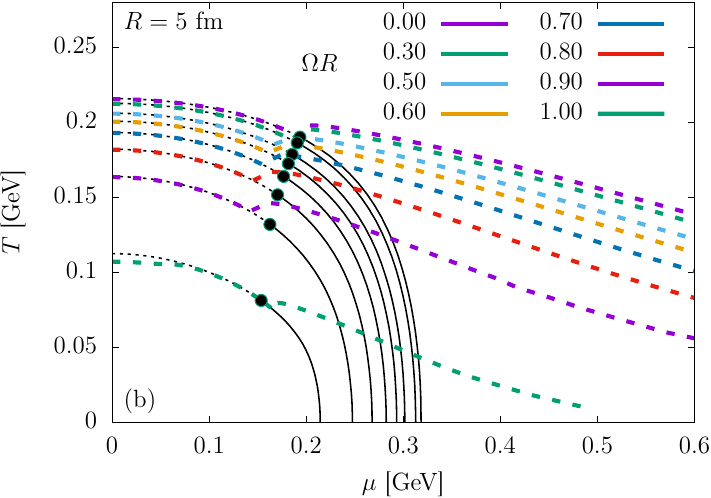}
\end{tabular}
\caption{Phase diagram for the deconfinement transition, shown with colored lines, computed in the $T{-}\mu$ plane at selected angular velocities, for the system sizes: (a) $R=2$ and (b) $5$ fm. The black lines represent the chiral transition lines shown also in Figs.~\ref{fig:Phase_diagramRot}(b) and (c), respectively.
}
\label{fig:Phase_diagramRot_L}
\end{figure}

We now consider the effect of rotation on the phase diagram of the model. Considering the system size $R$ and the rotation parameter $\Omega$ fixed, we constructed the phase diagram by considering the variations of the control parameters $\sigma$, $L$ and $L^*$ with respect to $G=\sqrt{T^2+\mu^2}$. We discuss the phase diagrams for the chiral restoration and deconfinement transitions for a small system ($R = 0.4$ fm) in panels (a) and (b) of Fig.~\ref{fig:PD_smallR}. As expected, increasing $\Omega$ promotes both transitions. Moreover, the splitting shown in Fig.~\ref{fig:smallR}(b) for the static case is decreased as $\Omega$ increases, from about $70$ MeV at $\Omega R = 0$ to about $50$ MeV at $\Omega R = 1$. Generally, rotation reduces the splitting between the chiral and deconfinement transitions, as reported in Ref.~\cite{Singha:2024tpo}.

Figure~\ref{fig:Phase_diagramRot} further presents the phase diagrams for the chiral transition in systems of sizes (a) $R = 1$, (b) $2$ and (c) $5$ fm. In all cases, as $\Omega R$ is increased, the transition line moves monotonically, without crossings, towards lower temperatures and chemical potentials. Even in the causal limit, when $\Omega R = 1$, all systems behave regularly, provided the system size $R$ is finite.

Rotation maintains the phase structure, the system exhibiting a crossover transition at small $\mu$ and a first-order phase transition at large $\mu$, separated by a critical point. The trajectory of the critical point in the $(T,\mu)$ plane as $\Omega R$ is varied at fixed $R$ is shown in Fig.~\ref{fig:Phase_diagramRot}(d), for various values of $R$. At small $R$, the critical point travels almost on rectangular segments: first towards decreasing temperatures at fixed chemical potential; then towards lower chemical potentials at fixed $T$. At intermediate values of $R$, this trajectory becomes more diagonal, until eventually, at $R = 10$ fm, it becomes an almost vertical line: the chemical potential at the critical point seems to become independent of rotation. Conversely, the temperature of the critical point in the causal limit $\Omega R \to 1$ migrates towards vanishing values. At the same time, as $R$ is increased, the critical point for the non-rotating case ($\Omega R = 0$) approaches that of the static, unbounded system, shown as a red dot in Fig.~\ref{fig:Phase_diagramRot}(d).

The phase diagram for the deconfinement transition is shown in Fig.~\ref{fig:Phase_diagramRot_L}. As already mentioned in Figs.~\ref{fig:rot0_PD}(b) and \ref{fig:PD}(a,b), the deconfinement transition is delayed with respect to the chiral one when the chemical potential exceeds some threshold value. This is confirmed also in the case of rotation: at low values of $\mu$, the deconfinement transition is either simultaneous with the chiral one (large $R$), or it takes place earlier (small $R$). At large $\mu$, the deconfinement transition takes place after the chiral one, being a crossover, even when the chiral transition is of first-order type.

\begin{figure*}
\centering
\begin{tabular}{cc}
\includegraphics[width=0.46\linewidth]{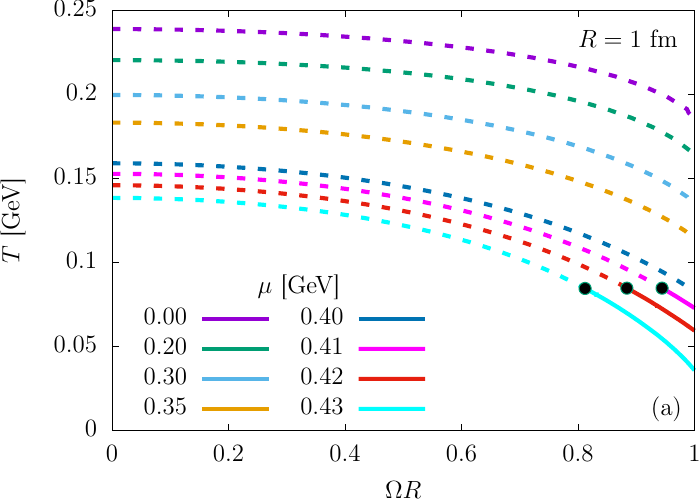} &
\includegraphics[width=0.46\linewidth]{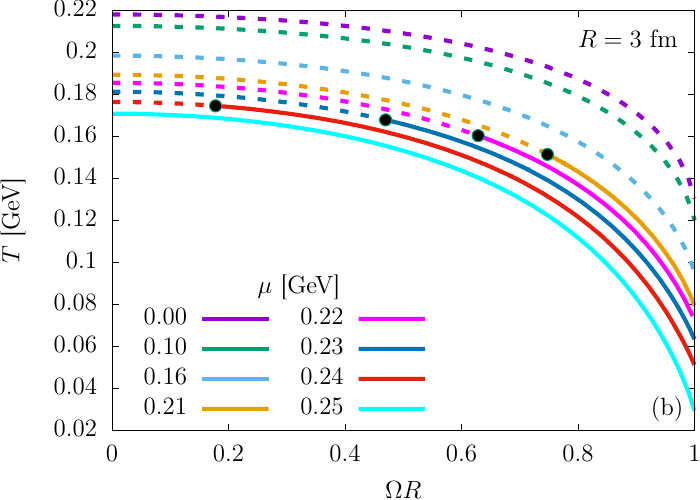} \\
\includegraphics[width=0.46\linewidth]{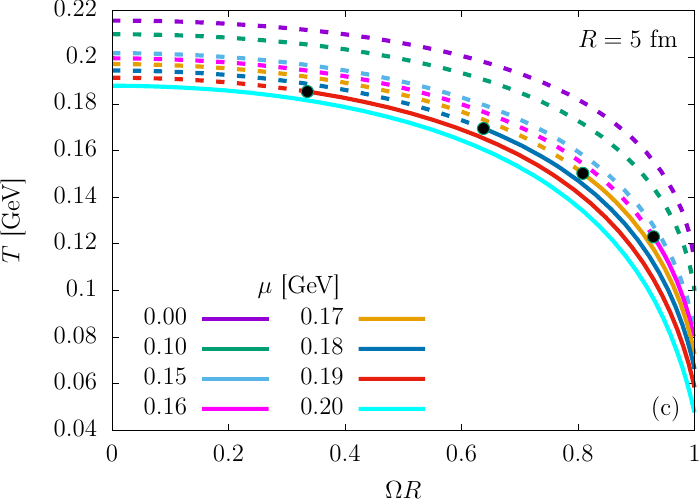} &
\includegraphics[width=0.46\linewidth]{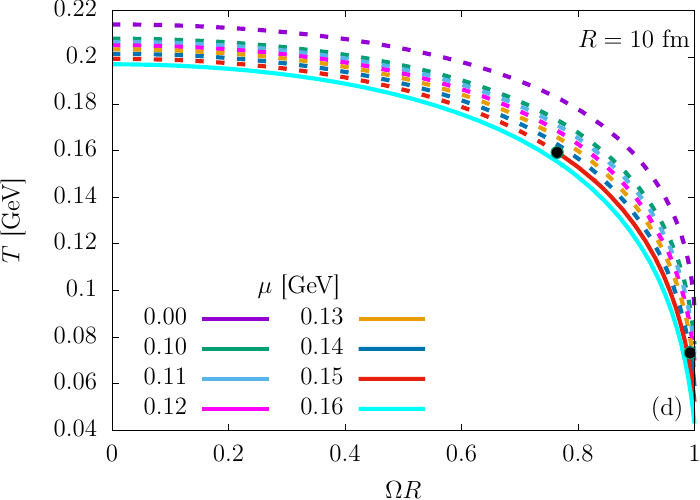}
\end{tabular}
\caption{Phase diagram in the $T{-}\Omega R$ plane for the chiral transition, computed at selected values of the chemical potential $\mu$, for systems of various sizes: (a) $R=1$; (b) $3$; (c) $5$ and (d) $10$ fm.}
\label{fig:Phase_diagramTOmega}
\end{figure*}

Coming back to panels (b)--(d) of Fig.~\ref{fig:sigmaRotmu0.25}, we have seen that increasing $\Omega R$ can turn a crossover transition into a first-order one. To systematically study this effect, we constructed in Fig.~\ref{fig:Phase_diagramTOmega} the phase diagram of the chiral phase transition in the $T$-$\Omega R$ plane, computed by considering variations of the control parameters $\sigma$, $L$ and $L^*$ with respect to $T$ and $\Omega R$, at various values of $\mu$, for (a) $R = 1$, (b) $2$, (c) $5$ and (d) $10$ fm. As the parameter range of $\Omega R$ is restricted between $0$ and $1$, the transition line has a finite extent also with respect to temperature, extending down to the transition temperature corresponding to $\Omega R = 1$. When the system undergoes a crossover transition, we compute the transition point by maximizing the slope along a line in the $T$-$\Omega R$ plane, namely
\begin{equation}
 \frac{dp}{d G}(\theta) = \cos\theta \frac{\partial p}{\partial T} + \alpha \sin\theta \frac{\partial p}{\partial \Omega R},
\end{equation}
where $G = \sqrt{T^2 + (\Omega R / \alpha)^2}$, with the constant $\alpha$ required for dimensional reasons. For definiteness, we take it as $\alpha = 1\ {\rm GeV}^{-1}$, while $\tan \theta = \Omega R /\alpha T$.

We now consider the trajectory of the critical point in the $T$-$\Omega R$ plane. It can be seen in Fig.~\ref{fig:Phase_diagramTOmega} that the critical point moves towards the lower $\Omega R$ as $\mu$ increases. Figure~\ref{fig:Phase_diagramTOmega}(c) shows that, when $R = 5$ fm, the critical point in the $T$-$\Omega R$ phase diagram exists for a very narrow range of chemical potentials, when $0.16$ GeV $\lesssim \mu \lesssim 0.2$ GeV. For values of $\mu$ smaller than this range, the transition is of crossover type for all values of $\Omega R$; while for larger $\mu$, the phase transition becomes of first order for all values of $\Omega R$. This property can be understood by going back to Fig.~\ref{fig:Phase_diagramRot}(d), where we showed the trajectory of the critical point as a function of $\Omega R$ for different values of $R$ in the $T$-$\mu$ plane. We observe that, as $R$ increases, the trajectories become almost parallel to the temperature axis, thus the range of chemical potential over which the critical point moves as a function of $\Omega R$, at fixed radius $R$, decreases as $R$ is increased.

\section{Comparison to Tolman-Ehrenfest law}\label{sec:TE}

We complete our study by clarifying to what extent the Tolman-Ehrenfest picture agrees with the LSM${}_q$ and PLSM${}_q$ results. We will discuss the Tolman-Ehrenfest prediction for a rigidly rotating system and present a comparative study to investigate the range of applicability of such classical ideas to study the phase structure of the (P)LSM${}_q$ models under rotation.

We calculate the chiral transition temperature under rotation within our effective model framework, under the assumption of validity of the TE law for the local thermodynamic state for a system under rotation. In Subsec.~\ref{sec:TE:kinetic}, we formulate the rotating (P)LSM$_q$ model within kinetic theory. We then apply our discussions to the LSM$_q$ and PLSM$_q$ models for vanishing chemical potential in Subsections~\ref{sec:TE:LSM} and \ref{sec:TE:PLSM}. We discuss the vanishing temperature case in Subsec.~\ref{sec:TE:T0}. Subsection~\ref{sec:TE:comparison} provides numerical validation of our TE predictions.

\subsection{Kinetic theory formulation of the rotating system}\label{sec:TE:kinetic}

As is well known, a rigidly rotating state is described by the following temperature four-vector \cite{Cercignani:2002}:
\begin{equation}
 \beta^\mu \partial_\mu = \frac{1}{T} (\partial_t + \Omega \partial_\varphi).
\end{equation}
Writing $\beta^\mu = \beta(x) u^\mu$, we can extract the local temperature $T_\rho = 1 / \sqrt{\beta^\mu \beta_\mu}$ and the four-velocity,
\begin{equation}
 T_\rho = \Gamma_\rho T, \quad u^\mu \partial_\mu = \Gamma_\rho (\partial_t + \Omega \partial_\varphi),
\end{equation}
where $\Gamma_\rho = (1 - \rho^2 \Omega^2)^{-1/2}$ is the Lorentz factor at distance $\rho = \sqrt{x^2 + y^2}$ from the rotation ($z$) axis.

At the level of statistical mechanics, rigid rotation can be modeled by replacing the density operator $\hat{\rho} = e^{-\beta (\widehat{H} - \mu \widehat{Q})} \to e^{-\beta(\widehat{H} - \Omega \widehat{J}^z - \mu \widehat{Q})}$, where $\widehat{H}$ is the Hamiltonian, $\widehat{Q}$ is the charge and $\widehat{J}^z$ is the total angular momentum along the rotation axis. For a classical particle with momentum $p^\mu$, the orbital angular momentum along the $z$ axis is $L^z = x p^y - y p^x$, such that in kinetic theory, implementing rigid rotation amounts to changing the Boltzmann factor as follows:
\begin{equation}
 e^{-\beta \mathcal{E}_\varsigma} \rightarrow e^{-\beta (p^0 +\Omega y p^x - \Omega x p^y - \varsigma \mu)} = e^{-\beta_\rho (u_\mu p^\mu - \varsigma \mu_\rho)},
\end{equation}
where $\beta_\rho = 1/T_\rho$ is the local inverse temperature, $\mu_\rho = \mu \Gamma_\rho$ is the local chemical potential (such that $\beta_\rho \mu_\rho = \beta \mu = {\rm const}$ \cite{Cercignani:2002}) and the four-vector $u^\mu$ is the velocity corresponding to rigid rotation. The above procedure is equivalent to the local thermal equilibrium approximation for the density operator $\hat{\rho}$, discussed in Refs.~\cite{Becattini:2014yxa,Becattini:2015nva,Buzzegoli:2018wpy}. We apply this prescription at the level of the quark grand potential:
\begin{multline}
F_q = -\frac{2N_f N_c T}{V} \int_V d^3x \sum_{\varsigma = \pm 1} \int \frac{d^3p}{(2\pi)^3} \\\times \left.F_{\varsigma}\right\rvert_{\beta \mathcal{E}_\varsigma \to \beta_\rho (u_\mu p^\mu - \varsigma \mu_\rho)},
\end{multline}
where $V$ represents a cylindrical volume of radius $R$, while $F_{\varsigma}$ is given in Eq.~\eqref{eq:static_Fs}.

We now perform a Lorentz transformation to the local rest frame, such that $u^\mu \rightarrow \delta^\mu_0$. The integration measure $d^3p$ is not Lorentz covariant, however, $d^3p / E$ is. Writing $u_\mu p^\mu = g_{\mu\nu} \Lambda^\mu{}_\alpha \Lambda^\nu{}_\rho u^\alpha p^\rho$, where $\Lambda^\mu{}_\alpha u^\alpha = \delta^\mu_0$ is the Lorentz transformation to the rest frame, we have
\begin{equation}
 F_q = -\frac{2N_f N_c T}{V} \int_V d^3x \sum_{\varsigma = \pm 1}  \int \frac{d^3p}{(2\pi)^3 E} p^\mu \Lambda_\mu{}^0 F^\rho_{\varsigma},
\end{equation}
where now $F^\rho_{\varsigma}$ is given in Eq.~\eqref{eq:static_Fs}, with $\beta \mathcal{E}_\varsigma \to \beta_\rho (E - \varsigma \mu_\rho)$. Substituting $p^\mu \Lambda_\mu{}^0 = \Gamma_\rho(E + \mathbf{v} \cdot \mathbf{p})$, the part involving the local three-velocity $\mathbf{v} = \mathbf{u} / \Gamma_\rho$ can be seen to make a vanishing contribution, being odd with respect to $\mathbf{p} \rightarrow -\mathbf{p}$, such that
\begin{equation}
 F_q = -\frac{4N_f N_c}{R^2} \int_0^R d\rho\, \rho  T_\rho \sum_{\varsigma = \pm 1} \int \frac{d^3p}{(2\pi)^3} F^\rho_{\varsigma},
 \label{eq:TE_Fq_aux}
\end{equation}
with $T_\rho = \Gamma_\rho T$. Introducing the integration variable $x = p / T_\rho$, we have
\begin{equation}
\hspace{-10pt} F_q = -\frac{4 N_f N_c}{R^2} \int_0^R d\rho\, \rho T^4_\rho \sum_{\varsigma = \pm 1} \int_0^\infty \frac{dx\, x^2}{2\pi^2} F_{\varsigma}(\mathcal{Y}^\rho_\varsigma),
 \label{eq:TE_Fq}
\end{equation}
where $F_\varsigma(\mathcal{Y}^\rho_\varsigma) = \frac{1}{3} \ln (1 + 3L_\varsigma e^{-\mathcal{Y}^\rho_\varsigma} + 3L_{-\varsigma} e^{-2\mathcal{Y}^\rho_\varsigma} + e^{-3\mathcal{Y}^\rho_\varsigma})$ and
\begin{equation}
 \mathcal{Y}^\rho_\varsigma = y_\rho - \varsigma \beta \mu\,,
 \,\,\,
  y_\rho = \frac{E}{T_\rho} = \sqrt{x^2 + z_\rho^2}\,,
  \,\,
  z_\rho = \frac{g \sigma}{T_\rho}\,.
  \label{eq:TE_yrho}
\end{equation}
It is instructive at this stage to write down the expressions for the local energy density $\epsilon_\rho$ and pressure $P_\rho$, measured at distance $\rho$ from the rotation axis, which follow by replacing $f_\varsigma \to f_\varsigma^\rho$ in Eq.~\eqref{eq_P}. Considering the spherical coordinates in momentum space and switching the integration variable to $x = p / T_\rho$, we arrive at 
\begin{align}
 P_\rho &= \frac{N_f N_c T_\rho^4}{3\pi^2} \sum_{\varsigma = \pm 1} \int_0^\infty \frac{dx\, x^4}{y_\rho} f_\varsigma(\mathcal{Y}^\rho_\varsigma), \nonumber\\ 
 \epsilon_\rho &= \frac{N_f N_c T_\rho^4}{\pi^2} \sum_{\varsigma = \pm 1} \int_0^\infty dx\, x^2 y_\rho f_\varsigma(\mathcal{Y}^\rho_\varsigma).
 \label{eq:TE_P_eps}
\end{align}
It can be seen that $F_q = -\frac{2}{R^2} \int_0^R d\rho\, \rho\, P_\rho$. 

We now apply the same prescription to the fermion condensate, $\bar{\psi} \psi_\rho = -\partial P_\rho / \partial (g\sigma) = (\epsilon_\rho - 3P_\rho) / (g\sigma)$, and $\partial F_q / \partial L_\pm$, keeping in mind that both the condensate $\sigma$ and the Polyakov-loop values $L_\pm$ are coordinate-independent quantities (note that $L_+ \equiv L$ and $L_- \equiv L^*$). The saddle-point Eqs.~\eqref{eq_dFdp} become
\begin{subequations} \label{eq:TE_dFdp}
\begin{align}
 \frac{\partial V_\mathcal{M}}{\partial \sigma} + \frac{2 N_f N_c g^2 \sigma}{\pi^2 R^2} \int_0^R \hspace{-5pt} d\rho\,\rho T_\rho^2 \sum_{\varsigma = \pm 1} \int_0^\infty \frac{dx\, x^2}{y_\rho} f_\varsigma(\mathcal{Y}^\rho_\varsigma) &= 0, \label{eq:TE_dFdp:sigma}\\
 \frac{\partial V_L}{\partial L_\pm} - \frac{2 N_f N_c}{\pi^2 R^2} \int_0^R \hspace{-5pt} d\rho\, \rho T_\rho^4 \sum_{\varsigma = \pm 1} \int_0^\infty dx\, x^2\, f_{\varsigma, \pm}(\mathcal{Y}^\rho_\varsigma) &= 0. \label{eq:TE_dFdp:L}
\end{align}
\end{subequations}
The purpose of this section is to study the effect of rotation, as captured by the TE law, on the phase transition of the (P)LSM$_q$ models. To simplify the analysis, we will consider only the limiting cases of vanishing quark chemical potential ($\mu = 0$), discussed in Subsections~\ref{sec:TE:LSM} and \ref{sec:TE:PLSM} for the LSM$_q$ and PLSM$_q$ models, respectively, and vanishing temperature, considered in Subsec.~\ref{sec:TE:T0}.

\subsection{Vanishing chemical potential: LSM\texorpdfstring{${}_q$}{q} model}\label{sec:TE:LSM}

We now consider the LSM$_q$ model at vanishing chemical potential. The omission of the Polyakov-loop degrees of freedom is equivalent to setting $L =L^* = 1$ and taking only the saddle-point equation for the $\sigma$ meson. Substituting the meson potential~\eqref{eq.LagMeson} into Eq.~\eqref{eq:TE_dFdp:sigma} gives us:
\begin{gather}
 \lambda(\sigma^2 - v^2)\sigma - h + g \langle \Bar{\psi} \psi \rangle_R = 0,
 \nonumber\\
 \langle \Bar{\psi} \psi \rangle_R = \frac{4 N_f N_c g \sigma}{\pi^2 R^2} \int_0^R d\rho\,\rho T_\rho^2 \int \frac{dx\, x^2}{y_\rho(e^{y_\rho} + 1)},
 \label{eq_TE_LSMq}
\end{gather}
where we took into account that $f_\varsigma(\mathcal{Y}^\rho_\varsigma) = (e^{y_\rho} + 1)^{-1}$ is the distribution for a neutral Fermi gas. The above form of the saddle-point equation serves to indicate that the difference between the rotating and non-rotating systems comes from the evaluation of the fermion condensate. At large temperatures, or equivalently, small $z_\rho = g\sigma / T_\rho$, we have $y_\rho \simeq x$ and the fermion condensate can be approximated by
\begin{align}
 \langle \Bar{\psi} \psi \rangle_R & \simeq
 \frac{N_f N_c g \sigma T^2}{\pi^2} \langle \Gamma_\rho^2 \rangle_R I_\sigma,
 \nonumber\\
 I_\sigma & = 2 \int_0^\infty \frac{dx\, x}{e^x + 1} = \frac{\pi^2}{6},
\end{align}
where we neglected terms of order $O[(g \sigma / T)^3]$. In the same limit of massless particles, the local energy density $\epsilon_\rho$ and pressure $P_\rho$ introduced in Eq.~\eqref{eq:TE_P_eps} become
\begin{equation}
 \epsilon_\rho = 3P_\rho = \frac{7\pi^2 N_f N_c T^4}{60} \Gamma_\rho^4.
 \label{eq:TE_P_eps_LSM_m0}
\end{equation}
In this approximation, the average over the cylindrical volume affects only the Lorentz factor and gives
\begin{equation}
 \langle \Gamma_\rho^2 \rangle_R = \frac{2}{R^2\Omega^2} \ln \Gamma_R, \qquad
 \Gamma_R = (1 - R^2 \Omega^2)^{-1/2}.
\end{equation}

Replacing the above results into Eq.~\eqref{eq_TE_LSMq}, and setting $\sigma \to \sigma_c$ corresponding to the value of the condensate at the chiral transition, we arrive at
\begin{equation}
 \lambda(\sigma_c^2 - v^2)\sigma_c - h + \tfrac{1}{6} N_f N_c g^2 \sigma_c (T^{\Omega R}_\sigma)^2 \langle \Gamma_\rho^2 \rangle_R = 0.
 \label{eq:TE_LSM_saddle}
\end{equation}
From the form of the above equation, it is reasonable to expect that the conditions for the crossover transition are met at a value $\sigma_c$ similar to that encountered in the static case (see Subsec.~\ref{sec:PLSM:saddle}), when the effective quark mass is about half of its constituent mass:
\begin{equation}
 m_{q;c} = g \sigma_c \simeq 0.45 g f_\pi \simeq 0.138\ {\rm  GeV}.
\end{equation}
Applying the same approximations to the static case, $\sigma_c$ satisfies the equation
\begin{equation}
 \lambda(\sigma_c^2 - v^2) \sigma_c - h + \tfrac{1}{6} N_f N_c g^2 \sigma_c (T^{\rm nr}_\sigma)^2 = 0,
 \label{eq:TE_LSM_saddle_static}
\end{equation}
with $T^{\rm nr}_\sigma = 0.147$ GeV being the pseudocritical temperature for the non-rotating (``nr'') system, as shown in the previous sections. 
Note that, since the ratio $m_{q;c} / T^{\rm nr}_\sigma$ is not negligible, Eq.~\eqref{eq:TE_LSM_saddle_static} is not an accurate approximation of the full saddle-point equation and would therefore give a transition temperature different from $0.147$ GeV. Keeping in mind that the same level of approximation was taken also in Eq.~\eqref{eq:TE_LSM_saddle}, we can employ this approximate form to establish a relation between the transition temperatures in the rotating and static systems.
Comparing Eqs.~\eqref{eq:TE_LSM_saddle} and \eqref{eq:TE_LSM_saddle_static}, we conclude that in the rotating case, the pseudocritical temperature $T_\sigma^{\Omega R}$ in a rotating system is linked to the pseudocritical temperature $T_\sigma^{\rm nr}$ in the static case via
\begin{equation}\label{eq:TE_LSM}
 T_\sigma^{\Omega R} = \frac{T_{\sigma}^{\rm nr}}{\sqrt{\langle \Gamma_\rho^2 \rangle_R}} = \frac{R \Omega T^{\rm nr}_\sigma}{\sqrt{2 \ln \Gamma_R}}.
\end{equation}
The above function is shown with the black dashed line in Fig.~\ref{fig:TE_unb}(a).

\subsection{Vanishing chemical potential: PLSM\texorpdfstring{${}_q$}{q} model}\label{sec:TE:PLSM}

\begin{figure}
\begin{center}
\begin{tabular}{c}
\includegraphics[width=0.92\linewidth]{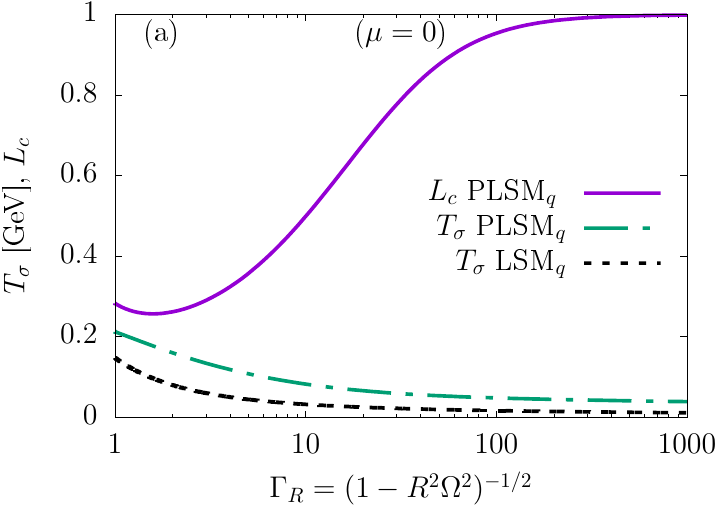} \\
\includegraphics[width=0.92\linewidth]{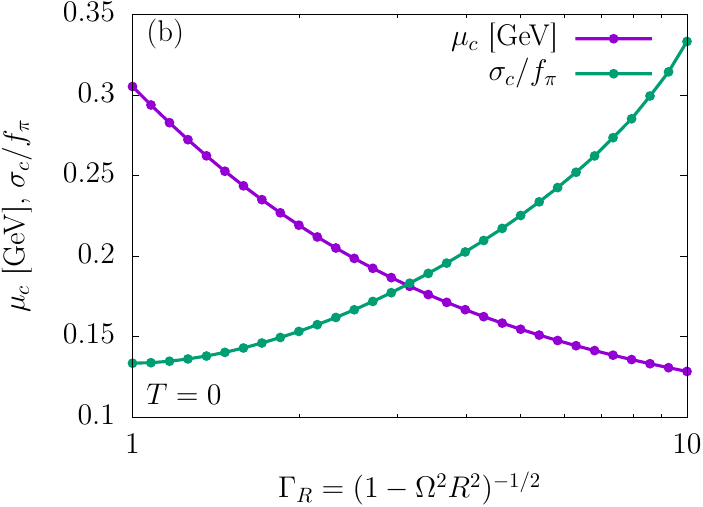}
\end{tabular}
\end{center}
\caption{Tolman-Ehrenfest prediction in the (P)LSM${}_q$ models. Panel (a) shows the pseudocritical temperatures $T_\sigma$ in the LSM${}_q$ (dashed line) and PLSM${}_q$ (dot-dashed line) models at $\mu = 0$. The Polyakov loop $L_c \equiv L(T_{\rm pc})$ calculated in the PLSM${}_q$ model at the transition point is shown by the solid purple line. Panel (b) shows the critical chemical potential $\mu_c$ and the sigma meson $\sigma_c / f_\pi$ measured after chiral symmetry restoration, for a system rotating with angular velocity $\Omega$ at $T = 0$. All quantities in both panels are shown for systems confined in a fictitious cylinder of radius $R$, as functions of the Lorentz factor $\Gamma_R = 1/(1 - R^2 \Omega^2)^{-1/2}$ of a co-rotating observer at the system edge.
}
\label{fig:TE_unb}
\end{figure}

Proceeding now to the PLSM$_q$ model, we consider the case of vanishing chemical potential, when $L = L^*$ and the distributions $f_\varsigma$ and $f_{\varsigma;\pm}$, introduced in Eqs.~\eqref{eq_fall}, satisfy
\begin{align}
 f_\varsigma(y) &= \frac{L e^{-y} + 2L e^{-2y} + e^{-3y}}{(e^{-y} + 1)(e^{-y-i \vartheta} + 1)(e^{-y+i \vartheta}  + 1)}, \nonumber\\
 \sum_{\varsigma = \pm 1} f_{\varsigma;\pm}(y) &= \frac{e^{-y}}{(e^{-y - i \vartheta} + 1)(e^{-y + i \vartheta} + 1)},
 \label{eq:TE_PLSM_f_aux}
\end{align}
where $y \equiv y_\rho$ is introduced in Eq.~\eqref{eq:TE_yrho} and the eigenvalue $\vartheta$ of the Polyakov loop matrix satisfies
\begin{equation}
 (e^{-y} + e^{i \vartheta})(e^{-y} + e^{-i \vartheta}) = e^{-2y} + (3L - 1) e^{-y} + 1,
\end{equation}
such that \cite{Endrodi:2019whh,Haensch:2023sig}
\begin{equation}
 \sin \vartheta = \frac{\sqrt{3}}{2}\sqrt{(1 - L) ( 1 + 3L)},
 \quad
 \cos\vartheta = \frac{3L - 1}{2}.
 \label{eq_def_varphi}
\end{equation}
The angle $\vartheta$ decreases from $2\pi / 3$ when $L = 0$ (confined state) towards $0$ as $L \rightarrow 1$ (deconfined state).
We can thus split the fractions appearing in Eqs.~\eqref{eq:TE_PLSM_f_aux} as
\begin{align}
 f_\varsigma(y) &= \frac{1}{3} \left(\frac{1}{e^{y} + 1} + \frac{1}{e^{y+i \vartheta} + 1} + \frac{1}{e^{y - i\vartheta} + 1}\right),\nonumber\\
 \sum_{\varsigma = \pm 1} f_{\varsigma;\pm}(y) &= \frac{i}{2 \sin \vartheta} \left(\frac{1}{e^{y + i\vartheta} + 1} - \frac{1}{e^{y - i\vartheta} + 1}\right).\hspace{-15pt}
\end{align}
Focusing now on the saddle-point equation for the $\sigma$ meson and repeating the steps discussed in Sec.~\ref{sec:TE:LSM}, we obtain:
\begin{align}
 \langle \bar{\psi} \psi \rangle_R &= \frac{N_f N_c g \sigma T^2}{\pi^2} \langle \Gamma_\rho^2 \rangle_R I_\sigma\,, \nonumber\\
 I_\sigma &= \sum_{\varsigma = \pm 1} \int_0^\infty \frac{dx x^2}{y_\rho} f_\varsigma(y_\rho).
\end{align}
The term $I_\sigma$ can be evaluated in the limit of negligible mass as
\begin{align}
 I_\sigma &\simeq \frac{2}{3} \int_0^\infty dx\, x \left(\frac{1}{e^{x} + 1} + \frac{1}{e^{x+i \vartheta} + 1} + \frac{1}{e^{x - i\vartheta} + 1}\right)\nonumber\\
 &= \frac{\pi^2}{18} - \frac{2}{3} \left[{\rm Li}_2(-e^{i\vartheta}) + {\rm Li}_2(-e^{-i\vartheta})\right],
\end{align}
where ${\rm Li}_n(z) = \sum_{k = 1}^\infty z^k / k^n$ is the polylogarithm function, satisfying
\begin{equation}
 \int_0^\infty \frac{dx\, x^n}{e^{x + a} + 1} = -n! {\rm Li}_{n+1}(-e^{-a}),
\end{equation}
Using the property ${\rm Li}_2(-e^{a}) + {\rm Li}_2(-e^{-a}) = -\frac{\pi^2}{6} - \frac{a^2}{2}$, with $a =i \vartheta$, we find
\begin{equation}
 I_\sigma = \frac{\pi^2}{6} \left(1 - \frac{2\vartheta^2}{\pi^2}\right),
\end{equation}
such that the local fermion condensate evaluates to
\begin{equation}
 \bar{\psi} \psi_\rho = \frac{\epsilon_\rho - 3P_\rho}{g\sigma} \simeq N_f N_c g \sigma \frac{T_\rho^2}{6} \left(1 - \frac{2 \vartheta^2}{\pi^2}\right).
 \label{eq:TE_FC_PLSM_m0}
\end{equation}
For completeness, we evaluate below the local energy density $\epsilon_\rho$ and pressure $P_\rho$ introduced in Eq.~\eqref{eq:TE_P_eps}, extending the result \eqref{eq:TE_P_eps_LSM_m0} obtained in the LSM$_q$ model to
\begin{align}
 \epsilon_\rho &= 3P_\rho = \frac{N_f N_c T^4}{\pi^2} \Gamma_\rho^4 I_\epsilon, \nonumber\\ 
 I_\epsilon &\simeq \frac{2}{3} \int_0^\infty dx\, x^3 \nonumber\\
 &\qquad\times \left(\frac{1}{e^{x} + 1} + \frac{1}{e^{x+i \vartheta} + 1} + \frac{1}{e^{x - i\vartheta} + 1}\right).
\end{align}
Evaluating the above integral in terms of the polylogarithm function and using ${\rm Li}_4(-e^a) + {\rm Li}_4(-e^{-a}) = -\frac{a^4}{24} - \frac{\pi^2 a^2}{12} - \frac{7\pi^4}{360}$, we arrive at
\begin{equation}
 \epsilon_\rho = 3P_\rho = \frac{7\pi^2 N_f N_c T^4}{60} \Gamma_\rho^4 \left(1 - \frac{20 \vartheta^2}{7\pi^2} + \frac{10\vartheta^4}{7\pi^4} \right).
 \label{eq:TE_P_eps_PLSM_m0}
\end{equation}

Compared to the LSM$_q$ model, the saddle-point equation for $\sigma$ \eqref{eq:TE_LSM_saddle} gets modified to
\begin{multline}
 \lambda(\sigma_c^2 - v^2)\sigma_c - h + \frac{1}{6} N_f N_c g^2 \sigma_c (T^{\Omega R}_\sigma)^2 \langle \Gamma_\rho^2 \rangle_R \\\times 
 \left(1 - \frac{2\vartheta_{\Omega R}^2}{\pi^2}\right) = 0\,,
\end{multline}
where the angle $\vartheta_{\Omega R}$ sets the value of the Polyakov loop~\eqref{eq_def_varphi} in the rotating system, at the point of chiral restoration.
The above equation establishes a connection between the pseudocritical temperature and the Polyakov loop in the PLSM$_q$ model, compared to the pseudocritical temperature of the LSM$_q$ model:
\begin{equation}
 (T_\sigma^{\rm nr})_{\rm PLSM_q} = \frac{(T_\sigma^{\rm nr})_{\rm LSM_q}}{\sqrt{1 - 2\vartheta_{\rm nr}^2 /\pi^2}}.
 \label{eq:TE_TPLSM}
\end{equation}
Knowing that $T_\sigma^{\rm nr} =  0.147$ GeV in the LSM$_q$ model and $T_\sigma^{\rm nr} =  0.212$ GeV in the Polyakov-loop enhanced model, we can estimate
\begin{equation}
\hspace{-3pt} \vartheta_{\rm nr} = \frac{\pi}{\sqrt{2}} \sqrt{1 - \left(\frac{(T_\sigma^{\rm nr})_{\rm LSM_q}}{(T_\sigma^{\rm nr})_{\rm PLSM_q}}\right)^2} \simeq
  1.6\ {\rm rad},
\end{equation}
by which $L_{\rm nr} \simeq 0.313$.
The value estimated above differs from the actual value $L_{\rm nr} \simeq 0.289$ by less than $10\%$. This discrepancy can be expected to arise due to the major difference between the ratio $m_q / T_{\rm pc}$ of the effective quark mass $m_q = g \sigma f_\pi$ and the pseudocritical temperature $T_{\rm pc}$, as $\sigma_c$ has roughly the same value in both the LSM$_q$ and PLSM$_q$ models, while $T_{\rm pc}$ exhibits a significant variation. In any case, it becomes clear that the estimation of the pseudocritical temperature $T_\sigma$ for the chiral phase transition depends crucially on estimating the Polyakov loop $L$, for which we must employ Eq.~\eqref{eq:TE_dFdp:L}.

For the Polyakov loop part [Eq.~\eqref{eq:TE_dFdp:L}], we can compute the first mass correction by writing $f_\varsigma(y_\rho) = f_\varsigma(x) + (df_\varsigma / dy_\rho)_{y_\rho = x} \times (y_\rho - x) + \dots$, with $y_\rho - x \simeq z_\rho^2 / 2x$, i.e.:
\begin{equation}
 \int dx\, x^2 f_\varsigma(y_\rho) = \int dx\, \left[x^2 f_\varsigma(x) - \frac{z_\rho^2}{2} f_\varsigma(x) + O(z_\rho^4)\right].
\end{equation}
Thus, Eq.~\eqref{eq:TE_dFdp:L} becomes
\begin{equation}
 \frac{\partial V_L}{\partial L_\pm} - \frac{N_f N_c T^4}{\pi^2} \left(\langle \Gamma_\rho^4 \rangle_R I_{L;2} - \frac{g^2 \sigma^2}{2 T^2 } \langle \Gamma_\rho^2 \rangle_R I_{L;0}\right) = 0,
\end{equation}
where
\begin{subequations}
\begin{align}
 I_{L;2} &= \sum_{\varsigma = \pm 1} \int_0^\infty dx\, x^2 f_{\varsigma;\pm} \nonumber\\
 &= -\frac{i}{\sin\vartheta} \left[{\rm Li}_3(-e^{-i \vartheta}) - {\rm Li}_3(-e^{i\vartheta})\right],\\
 I_{L;0} &= \sum_{\varsigma = \pm 1} \int_0^\infty dx\, f_{\varsigma;\pm} \nonumber\\
 &= -\frac{i}{2\sin\vartheta} \left[{\rm Li}_1(-e^{-i \vartheta}) - {\rm Li}_1(-e^{i\vartheta})\right].
\end{align}
\end{subequations}
Using the relations
${\rm Li}_3(-e^{-a}) - {\rm Li}_3(-e^a) = \frac{1}{6} \pi^2 a + \frac{1}{6} a^3$ and 
${\rm Li}_1(-e^{-a}) - {\rm Li}_3(-e^a) = a$,
% \begin{align}
%  {\rm Li}_3(-e^{-a}) - {\rm Li}_3(-e^a) &= \frac{\pi^2 a}{6} + \frac{a^3}{6}, \\
%  {\rm Li}_1(-e^{-a}) - {\rm Li}_3(-e^a) &= a,
% \end{align}
we find
\begin{equation}
 I_{L;2} = \frac{\pi^2 \vartheta}{6\sin\vartheta} \left(1 - \frac{\vartheta^2}{\pi^2}\right), \qquad
 I_{L;0} = \frac{\vartheta}{2\sin\vartheta}.
\end{equation}

In the PLSM$_q$ model, both the transition temperature $T_\sigma^{\Omega R}$ and the Polyakov loop $L$ must be determined simultaneously and self-consistently. This amounts to solving the coupled system of equations
\begin{align}
 T_\sigma^{\Omega R} &= \frac{T_\sigma^{\rm nr} \sqrt{1 - 2\vartheta_{\rm nr}^2 /\pi^2}}{\sqrt{\langle \Gamma_\rho^2\rangle_R} \sqrt{1 - 2\vartheta_{\Omega R}^2 /\pi^2}},\nonumber\\
 \left.
 \frac{\partial V_L}{\partial L_\pm}
 \right\rvert_{\substack{L_\pm = L_{\Omega R} \\ 
 T = T_\sigma^{\Omega R}}} 
 &= \frac{N_f N_c \vartheta_{\Omega R} (T_\sigma^{\Omega R})^4}{6 \sin\vartheta_{\Omega R}} \left[  \left(1 - \frac{\vartheta_{\Omega R}^2}{\pi^2}\right) \langle \Gamma_\rho^4 \rangle_R \right.\nonumber\\
 & \left. - \frac{3 g^2 \sigma_c^2}{2\pi^2 (T_\sigma^{\Omega R})^2} \langle \Gamma_\rho^2 \rangle_R\right],
 \label{eq:TE_PLSM} 
\end{align}
where $\langle \Gamma_\rho^2 \rangle_R = 2 \ln \Gamma_R / (R^2 \Omega^2)$ and $\langle \Gamma_\rho^4 \rangle_R = \Gamma^2_R$.
We employ as input parameters the transition temperature $T_\sigma^{\rm nr} = 0.212$ GeV in the non-rotating case and the sigma meson $\sigma_c = 0.45 f_\pi$ at the crossover transition. We determine the non-rotating Polyakov loop parameter self-consistently, giving $L_{\rm nr} = 0.283$, in good agreement with the numerical results obtained in Sec.~\ref{sec:PLSM}. The derivative $\partial V_L / \partial L_\pm$ can be found in Eq.~\eqref{eq_dVdL}.

Figure~\ref{fig:TE_unb}(a) shows, in addition to the pseudocritical temperature $T_\sigma$ in the LSM${}_q$, the dependence of the pseudocritical chiral restoration temperature $T_\sigma$ and the Polyakov loop $L_c$ computed at $T = T_\sigma$ in the PLSM${}_q$, thus allowing us to make a comparison between the predictions of these two models below. The quantities are shown as functions of the Lorentz factor $\Gamma_R = (1 - R^2 \Omega^2)^{-1/2}$. When $\Gamma_R \lesssim 10$, the decrease of $T_\sigma$ with $\Gamma_R$ is visibly slower in the PLSM${}_q$ model than in the LSM${}_q$ model. This slowing down occurs because $L_c$ increases from its value $L_{\rm nr}$ without rotation towards $1$, while the angle $\vartheta_{\Omega R}$ decreases towards $0$.
It is interesting to remark that, when $\Gamma_R$ increases above $\simeq 5$, the Polyakov loop $L_c$ at the chiral transition becomes increasingly large, signaling that the system is already deconfined when chiral restoration takes place. The decrease of the deconfinement temperature below the chiral restoration one for fast rotation is consistent with the results reported in Fig.~6 of Ref.~\cite{Singha:2024tpo} in the case when $\Omega R = 1$.

\subsection{Finite-density matter at vanishing temperature}
\label{sec:TE:T0}

As already discussed in Sections~\ref{sec:PLSM:T0} and \ref{sec:rot:T0}, the Polyakov sector plays no role when $T {=} 0$. Therefore, both PLSM${}_q$ and LSM${}_q$ give the same predictions at vanishing temperature. In this case, the local fermion condensate can be evaluated as it was done in Eq.~\eqref{eq_unb_T0}. Hence, the averaged fermion condensate is
\begin{multline}
 \langle \Bar{\psi} \psi \rangle_R = \frac{N_f N_c g\sigma}{\pi^2 R^2} \int_0^R d\rho\, \rho \theta(|\mu_\rho| - g\sigma)\\ \times \left[p^\rho_f |\mu_\rho| - (g\sigma)^2 {\rm artanh}\frac{p^\rho_f}{|\mu_\rho|}\right],
\end{multline}
where $\mu_\rho = \Gamma_\rho \mu$ and $p_f^\rho = \sqrt{\mu_\rho^2 - g^2 \sigma^2}$ is the local Fermi momentum. Switching the integration variable to $\zeta = (g\sigma / \mu_\rho)^2$ and using $d\zeta = -2(g \sigma \Omega / \mu)^2 \rho d\rho$ leads to
\begin{multline}
 \langle \Bar{\psi} \psi \rangle_R = \frac{N_f N_c \mu^2 g \sigma}{2\pi^2 \Omega^2R^2} \theta(1 - \zeta_R) \\ \times \int_{\zeta_R}^{\zeta_{\rm max}} d\zeta \left(\frac{1}{\zeta} \sqrt{1 - \zeta} - {\rm artanh} \sqrt{1 - \zeta}\right),
\end{multline}
with
\begin{equation}
 \zeta_{\rm max} = \begin{cases}
  g^2 \sigma^2 / \mu^2, & \mu > g \sigma, \\
  1, & \text{otherwise}.
 \end{cases}
\end{equation}
The above integral can be performed analytically, leading to the following saddle-point equation:
\begin{multline}
\lambda(\sigma^2 - v^2) \sigma = h -
\theta(1 - \zeta_R) \frac{N_f N_c \mu^2 g^2 \sigma}{2\pi^2 \Omega^2 R^2} \\
\times\left(3\sqrt{1 - \zeta} - (2+\zeta) {\rm artanh}\sqrt{1 - \zeta}\right)_{\zeta_R}^{\zeta_{\rm max}}.
 \label{eq:TE_T0_saddle}
\end{multline}
As the system undergoes a first-order phase transition, the above equation typically allows for three simultaneous roots. The physical solution must minimize the total free energy $F = \frac{\lambda}{4}(\sigma^2 - v^2) - h\sigma + F_q$, where the contribution from the quark sector $F_q$ can be computed by inserting the pressure $P_q = -F_q(\rho)$ given in Eq.~\eqref{eq_unb_T0} into Eq.~\eqref{eq:TE_Fq_aux}:
\begin{align}
 F_q &= -\frac{N_f N_c}{12 \pi^2 R^2} \int_0^R d\rho\, \rho \theta(|\mu_\rho| - g\sigma) \nonumber\\
& \times \left[|\mu_\rho| p^\rho_f (2\mu_\rho^2 - 5g^2 \sigma^2) + 3 (g \sigma)^4 {\rm artanh} \frac{p^\rho_f}{|\mu_\rho|} \right] \nonumber\\
 &= -\theta(1 - \zeta_R) \frac{N_f N_c \mu^2 g^2 \sigma^2}{24\pi^2 \Omega^2 R^2} \nonumber\\ 
 &\times \left[3(4 + \zeta){\rm artanh}\sqrt{1 - \zeta} - \frac{\sqrt{1-\zeta}}{\zeta}(2 + 13\zeta)\right]_{\zeta_R}^{\zeta_{\rm max}}.
 \label{eq:TE_T0_Fq}
\end{align}
Figure~\ref{fig:TE_unb}(b) shows the dependence of the critical chemical potential $\mu_c$ at which the chiral symmetry gets restored along with the mean-field value of the scalar condensate $\sigma_c$, computed after chiral restoration, with respect to the Lorentz factor $\Gamma_R = (1 - R^2 \Omega^2)^{-1/2}$. It can be seen that the scalar condensate $\sigma_c$ increases as $\Omega R$ becomes larger according to our Tolman-Ehrenfest prediction.

\subsection{Tolman-Ehrenfest predictions vs. numerical results}
\label{sec:TE:comparison}

\begin{figure}
\begin{center}
\begin{tabular}{c}
 \includegraphics[width=0.92\linewidth]{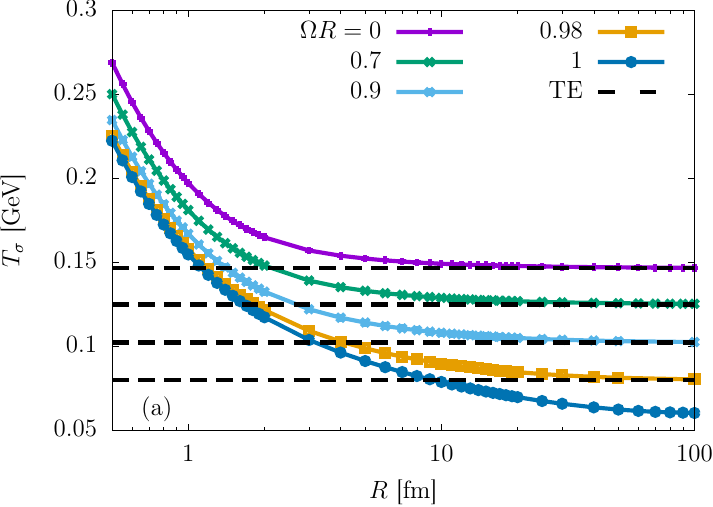} \\
 \includegraphics[width=0.92\linewidth]{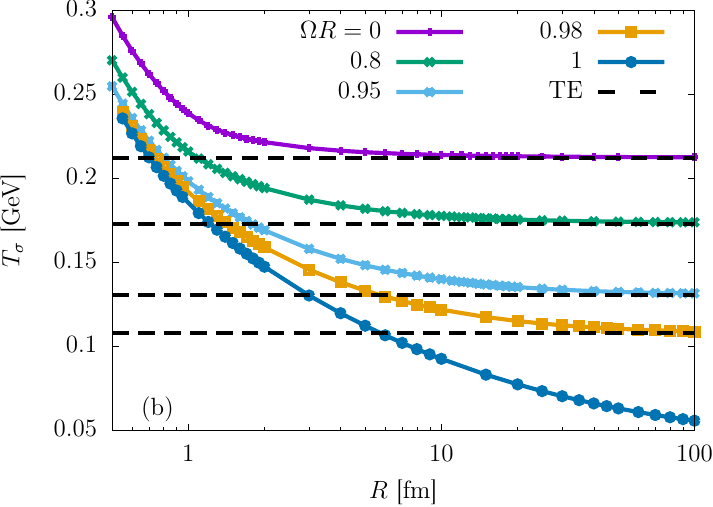}
\end{tabular}
\end{center}
\caption{The pseudocritical temperature $T_\sigma$ of the chiral crossover transition for (a) LSM${}_q$ and (b) PLSM${}_q$, as a function of the system radius $R$. The dashed black lines represent the Tolman-Ehrenfest prediction~\eqref{eq:TE_LSM} and \eqref{eq:TE_PLSM}.
\label{fig:b_TE}
}
\end{figure}

\begin{figure}
\centering
\includegraphics[width=0.92\linewidth]{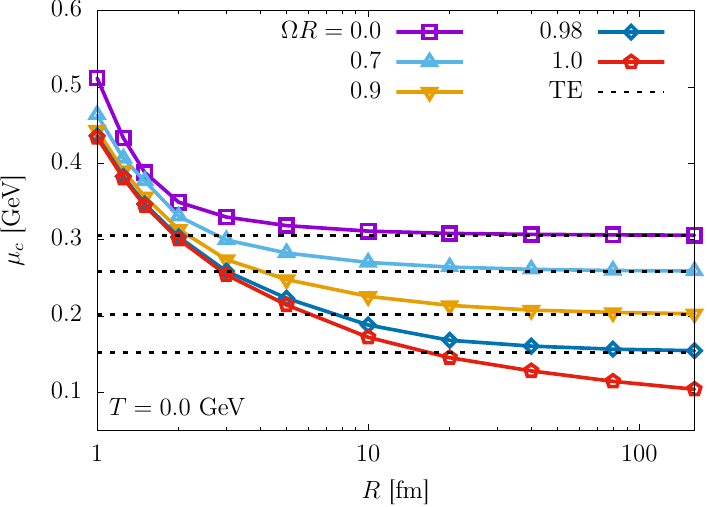}
\caption{The critical chemical potential $\mu_c$ for the chiral transition at $T = 0$, shown as a function of system size~$R$, for various values of $\Omega R$, in both LSM${}_q$ and PLSM${}_q$. The dashed lines correspond to the TE prediction obtained in Subsec.~\ref{sec:TE:T0}.}
\label{fig:TE_mumin}
\end{figure}

We now address the validity of the TE law. As highlighted in Sec.~\ref{sec:res:rot0}, boundary effects become significant when $R T_c \lesssim 1$. Therefore, we can expect the TE law to become applicable when boundary effects are negligible, i.e., when $R \rightarrow \infty$. In this limit, a meaningful comparison with the equivalent system introduced in Sec.~\ref{sec:TE:kinetic} can be drawn by enforcing that the product $\Omega R$ is kept constant. This implies that the angular velocity decreases, $\Omega \sim R^{-1} \rightarrow 0$. Consequently, the quantum corrections to the thermal expectation values in the rigidly rotating system, which are of order $O(\Omega^2)$ (see, e.g., Refs.~\cite{Ambrus:2014uqa, Becattini:2015nva}), become negligible.

Figure~\ref{fig:b_TE} shows ---with the horizontal dashed lines--- the TE prediction for the LSM$_q$ (panel a) and the PLSM$_q$ (panel b) models, obtained via Eqs.~\eqref{eq:TE_LSM} and \eqref{eq:TE_PLSM}, respectively. It can be seen that the numerical results for the pseudocritical temperature $T_\sigma$ of the chiral symmetry restoration transition at $\mu = 0$ and fixed $\Omega R$, indeed converge, in both models, to the TE predictions as the system size $R$ is increased.

It is worth noting that, as $\Omega R$ increases, the radius of the system must be increased in order for the restoration temperature to reach the TE prediction. This can be understood by noting that the local temperature, $T_\rho = \Gamma_\rho T$, increases due to the TE law only up to the vicinity of the boundary. In the boundary layer, having a thickness $\delta$, the fermion condensate (and all fermion thermal expectation values) are damped, therefore leading to a decrease of the local temperature (see, e.g., discussion in Ref.~\cite{Ambrus:2015lfr}). We can therefore expect that the maximum temperature achieved inside the cylinder is $T_{\rm max} = T_{R -\delta} = \Gamma_{R - \delta} T$. Let us consider the function $\Gamma_{R - \delta}$ for the case when the boundary layer is small compared to the system size:
\begin{equation}
 \Gamma_{R - \delta} \simeq \Gamma_R - \frac{\delta}{R} \Gamma_R (\Gamma_R^2 -1).
\end{equation}
Considering $\Omega R$ fixed, the second term becomes inversely proportional to $R$, therefore the damping effect on the local temperature decreases with $R$. Imposing now a relative difference $(\Gamma_R - \Gamma_{R - \delta})/ \Gamma_R = \varepsilon$ leads to the expression
\begin{equation}
 R_\varepsilon = \frac{\delta}{\varepsilon} (\Gamma_R^2 - 1).
\end{equation}
The above expression shows that, when the boundary thickness $\delta$ is independent of $R$, the maximum system temperature approaches the maximum TE temperature when $R$ increases proportionally to $\Gamma_R^2 - 1$. Hence, larger values of $\Omega R$ require larger system sizes to exhibit the convergence to the TE prediction.

Finally, Fig.~\ref{fig:TE_mumin} shows the critical chemical potential $\mu_c$, computed at vanishing temperature and for various values of $\Omega R$, as a function of the system size $R$. In this limit, both models give the same results. It can be seen that $\mu_c$ approaches the TE prediction (shown with horizontal dashed lines) as $R$ increases.

\section{Mechanical properties of rotating plasma}
\label{sec:res:I}

\subsection{Moment of inertia}

\begin{figure*}
\centering
\begin{tabular}{cc}
\includegraphics[width=0.46\linewidth]{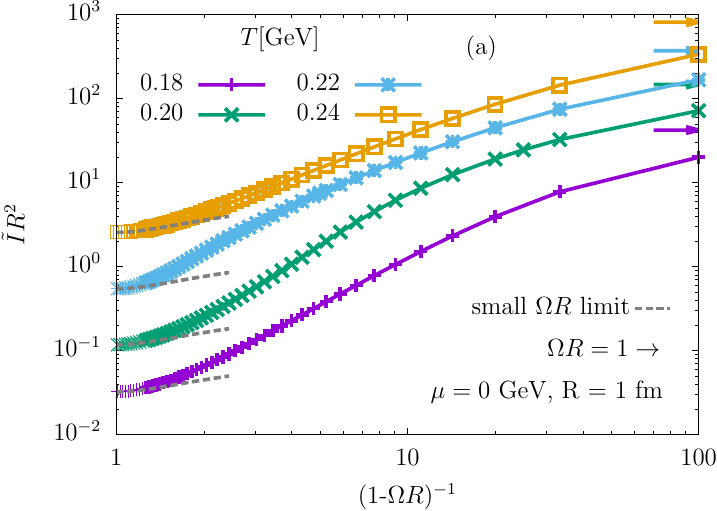} &
\includegraphics[width=0.46\linewidth]{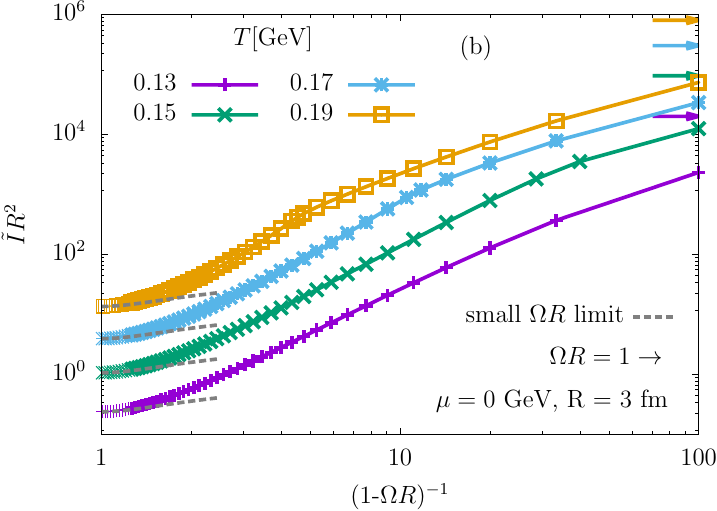}\\
\includegraphics[width=0.46\linewidth]{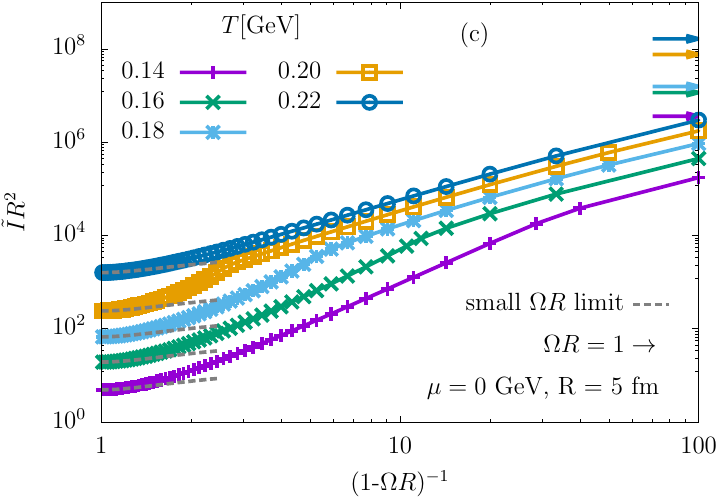} &
\includegraphics[width=0.46\linewidth]{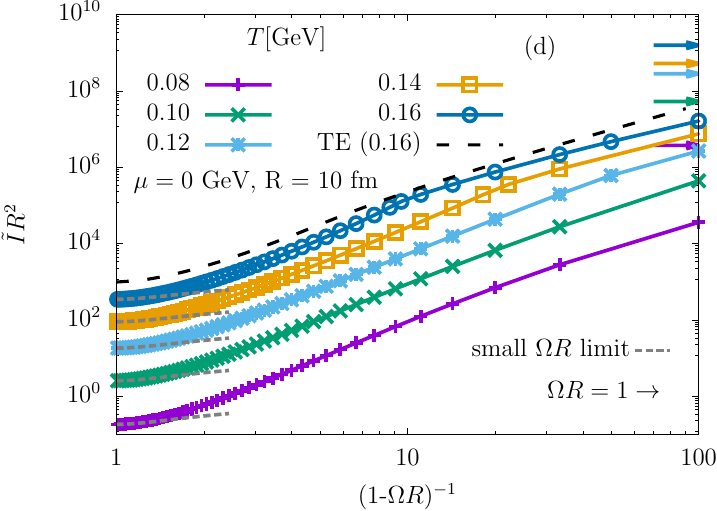}
\end{tabular}
\caption{ Scaled moment of inertia, $\tilde{I} R^2$, as a function of $(1 - \Omega R)^{-1}$, for systems of various radii: (a) $R=1$, (b) $3$, (c) $5$, and (d) $10$ fm; at vanishing chemical potential $\mu = 0$. The horizontal arrows indicate the $\Omega R \to 1$ limits, while the dashed lines indicate the small-$\Omega R$ limit in Eq.~\eqref{eq.MItaylor1}.   In panel (d), the TE prediction \eqref{eq:IR2_large_gamma} for $T=0.16$ GeV is shown with black dashed lines for comparison, see text for details.}
\label{fig:MImu0}
\end{figure*}

\begin{figure*}
\centering
\begin{tabular}{cc}
\includegraphics[width=0.46\linewidth]{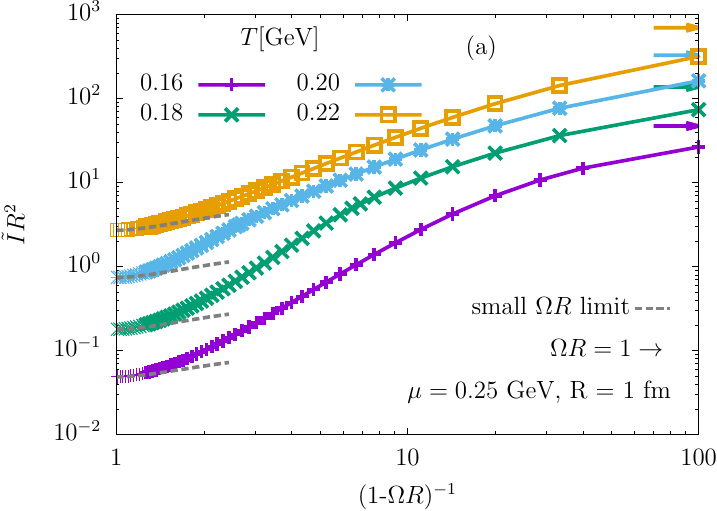} &
\includegraphics[width=0.46\linewidth]{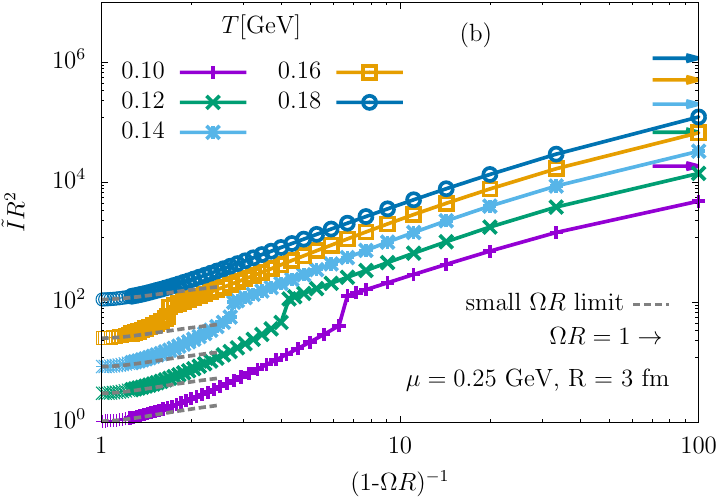}\\
\includegraphics[width=0.46\linewidth]{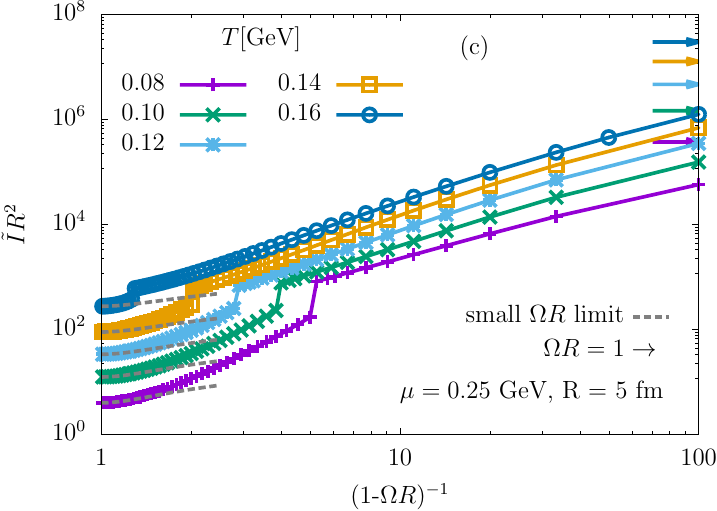}&
\includegraphics[width=0.46\linewidth]{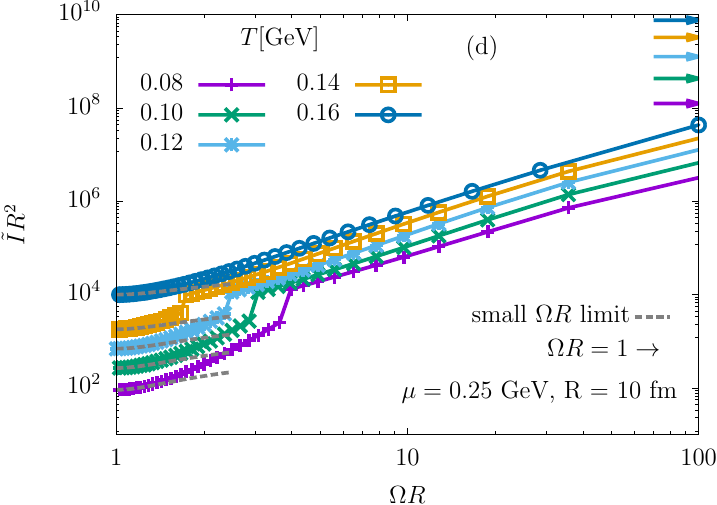}
\end{tabular}
\caption{Same as Fig.~\ref{fig:MImu0} for a system with finite quark chemical potential $\mu = 0.25$ GeV.
}
\label{fig:MImu25}
\end{figure*}

The mechanical response of the system at finite angular velocity $\Omega$ can be analyzed in terms of the moment of inertia associated with the volume-averaged angular momentum $\widetilde{L}$, defined as
\begin{align}
 \widetilde{L} = -\frac{\partial \widetilde{F}}{\partial \Omega} = \widetilde{I}(\Omega)\Omega\,,
 \label{eq_tilde_L}
\end{align}
where $\widetilde{F}$ is the volume-averaged grand potential of the rotating system.
In our notations, we use tildas ``$\, {\widetilde{~}}\, $'' to denote the quantities computed on the basis of the eigenmode series in the rotating cylinder. These results will later be compared with the prediction of the Tolman-Ehrenfest law, which are used without tildas.
The quantity $\widetilde{I}$ corresponds to the volume-averaged moment of inertia, that is the moment of inertia divided by the volume of the cylinder. Therefore, $\widetilde{I}$ does not contain the factor associated with the length of the cylinder in the longitudinal direction.

Since both the meson ($V_{\mathcal{M}}$) and the Polyakov ($V_L$) potentials do not depend explicitly on the angular frequency $\Omega$, the angular momentum of the system receives contributions only from the quark part $\widetilde{F}_q$.
Therefore, the definition of the angular momentum~\eqref{eq_tilde_L} in particular and the mechanical response of the system to the state of rotation in general involves only the density of the quark grand potential, $\widetilde{F}_q$, where the dependence on the angular frequency $\Omega$ enters explicitly via Eqs.~\eqref{eq:rot_Fq}--\eqref{eq:rot_Em_refl}.

In the last section, we observed the rotation-induced effects on the phase transition properties of the system. One can then naturally expect the signature of the phase transition to be reflected in the behaviour of the volume-averaged moment of inertia, $\widetilde{I}$. We can evaluate $\widetilde{I}$ as a function of $T$, $\mu$ and $\Omega$ from its definition:
\begin{equation}
 \widetilde{I}(\Omega) = -\frac{1}{\Omega} \frac{\partial \widetilde{F}_q}{\partial \Omega} = \frac{2N_f N_c}{\pi^2 R^2\Omega}\sum_{\mathrm{b},\varsigma = \pm 1} m \int_{0}^\infty dk\, \tilde{f}_\varsigma\,,
 \label{eq.MI}
\end{equation}
where $\tilde{f}_\varsigma$ denotes the Fermi-Dirac distribution for the rotating state~\eqref{eq_rot_f} and the sum over the collective quantum number ``b'' is defined in Eq.~\eqref{eq:rot_int_measure}.

Figures~\ref{fig:MImu0} ($\mu = 0$) and \ref{fig:MImu25} ($\mu = 0.25$ GeV) show the dependence of the dimensionless moment of inertia $\widetilde{I} R^2$ with respect to $(1 - \Omega R)^{-1}$, for various system sizes [$R = 1$, $3$, $5$ and $10$ fm, shown in the panels (a)--(d)], and for a selection of temperatures (shown as different lines with points), chosen such that the system undergoes the chiral symmetry restoration as $\Omega R$ increases (see Figs.~\ref{fig:Phase_diagramRot} and \ref{fig:Phase_diagramTOmega}). There is a monotonic increase of $\widetilde{I} R^2$ with both $T$ and $\Omega R$ and $\widetilde{I} R^2$ varies over several orders of magnitude as the causal limit is approached (shown with horizontal arrows). In the case of $\mu = 0$ (Fig.~\ref{fig:MImu0}), the phase transition appears as an inflection point, beyond which the increase of $\widetilde{I} R^2$ with respect to $(1 - \Omega R)^{-1}$ becomes milder. At $\mu = 0.25$ GeV (Fig.~\ref{fig:MImu25}), $\widetilde{I} R^2$ exhibits a jump to a larger value when the system undergoes a first-order phase transition. This does not happen in panel (a), corresponding to a small system size $R = 1$ fm, where the transition is of crossover type and the situation is similar to that seen in Fig.~\ref{fig:MImu0}(a). These two figures also show the moment of inertia in the causal limit, $\Omega R = 1$, as horizontal arrows. In the case when $R$ is small, the difference between the causal limit and the value corresponding to the last visible point, where $(1 - \Omega R)^{-1} = 100$, corresponds to a factor of $2{-}10$. At $R = 10$ fm, this factor already increases to about $10^2$.

To gain more insight into the large-$\Omega R$ behavior of the volume-averaged moment of inertia $\widetilde{I}$, let us compute this quantity from the TE approximation for the quark potential, given in Eq.~\eqref{eq:TE_Fq}. We use the notation $I$ to denote the volume-averaged moment of inertia in the TE approach.

Taking into account the relation
\begin{equation}
 \frac{\partial F_\varsigma(\mathcal{Y}^\rho_\varsigma)}{\partial \Omega} = \frac{z_\rho^2}{y_\rho} \rho^2 \Omega \Gamma_\rho^2 f_\varsigma(\mathcal{Y}^\rho_\varsigma),
\end{equation}
we arrive at
\begin{align}
 I &= \frac{2 N_f N_c T^4}{\pi^2 R^2} \int_0^R d\rho\, \rho^3 \Gamma_\rho^6
 \int_0^\infty \frac{dx\, x^2}{y_\rho} \left(\frac{4x^2}{3} + z_\rho^2\right) f^\rho_\varsigma \nonumber\\
  &= \frac{2}{R^2} \int_0^R d\rho\, \rho^3 \Gamma_\rho^2 (\epsilon_\rho + P_\rho),
\end{align}
where the local energy density and pressure were introduced in Eq.~\eqref{eq:TE_P_eps}.
At large $\Gamma_R = (1 - R^2 \Omega^2)^{-1/2}$, the magnitude of the quantity $z_\rho$ becomes negligible and we can use the relation \eqref{eq:TE_P_eps_PLSM_m0} for the local energy density and pressure for massless particles. In this case, the coordinate dependence is contained in the $\rho^3 \Gamma_\rho^6$ factor, which can be integrated straightforwardly, leading to
\begin{equation}\label{eq:IR2_large_gamma}
 I R^2 = \frac{7\pi^2 N_f N_c}{90} (R \Gamma_R T)^4 \left(1 - \frac{20 \vartheta^2}{7\pi^2} + \frac{10 \vartheta^4}{7\pi^4}\right)\,,
\end{equation}
where the angle $\vartheta$ is related to the mean-field value of the Polyakov loop through Eq.~\eqref{eq_def_varphi}. If the system is deconfined ($L = 1$), then $I R^2 \simeq \frac{7\pi^2 N_f N_c}{90} (R \Gamma_R T)^4$.
The quantity $IR^2$, as obtained from Eq.~\eqref{eq:IR2_large_gamma} for $R=10$ fm and $T = 0.16$ GeV, is shown by the black dashed line in Fig.~\ref{fig:MImu0}(d).

\subsection{Shape coefficients}

For slow rotation, one can expand the volume-averaged quark grand potential in a power series with respect to the velocity at the boundary $v_R = \Omega R$ as follows~\cite{Ambrus:2023bid}:
\begin{align}
\hspace{-10pt} \widetilde{F}_q = \widetilde{F}_q^{(0)}\sum_{n=0}^\infty \frac{v_R^{2n}}{(2n)!} \widetilde{K}_{2n}~, \,\,
 \widetilde{K}_{2n}=\frac{1}{\widetilde{F}_q^{(0)}} \left.\frac{\partial^{2n} \widetilde{F}_q}{\partial v_R^{2n}}\right\rvert_{\Omega=0},\hspace{-5pt}
 \label{eq_tilde_F_Kn}
\end{align}
with $\widetilde{F}_q^{(0)}$ being the quark grand potential in the absence of rotation. In the thermodynamic ground state, this system respects the invariance under the time-reversal transformation, $t \to - t$, which also flips the direction of rotation, $\Omega \to -\Omega$. This invariance, together with analyticity, implies that the quark potential~\eqref{eq_tilde_F_Kn} is an even function of the angular frequency $\Omega$.

At small $\Omega R$, the volume-averaged moment of inertia can be expressed in terms of the shape coefficients $K_{2n}$ via
\begin{equation}
 \widetilde{I} = -R^2 \widetilde{F}^{(0)}_q \left[\widetilde{K}_2+\frac{1}{6} \Omega^2 R^2 \widetilde{K}_4 + O(\Omega^4 R^4)\right]\,,
\label{eq.MItaylor1}
\end{equation}
where we used Eq.~\eqref{eq_tilde_F_Kn} together with the definition of $\widetilde{I}$ in Eq.~\eqref{eq.MI}. We remind that $\widetilde{F}^{(0)}_q  < 0$ in Eq.~\eqref{eq.MItaylor1}.

The terms in the series~\eqref{eq.MItaylor1} can be intuitively understood. The coefficient $K_2$ can be associated with a normalized moment of inertia in the limit of vanishing rotation, $\widetilde{I}_{\Omega = 0} = -R^2 \widetilde{F}_q^{(0)} K_2$. The coefficient $K_4$ determines the first nontrivial rotation-dependent contribution to the moment of inertia that arises due to rotation. One can understand the appearance of a nonzero $K_4$ coefficient as a result of the inertial (centrifugal) forces that arise in the rotating system: the rigid rotation tends to pull the rotating matter (for example, an element of the fluid) outwards, thus increasing the density of matter at the periphery of rotation and making the moment of inertia larger. The leading contribution of this effect to the moment of inertia is represented by a non-vanishing $K_4$ coefficient in Eq.~\eqref{eq.MItaylor1}. In a free rotating system, the shape coefficients are $K_{2n} = (2n)!$~\cite{Ambrus:2023bid}.
The limit of slow rotation~\eqref{eq.MItaylor1} is shown with the dashed gray lines in Figs.~\ref{fig:MImu0} and \ref{fig:MImu25}, being in agreement with the results obtained through the direct evaluation of Eq.~\eqref{eq.MI}.

\begin{figure*}
\centering
\begin{tabular}{cc}
\includegraphics[width=0.46\linewidth]{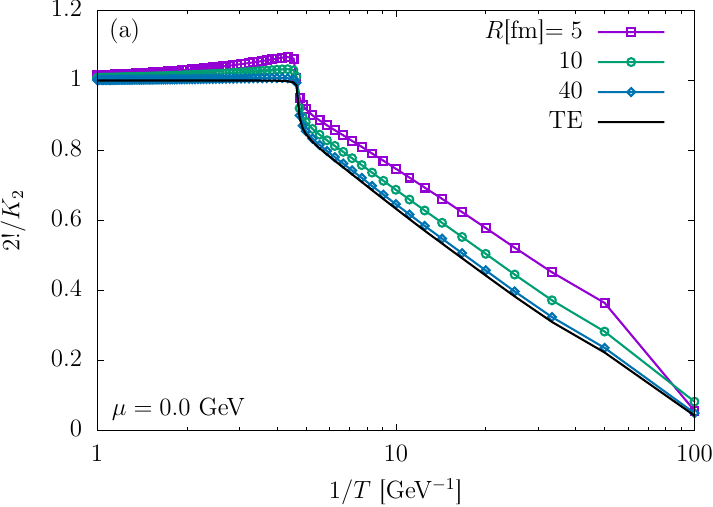} &
\includegraphics[width=0.46\linewidth]{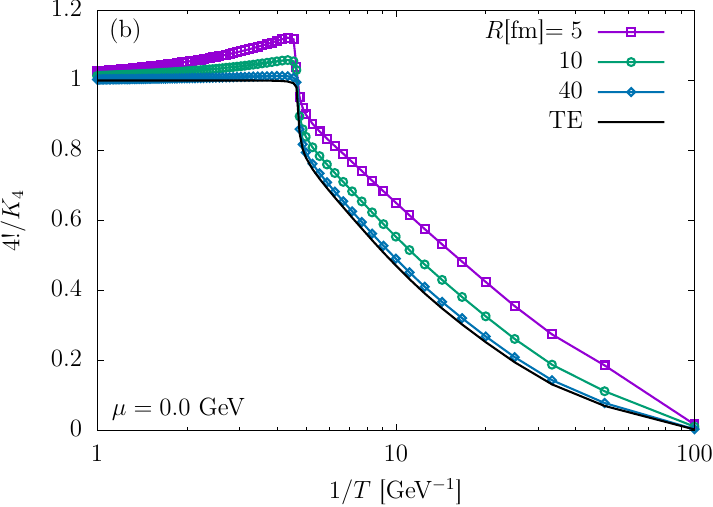}\\
\includegraphics[width=0.46\linewidth]{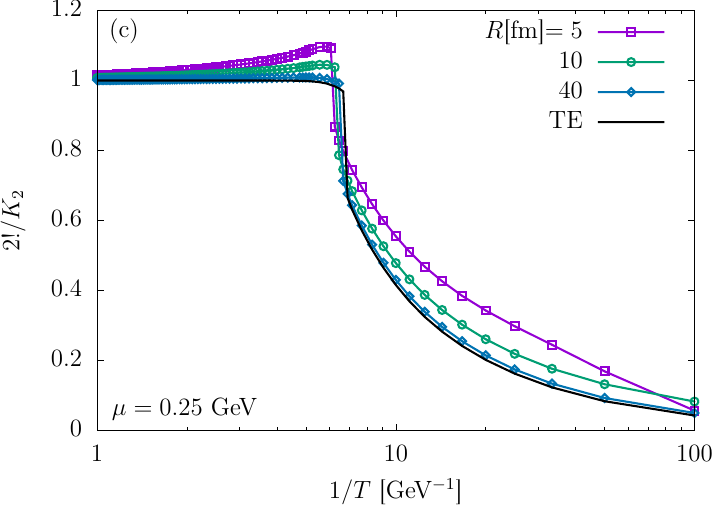}&
\includegraphics[width=0.46\linewidth]{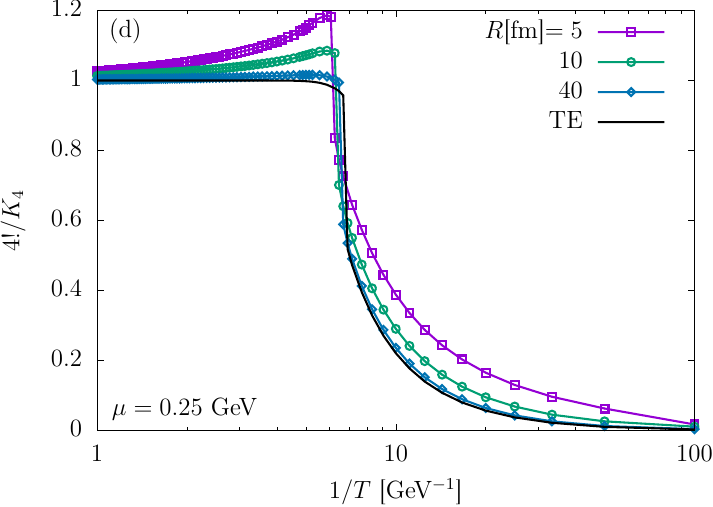}
\end{tabular}
\caption{Scaled shape coefficients $2!/\widetilde{K}_2$ (left) and $4!/\widetilde{K}_4$ (right), shown as functions of temperature for $\mu = 0$ (top) and $\mu=0.25$ GeV (bottom), for various system sizes $R = 5$, $10$ and $40$ fm, along with the TE prediction \eqref{eq:TE_F0K2K4}, shown using black lines. When $T \to \infty$, all curves tend to $1$, thus confirming the free-gas limit $K_{2n} = (2n)!$.
}
\label{fig:k2k4_vs_T_varR}
\end{figure*}

\begin{figure}
\centering
\begin{tabular}{c}
\includegraphics[width=0.92\linewidth]{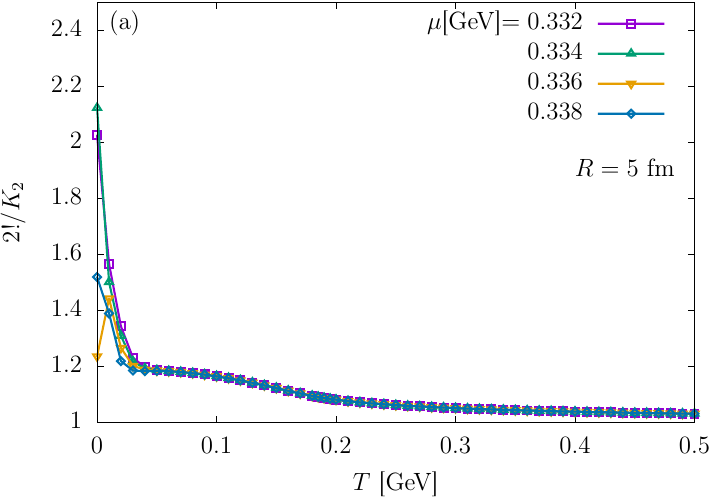} \\
\includegraphics[width=0.92\linewidth]{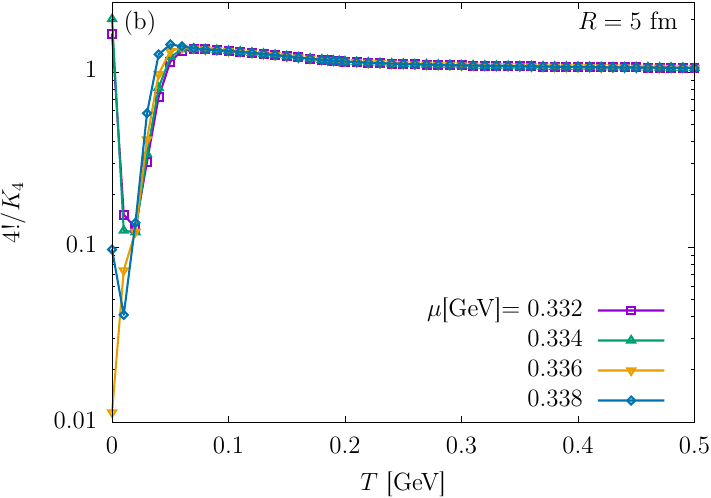}
\end{tabular}
\caption{Scaled shape coefficients, $2!/\widetilde{K}_2$ (a) and $4!/\widetilde{K}_4$ (b), as functions of temperature $T$, for fixed system size ($R=5$ fm), computed at various chemical potentials $\mu\in\{ 0.332, 0.334, 0.336,0.338\}$.
}
\label{fig:k2k4_vs_T_varmu}
\end{figure}

\begin{figure}
\centering
\begin{tabular}{c}
\includegraphics[width=0.92\linewidth]{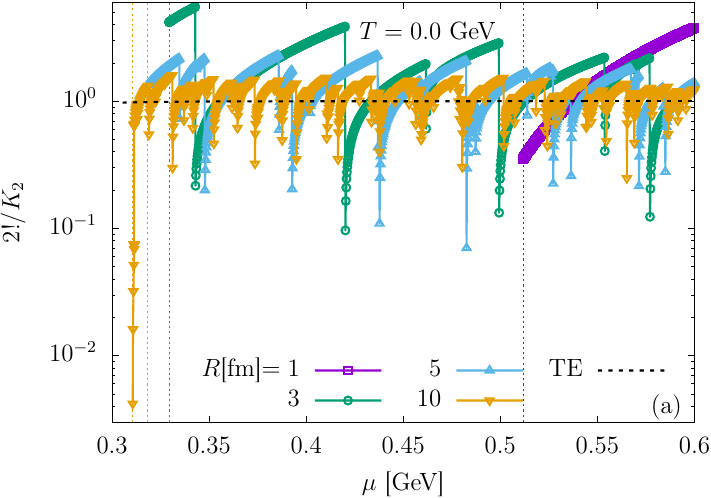} \\
\includegraphics[width=0.92\linewidth]{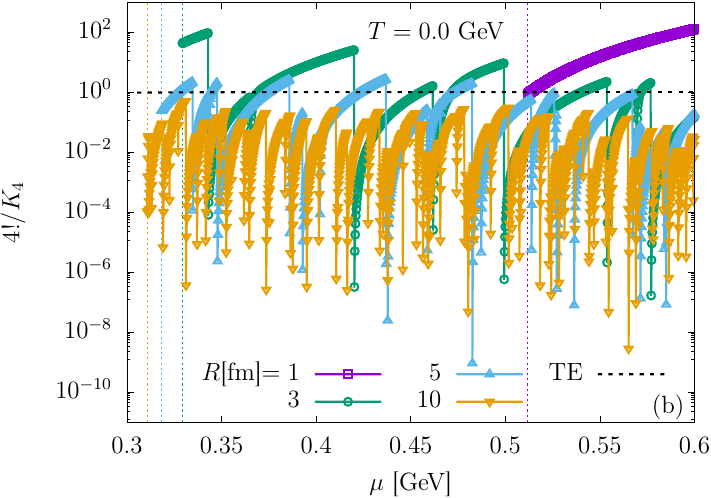}
\end{tabular}
\caption{Scaled shape coefficients, $2!/\widetilde{K}_2$ and $4!/\widetilde{K}_4$, as functions of the chemical potential $\mu$ at $T=0$ for various system sizes $R = 1$, $3$, $5$ and $10$ fm.
The black dashed lines correspond to the TE expressions in Eq.~\eqref{eq:TE_F0K2K4_T0}, with $\sigma$ obtained as described in Sec.~\ref{sec:PLSM:T0}. In the domain shown, $K_2$ and $K_4$ approach the free, conformal gas limit, $K_{2n} \simeq (2n)!$. The vertical lines mark the $R$-dependent critical chemical potential above which the system resides in the chirally restored phase. }
\label{fig:k2k4_vs_mu_T0_varR}
\end{figure}

In the (P)LSM${}_q$ model, the coefficients $K_{2n}$ can be obtained by differentiating the quark potential, given in Eq.~\eqref{eq:rot_Fq}, with respect to $v_R = \Omega R$. This derivative is understood to be taken with respect to the angular velocity, as we are looking for the system's response to the increase of angular velocity, $\Omega$, while keeping the system size $R$ constant. Since $\widetilde{F}_q$ depends on $\Omega$ only through the argument $\widetilde{\mathcal{E}}_\varsigma = \mathcal{E}_\varsigma - \Omega m$, these derivatives are simply
\begin{equation}
 \widetilde{F}^{(0)}_q \widetilde{K}_{2n} = \frac{2 N_f N_c}{\pi^2 R^{2n+2} T^{2n - 1}} \sum_{\mathrm{b}, \varsigma = \pm 1} m^{2n} \int_0^\infty dk\, f^{(2n-1)}_\varsigma,
\end{equation}
where $f^{(n)}_\varsigma \equiv \partial^{n} f_\varsigma / \partial(\beta \mathcal{E}_\varsigma)^{n}$ is the $n$-th derivative, with $n$ a non-negative integer. Note that the above expression is evaluated at vanishing angular velocity. Using Eq.~\eqref{eq_df} to compute $f'_\varsigma$, as well as the following relations,
\begin{equation}
 f_{\varsigma\varsigma'}' = \frac{\partial f_{\varsigma\varsigma'}}{\partial (\beta\mathcal{E}_\varsigma)} = -f_{\varsigma\varsigma'}(2 - \delta_{\varsigma\varsigma'} - 3f_\varsigma),
 \label{eq.higher_deriv}
\end{equation}
we obtain
\begin{align}
 \widetilde{F}_q^{(0)} \widetilde{K}_2 &= \frac{2N_f N_c}{\pi^2 R^4 T}\sum_{\mathrm{b}, \varsigma} m^2\int_0^\infty dk f'_\varsigma~,
 \label{eq_K2_int}\\
 \widetilde{F}^{(0)}_{q} \widetilde{K}_4 &= \frac{2N_f N_c}{\pi^2 R^6 T^3}\sum_{\mathrm{b}, \varsigma = \pm 1} m^4 \int_{0}^\infty dk \nonumber\\
 & \times  \bigg[2 \sum_{\varsigma'=\pm} L_{\varsigma'} f_{\varsigma\varsigma'} (2-\delta_{\varsigma\varsigma'} - 3f_\varsigma)(5-\delta_{\varsigma\varsigma'} - 9 f_\varsigma) \nonumber\\
 & + 9 f'_\varsigma(1 - 3f_\varsigma + 3f_\varsigma^2 + f_\varsigma') \bigg].
 \label{eq_K4_int}
\end{align}
The coefficients $K_2$ and $K_4$, determined by Eqs.~\eqref{eq_K2_int} and \eqref{eq_K4_int}, are presented in Figs.~\ref{fig:k2k4_vs_T_varR}--\ref{fig:k2k4_vs_mu_T0_varR}.

Figure~\ref{fig:k2k4_vs_T_varR} shows that $K_2$ and $K_4$ quickly approach the free, non-interacting limits $K_2 = 2!$ and $K_2 = 4!$~\cite{Ambrus:2023bid} when the temperature increases and the radius becomes large, $R \to \infty$, so that the system reaches the thermodynamic limit. As the temperature is decreased, the coefficients $K_2$ and $K_4$ decrease until the system reaches the chiral phase transition, after which they start increasing. For $\mu = 0$ (upper panels) and $\mu = 0.25$ GeV (lower panels), both $K_2$ and $K_4$ diverge as $T \to 0$ and the system goes to the chirally broken, confined phase.

To gain further insight into the coefficients $K_2$ and $K_4$, we compute them, as well as the quark potential, using the TE approximation, by differentiating $F_q$ in Eq.~\eqref{eq:TE_Fq} with respect to $\Omega R$:
\begin{align} \label{eq:TE_F0K2K4}
F_q^{(0)} &= -\frac{N_f N_c T^4}{3\pi^2} \sum_{\varsigma = \pm 1} \int_0^\infty \frac{dx\, x^4}{y} f_\varsigma, \\
 F_q^{(0)} K_2 &=-\frac{N_f N_c T^4}{2\pi^2} \sum_{\varsigma = \pm 1} \int_0^\infty \frac{dx\, x^2}{y} \left(\frac{x^2}{3} + y^2\right) f_\varsigma, \nonumber\\
 F_q^{(0)} K_4 &=-\frac{N_f N_c T^4}{\pi^2} \sum_{\varsigma = \pm 1} \int_0^\infty \frac{dx}{y} (8x^2 y^2 + z^4) f_\varsigma,\nonumber
\end{align}
where $y = \sqrt{x^2 + z^2}$ and $z = g\sigma / T$ [see Eq.~\eqref{eq:TE_yrho} regarding the notation]. It is not difficult to identify $F^{(0)}_q = -P_0$ and $F_q^{(0)} K_2 = -(\epsilon_0 + P_0) / 2$, with the energy density and pressure given by Eqs.~\eqref{eq:TE_P_eps}, evaluated on the rotation axis. Similarly, noting that $\bar{\psi} \psi = (\epsilon - 3P) / (g\sigma)$, we see that 
\begin{equation}
 F^{(0)}_q K_4 = -8 \epsilon_0 + (g\sigma)^3 \frac{d}{d(g\sigma)} \left(\frac{\bar{\psi} \psi_0}{g \sigma}\right).
\end{equation}

In the high-temperature limit, where $z \to 0$, it is not difficult to see that $K_2 \to 2!$ and $K_4 \to 4!$, as expected~\cite{Ambrus:2023bid}. This property is clearly seen in Fig.~\ref{fig:k2k4_vs_T_varR}, where these free-field expressions are gradually reached in the large-temperature limit. The first mass correction can be evaluated by writing $\epsilon_0 = 3P_0 + g\sigma \bar{\psi}\psi_0$. Using Eqs.~\eqref{eq:TE_FC_PLSM_m0} and \eqref{eq:TE_P_eps_PLSM_m0}, we arrive at
\begin{align}
    K_2  &= 2! + \frac{15 }{7\pi^2} \left(\frac{g\sigma}{T}\right)^2 \Phi(\vartheta), \nonumber\\
    K_4 &= 4! + \frac{240}{7\pi^2} \left(\frac{g\sigma}{T}\right)^2 \Phi(\vartheta),
 \label{eq_K2_K4_L}
\end{align}
where the higher-order corrections are not shown.
The function
\begin{align}
   \Phi(\vartheta) = \frac{1 - \frac{2}{\pi^2} \vartheta^2}{1 - \frac{20}{7\pi^2} \vartheta^2 + \frac{10}{7\pi^4} \vartheta^4}\,,
 \label{eq_Phi_L}
\end{align}
is expressed via the variable $\vartheta$ that encodes the value of the expectation value of the Polyakov loop~\eqref{eq_def_varphi}.
Thus, the chiral symmetry breaking, encoded in the condensate $\sigma$, and the effect of confinement of color affect both shape coefficients~\eqref{eq_K2_K4_L}.

The angle $\vartheta$ vanishes, $\vartheta = 0$, as the system reaches the asymptotic deconfined state at $L \rightarrow 1$ and $\vartheta = 2\pi / 3$ in the confined state ($L = 0$). The function~\eqref{eq_Phi_L} is a rising, positive-valued function of $\vartheta$, with $1 \leqslant \Phi \leqslant 9$ in this segment of values. Therefore, the chiral symmetry breaking (characterized by a nonzero condensate $\sigma$) and confinement of color (characterized by nonzero $\vartheta$) imply that both $K_2$ and $K_4$, presented in Eq.~\eqref{eq_K2_K4_L}, are larger than their free-field values. Both effects are readily understood at the intuitive level, since they imply the mass gap generation in the medium, which enhances its inertial mechanical response.

In the opposite limit of vanishing temperature, $z \gg 1$, the shape coefficients $\widetilde{K}_2$ and $\widetilde{K}_4$ can be derived from the grand potential of a rotating system at vanishing temperature, given in Eq.~\eqref{eq.F0T}:
\begin{subequations}\label{eq_K2K4_low_T}
\begin{align}
    \widetilde{F}_q^{T = 0}(0) \widetilde{K}_2^{T=0} &= -\frac{2N_f N_c}{\pi^2 R^4}\sum_{\mathrm{b}} m^2 \frac{E^f_{ml}}{k_{ml}^f} \theta(|\mu| - M_{ml}), \\
    \widetilde{F}_q^{T = 0}(0) \widetilde{K}^{T=0}_4 &= -\frac{6 N_f N_c}{\pi^2 R^6}\sum_{\mathrm{b}} m^4 \frac{M_{ml}^2 E^f_{ml}}{(k_{ml}^{f})^5} \theta(|\mu| - M_{ml}),
\end{align}
\end{subequations}
where $E^f_{ml} = \mu$, $M_{ml} = \sqrt{q_{ml}^2 + g^2 \sigma^2}$ and $k_{ml}^f = \sqrt{\mu^2 - M_{ml}^2}$ represent the $\Omega = 0$ limit of the same quantities introduced in Sec.~\ref{sec:rot:T0}.
From the TE expressions in Eqs.~\eqref{eq:TE_F0K2K4}, we have
\begin{subequations}\label{eq:TE_F0K2K4_T0}
\begin{align}
 F_q^{T = 0}(0) K_2 &= -\frac{N_f N_c}{6\pi^2} p_f^3 E_f \theta(\mu - g\sigma), \\
 F_q^{T = 0}(0) K_4 &= -\frac{N_f N_c}{\pi^2} p_f E_f (p_f^2 + E_f^2) \theta(\mu - g\sigma),
\end{align}
\end{subequations}
where $E_f = |\mu|$ and $p_f= \sqrt{E_f^2 - g^2 \sigma^2}$ are the Fermi energy and momentum, respectively.

Figure~\ref{fig:k2k4_vs_T_varmu} shows the behaviour of the shape coefficients as functions of temperature at finite chemical potential and finite radius of the system, $R = 5$ fm. At low temperatures, both coefficients demonstrate substantial oscillations in their values as the chemical potential varies in a rather small interval. This seemingly irregular behaviour disappears quickly as the temperature increases.

The zero-temperature oscillations can clearly be seen in Fig.~\ref{fig:k2k4_vs_mu_T0_varR}, where both coefficients, computed from Eqs.~\eqref{eq_K2K4_low_T}, are shown as functions of the chemical potential $\mu$ for various radii $R$ of the system. They irregularly oscillate around the corresponding TE values (shown by the horizontal dashed lines), computed using Eqs.~\eqref{eq:TE_F0K2K4_T0}. It is noteworthy that both $K_2 \simeq 2! + 3 g^2 \sigma^2 / \mu^2$ and $K_4 \simeq 4! + 48 g^2 \sigma^2 / \mu^2$ quickly approach their massless limit, such that the TE lines appear almost horizontal.
The reason for the oscillations seen in the bounded results is rooted in an analogue of the Shubnikov-de Haas effect, which arises due to the modulation of the density of states at the Fermi level~\cite{Schubnikow1930}.
In our system, the irregular oscillations arise when the chemical potential crosses the next transverse level determined by the transverse mass eigenvalues $M_{ml}$ and associated with the radial wavefunctions in the cylinder. Mathematically, the irregular behaviour appears due to the presence of Heaviside functions $\theta(|\mu| - M_{ml})$, that emerge as a low-temperature limit of the Fermi-Dirac distribution in Eq.~\eqref{eq_K2K4_low_T}, as well as due to the negative powers of the longitudinal momentum $k_{ml}$. As the temperature increases, the sharp Fermi surfaces with $|\mu| = M_{ml}$ erode, and the oscillations disappear, bringing the system back to the smooth behaviour of Fig.~\ref{fig:k2k4_vs_T_varmu}.

\section{Conclusions} \label{sec:conc}

In this paper, we studied the effects of the uniform steady rotation on the phase structure of strongly interacting matter. We work in the effective model approach represented by the linear sigma model coupled to quarks and to the Polyakov loop (PLSM${}_q$). The attractive features of this model are its renormalizability and the presence of the coupling between chiral and confining degrees of freedom. We fixed the coupling constants of the model to their vacuum values, without allowing for heuristic modification of the model parameters that would fine-tune effects of rotation or temperature. We put the system in a cylinder of a finite radius $R$, imposing the spectral boundary conditions on the quark degrees of freedom, in order to protect the system from breaking causality. We adopted the uniform approximation to the ground state, described by the scalar condensate that leads to the breaking of the chiral symmetry, and the Polyakov loop expectation value that encodes the confinement of color.

Our main focus was to study both the boundary and the rotation effects on the phase diagram of the PLSM$_q$ model.
We explored a wide parameter range, covering regimes corresponding to different temperatures, various radii of the rotating plasma and different values of the quark chemical potential.
We found that, in the studied range of parameters, the model predicts that the rotation decreases the (pseudo)critical temperatures of the transitions corresponding to both color deconfinement and chiral symmetry breaking.

For small system sizes, we showed that the suppression of the fermionic sector allows the deconfinement transition to proceed as in the pure-glue case, at the temperature $T = T_0 = 0.27$ GeV. This is a consequence of the fact that the Polyakov sector is not influenced directly by the presence of the boundary. Since the Polyakov potential is also insensitive to the chemical potential $\mu$, as we increased $\mu$ and the fermionic sector became more important, we showed that the first-order pure-glue transition turned into a crossover transition. Thus, we found a small-$\mu$ critical point occurring for small enough systems, separating the first-order and crossover regimes for the deconfinement phase transition. On the other hand, we showed that the chiral transition takes place at a larger temperature, eventually merging into the deconfinement transition when $\mu$ is sufficiently increased. We also found that rotation inhibits the small-$\mu$ endpoint appearing in small systems.

At finite volume, we found that rotation promotes both chiral restoration and deconfinement, leading to a melting of the $\sigma$ meson condensate and an enhancement of the Polyakov loop expectation values. We performed an extensive investigation of the effects of both boundary and rotation on the phase diagram, from small to large systems, while varying the angular velocity between $0$ and the causal limit $\Omega R = 1$. In all cases, we found that increasing $\Omega R$ promotes both the deconfinement and the chiral restoration transitions, leading to smaller (pseudo)critical values of the temperature and chemical potential at the transition point. We considered the trajectory of the critical point for the chiral restoration transition as a function of $\Omega R$, for various system sizes. As the system size is increased and $R \to \infty$, the critical point for the static (non-rotating) case smoothly tends to that for the unbounded system. Surprisingly, when $R$ becomes large, the chemical potential corresponding to the critical point becomes independent of $\Omega$ and, as $\Omega R$ is increased, the critical point travels in the $T$-$\mu$ plane on an almost vertical line, towards lower temperatures.

In the large system-size limit, we considered a comparison of the numerical QFT results with the Tolman-Ehrenfest prediction of the uniformly rotating systems, by which the local temperature increases as the distance from the axis of rotation increases. At vanishing chemical potential, we were able to find an analytical relation between the pseudocritical chiral restoration temperatures for the rotating and non-rotating systems, validated by comparison to the numerical results in the large volume limit. Within this setup, we were able to reveal that the expectation value of the Polyakov loop $L$ at the pseudocritical temperature of the chiral transition increases to $1$ towards the causal limit, as $\Omega R \to 1$. Thus, in the causal limit, the system is already deconfined when the chiral restoration takes place. We also considered within our TE model the dependence of the critical chemical potential in the limit of vanishing temperature. All of the aforementioned results were supported by the numerical data obtained for large system sizes, where boundary effects become subleading.

While the TE picture is consistent with the PLSM${}_q$ results for all studied parameter ranges, it is conflicting with lattice results for the purely gluonic plasma~\cite{Braguta:2020biu, Braguta:2021jgn,Braguta:2023tqz}, as well as for the theory with dynamical fermions \cite{Braguta:2022str,Braguta:2023iyx}. We stress that at high temperatures, where the QGP becomes weakly coupled, the TE behavior should be restored, with only a minor non-perturbative pressure contribution from magnetic gluons~\cite{Kajantie:2002wa}. Thus, some of our findings may hold far above the deconfinement transition.

In addition, we analyzed the mechanical properties of rotating quark-gluon matter within PLSM${}_q$. We found that the dimensionless moment of inertia experiences a substantial increase at the phase transition line and a strong enhancement deeper in the deconfined phase. These properties can be attributed to the liberation of the degrees of freedom above the deconfinement temperature and a rapid increase in the energy density as temperature increases in the deconfinement phase. The Tolman-Ehrenfest law predicts a rapid rise of the form $I \sim \Gamma_R^4$, with $\Gamma_R = (1 - \Omega^2 R^2)^{-1/2}$ being the Lorentz factor of the boundary, which is confirmed by the numerical results as the velocity of the boundary $\Omega R$ approaches the causal limit.

We also studied the shape coefficients $K_2$ and $K_4$ of the rotating medium. The first quantity, $K_2$, corresponds to the ratio of the moment of inertia to the free energy of the system at vanishing rotation. The second quantity, $K_4$, effectively encodes the change of the shape of the rotating plasma with increasing rotation, due to the mass re-distribution within the system. Due to their particular normalization, these dimensionless quantities characterize the relative significance of the mechanical moment of inertia and the shape of the rotating plasma, factoring out the number of degrees of freedom. In the large-temperature and large-volume limit, both dimensionless coefficients approach their free-field values, $K_2 = 2!$ and $K_4 = 4!$. While in the deconfinement phase, these quantities are slowly varying functions of temperature, they experience a knee-like structure at the transition point and rise significantly as the temperature decreases. The latter fact implies that the number of thermal degrees of freedom (represented by the magnitude of the free energy) decreases faster than these mechanical characteristics of the rotating plasma. Expectedly, as the radial size of the plasma increases, both these quantities approach the TE limit of the PLSM${}_q$ model in both phases. The latter limit also includes the effects of the fermionic mass and a finite mean-field value of the Polyakov loop that are nonvanishing on both sides of the (pseudo-)critical transition.

Interestingly, we showed analytically that the TE law implies that both the chiral symmetry breaking and the quark confinement lead to the increase of the $K_2$ and $K_4$ coefficients in the zero-density plasma ($\mu = 0$). On the contrary, at vanishing temperature, the varying quark chemical potential leads to quasi-periodic oscillations of the mechanical coefficients $K_2$ and $K_4$, that appear as a result of the modulation of the density of states at discrete thresholds of the nested energy levels that resemble the Shubnikov-de Haas oscillations~\cite{Schubnikow1930}.

The raw data and scripts to generate the plots shown in
this paper are available as ancillary files to Ref.~\cite{Singha:2025zvh}.

\begin{acknowledgments}
We thank P. Aasha for fruitful discussions.
This work was funded by the EU’s NextGenerationEU instrument through the National Recovery and Resilience Plan of Romania - Pillar III-C9-I8, managed by the Ministry of Research, Innovation and Digitization, within the project entitled ``Facets of Rotating Quark-Gluon Plasma'' (FORQ), contract no. 760079/23.05.2023 code CF 103/15.11.2022.
\end{acknowledgments}

\section*{Data availability}
The data that support the findings of this article are openly available to Ref. \cite{Singha:2025zvh}.

\appendix
\section{The Polyakov potential}\label{app:poly}

\begin{figure}
\centering
\begin{tabular}{c}
\includegraphics[width=0.92\linewidth]{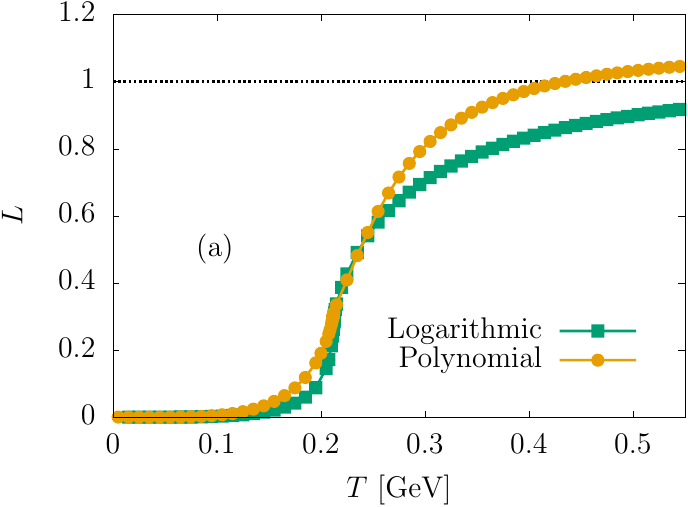} \\
\includegraphics[width=0.92\linewidth]{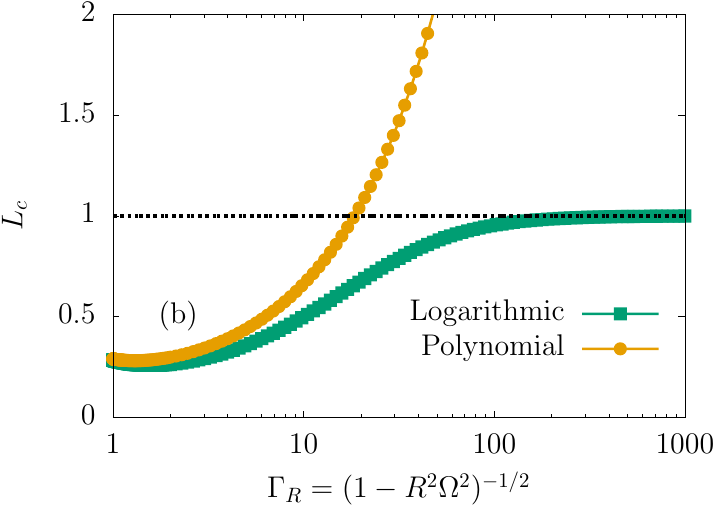}
\end{tabular}
    \caption{(a) Expectation value of the Polyakov loop as a function of temperature at zero chemical potential for the cases of the logarithmic and polynomial Polyakov potentials. (b) The expectation value of the Polyakov loop at the chiral transition temperature, as a function of $\Gamma_R$, according to the Tolmann-Ehrenfest prediction.
    }
    \label{fig:diffpot}
\end{figure}

In this appendix, we briefly review the polynomial model for the effective Polyakov potential $V_L$, proposed in Ref.~\cite{Ratti:2005jh}. Compared to the logarithmic potential \cite{Roessner:2006xn} considered in the main text, the qualitative outcomes from the polynomial potential are mostly similar, since both models are constructed based on the ${\mathbb Z}_3$ symmetry properties of ${\rm SU}(3)$ pure gauge theory, their parameters being fine-tuned to reproduce lattice pure gauge thermodynamics. We also discuss the extension allowing for the transition temperature $T_0$ appearing in the Polyakov potential to become a function of the number of fermion flavors and of the fermion chemical potential, as described in Ref.~\cite{Schaefer:2007pw}.

In Subsec.~\ref{app:poly:poly}, we describe the structure and features of the polynomial potential in the pure-glue theory. In Subsec.~\ref{app:poly:T0}, we discuss the models for the transition temperature $T_0$.

\subsection{Polynomial Polyakov potential} \label{app:poly:poly}

In the polynomial parametrization, the potential can be written as follows,
\begin{equation}
 \frac{V^{\rm p.}_L}{T^4} = -\frac{b_2(T)}{2} L^* L - \frac{b_3}{6}[(L^*)^3 + L^3] + \frac{b_4}{4} (L^* L)^2,
 \label{eq:polynomial}
\end{equation}
where $b_3$ and $b_4$ are constants, while $b_2(T)$ reads
\begin{equation}
 b_2(T) = a_0 + a_1 \left(\frac{T_0}{T}\right) + a_2 \left(\frac{T_0}{T}\right)^2 + a_3 \left(\frac{T_0}{T}\right)^3.
\end{equation}
Imposing the Stefan-Boltzmann limit (assuming $L,L^* \rightarrow 1$ when $T \rightarrow \infty$) implies
\begin{equation}
 \frac{a_0}{2} + \frac{b_3}{3} - \frac{b_4}{4} = \frac{8\pi^2}{45}.
\end{equation}
Demanding that $V^{\rm p.}_L$ is minimized at a fixed temperature with respect to both $L$ and $L^*$ gives $L = L^*$ and
\begin{equation}
 \left. \frac{\partial (V^{\rm p.}_L / T^4)}{\partial L} \right|_{L^* = L} = -\frac{L}{2} [b_2(T) + b_3 L - b_4 L^2] = 0.
 \label{eq:polynomial_dVdL}
\end{equation}
Imposing that $L = 1 - \delta L$, with $\delta L \rightarrow 0$ when $T \rightarrow \infty$, we find another constraint involving $a_0$, $b_3$ and $b_4$, as well as the large-temperature behaviour of $L$:
\begin{equation}
 a_0 + b_3 - b_4 = 0, \qquad
 L \simeq 1 - \frac{a_1}{b_3 - 2b_4} \frac{T_0}{T}.
\end{equation}
As opposed to the case of the logarithmic potential, $L$ approaches $1$ at a slower rate. The approach is made from below if $a_0 / (b_3 - 2b_4) > 0$, which is true for the parametrization that we will consider here (see below).

We can see that Eq.~\eqref{eq:polynomial_dVdL} gives $L = 0$ as a root, meaning that the confined state always corresponds to a local minimum of the potential. Above a threshold temperature $T_-$, a second minimum develops. This can be found by setting to $0$ the term between the square brackets:
\begin{equation}
 L^{(\pm)} = \frac{b_3}{2b_4} \left(1 \pm \sqrt{1 + \frac{4 b_2(T) b_4}{b_3^2}}\right).
\end{equation}
The temperature $T_-$ corresponds to the solution
\begin{equation}
 b_2(T_-) = -\frac{b_3^2}{4b_4},
\end{equation}
in which case $L^{(+)} = L^{(-)} = b_3 / 2b_4$. As a side remark, since $0 < L^{(-)} \le L^{(+)} < 1$, we must have $b_3 < 2b_4$ and then $a_1 < 0$ such that $L$ approaches $1$ from below. Substituting now $T = T_-$ and $L = L^* = L^{(\pm)}$ into Eq.~\eqref{eq:polynomial} gives
\begin{equation}
 V^{\rm p.}_L(T_-, L^{(\pm)}) = \frac{b_3^4}{192b_4^3} > 0,
\end{equation}
such that at $T = T_-$, the solution $L = 0$ still corresponds to the global minimum of the Polyakov potential. Demanding now that the first-order phase transition occurs as $T = T_0$ and denoting $b_c \equiv b(T_0) = a_0 + a_1 + a_2 + a_3$, we impose $V^{\rm p.}_L(T_0, L^{(+)}) = 0$, such that
\begin{equation}
 \frac{(L^{(+)})^2}{12} (3b_c + b_3 L^{(+)}) = 0.
\end{equation}
Substituting the solution for $L^{(+)}$, we arrive at
\begin{equation}
 b_c = a_0 + a_1 + a_2 + a_3 = -\frac{2 b_3^2}{9b_4},
\end{equation}
which gives the value $L^{(+)}_c = 2b_3 / 3b_4$ immediately after the phase transition.The remaining coefficients, namely $a_1$, $a_2$ and $a_3$, are found by fitting the lattice data. Following Refs.~\cite{Ratti:2005jh,Schaefer:2007pw,Fukushima:2017csk},
\begin{equation}
 a_1 = -1.95, \quad a_2 = 2.625, \quad
 a_3 = -7.44.
\end{equation}
This leads to the following system of equations:
\begin{align}
 \frac{a_0}{2} + \frac{b_3}{3} - \frac{b_4}{4} = \frac{8\pi^2}{45}, \quad
 a_0 + b_3 - b_4 = 0, \quad \nonumber\\
 a_0 + \frac{2 b_3^2}{9b_4} = -a_1 - a_2 - a_3 \simeq 6.765,
\end{align}
which is solved by $a_0 \simeq 6.75$, $b_3 \simeq 0.82$ and $b_4 \simeq 7.56$. This also implies a mere $L^{(+)}_c \simeq 0.0722$ after the phase transition. For $T_0$, we take the pure glue transition temperature, $T_0=0.27$ GeV, which is the same as in the case of the logarithmic potential.

One of the major drawbacks of the polynomial potential is that, under coupling to fermions, the expectation value of the Polyakov loop $L$ goes beyond 1 as $T \rightarrow \infty$ \cite{Schaefer:2007pw}, as can be seen from Fig.\ref{fig:diffpot}(a), which will effectively imply an unphysical negative free energy for the heavy quark. On the other hand, the logarithmic term in the logarithmic potential for the Polyakov loop ensures that $L$ never exceeds $1$, even in the presence of fermions, approaching it asymptotically as the temperature increases~\cite{Ghosh:2007wy}.

The situation worsens as we turn on the rotation. As can be seen in Fig.\ref{fig:diffpot}(b), the expectation value of the Polyakov loop ($L_c$) at the chiral symmetry restoration temperature ($T_\sigma$), estimated according to the Tolmann-Ehrenfest prediction in Eq.~\eqref{eq:TE_PLSM}, increases sharply with increasing $\Gamma_R$ for the polynomial potential, while for the logarithmic potential, the Polyakov loop $L$ asymptotes to $1$ as $\Gamma_R \rightarrow \infty$. We thus conclude that the polynomial potential for the Polyakov loop is not suitable for the study of systems undergoing fast rotation.

\begin{figure}
\centering
\begin{tabular}{c}
\includegraphics[width=0.92\linewidth]{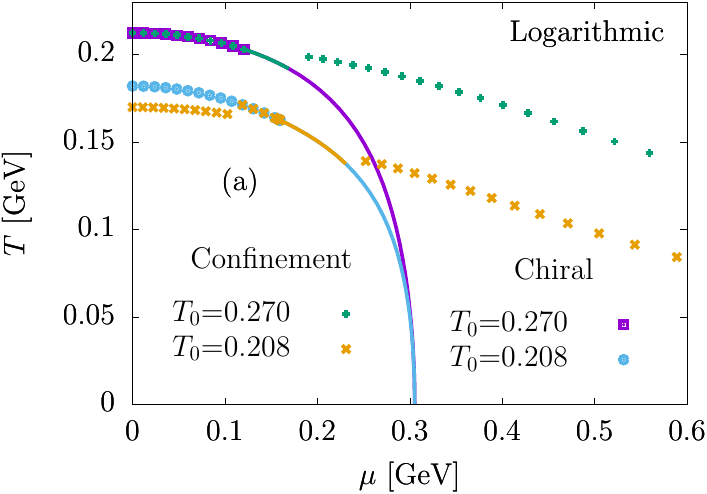} \\
\includegraphics[width=0.92\linewidth]{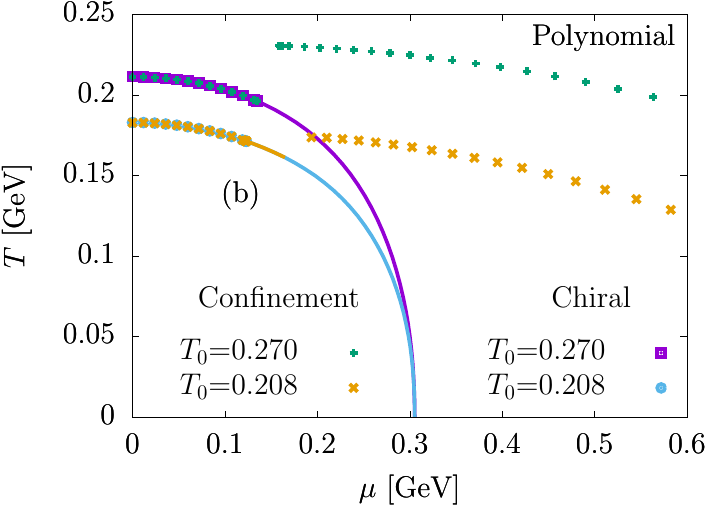}
\end{tabular}
    \caption{The chiral and deconfinement phase diagram for two different values of $T_0=0.270$ GeV and $0.208$ GeV for the (a) logarithmic and (b) polynomial Polyakov potentials. A solid line in both figures denotes the first-order transition.}
\label{fig:diffT0}
\end{figure}

\subsection{Choice of \texorpdfstring{${T}_0$}{T0} }\label{app:poly:T0}

So far, for both logarithmic and polynomial potentials, we have considered $T_0=0.27$ GeV, corresponding to the confinement-deconfinement transition, measured on the lattice for the ${\rm SU}(3)$ pure gauge theory. However, in presence of dynamical quarks, one can expect this transition temperature to be modified. In this section, we will discuss the model proposed in Ref.~\cite{Schaefer:2007pw}, starting with an $N_f$-dependent, but $\mu$-independent transition temperature in Subsec.~\ref{app:poly:T0:const}, and incorporating the $\mu$-dependence in Subsec.~\ref{app:poly:T0:mu}.

\subsubsection{Constant \texorpdfstring{${T}_0$}{T0}}\label{app:poly:T0:const}

First, we will consider a constant $T_0=0.208$ GeV, as suggested in Ref.~\cite{Schaefer:2007pw} for the case of $N_f = 2$ quark flavors. In Fig.~\ref{fig:diffT0}, we can see that, with $T_0=0.208$ GeV, the temperatures for the chiral and deconfinement transitions decrease for both the logarithmic and the polynomial potentials, compared to the $T_0 = 0.270$ GeV case. However, there is an unphysical splitting between the chiral and deconfinement transitions in the crossover region for the logarithmic potential at $T_0 = 0.208$ GeV. This behaviour, with the chiral temperature being higher than the deconfinement temperature, is not expected for the non-rotating system. On the other hand, the phase diagram at $T = 0.270$ GeV agrees qualitatively with the phenomenological expectations and, therefore, it is used in the main text of this paper.

For the polynomial potential, no such unphysical splitting can be seen, for either $T_0 = 0.270$ GeV or $0.208$ GeV. However, we still do not consider the polynomial potential in our numerical analysis, due to the limitation that $L$ and $L^*$ are free to increase without bounds.

\subsubsection{Chemical-potential-dependent T\texorpdfstring{${}_0$}{0}} \label{app:poly:T0:mu}

\begin{figure}
\centering
\begin{tabular}{c}
\includegraphics[width=0.92\linewidth]{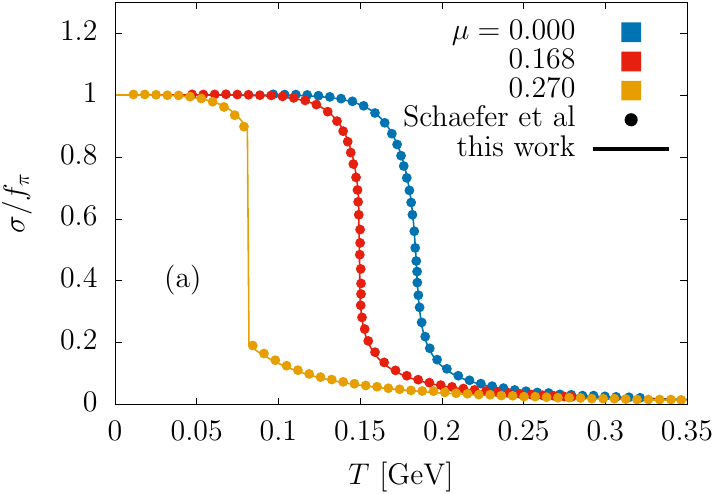} \\
\includegraphics[width=0.92\linewidth]{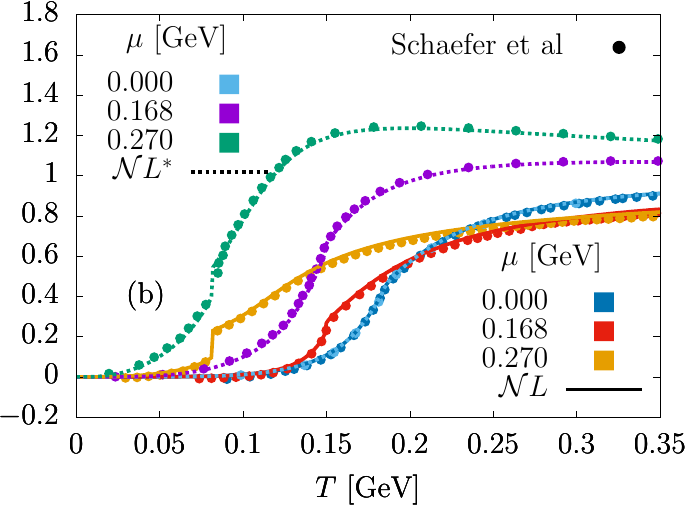}
\end{tabular}
\caption{Temperature dependence of (a) scaled $\sigma$ and (b)  $L$ and $L^*$ scaled by $\mathcal{N}=1.1$ for different values of chemical potential with polynomial Polyakov potential as discussed in Ref.~\protect\cite{Schaefer:2007pw}.
}
\label{fig:wbcompLsig}
\end{figure}

\begin{figure*}
\centering
\begin{tabular}{cc}
\includegraphics[width=0.45\linewidth]{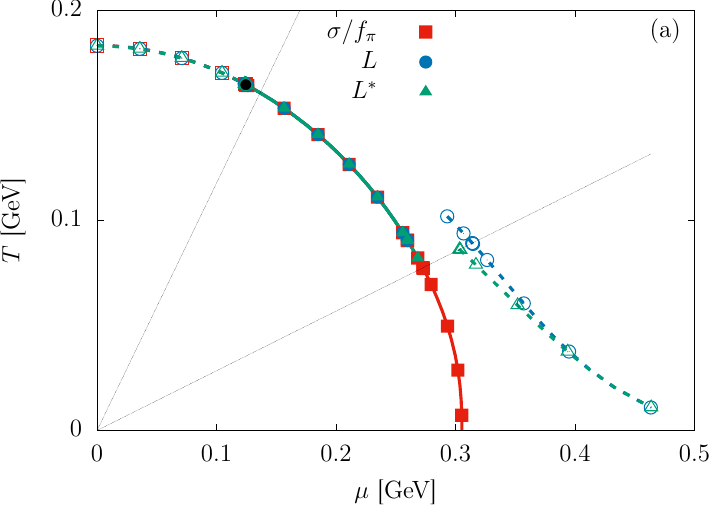} &
\includegraphics[width=0.45\linewidth]{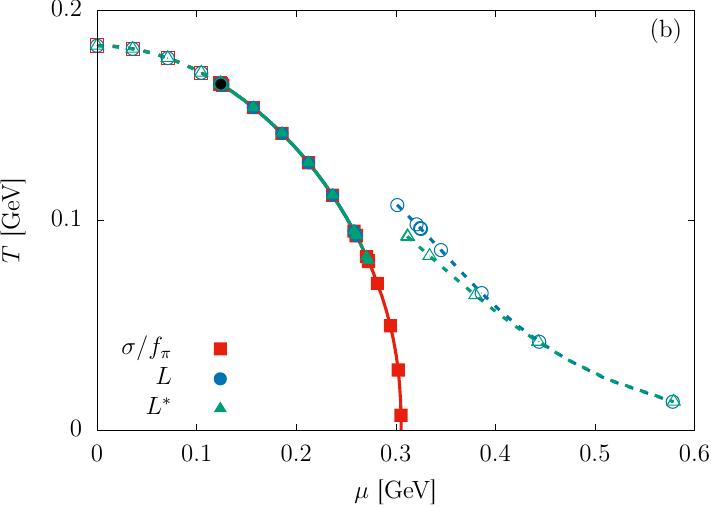} \\
\includegraphics[width=0.41\linewidth]{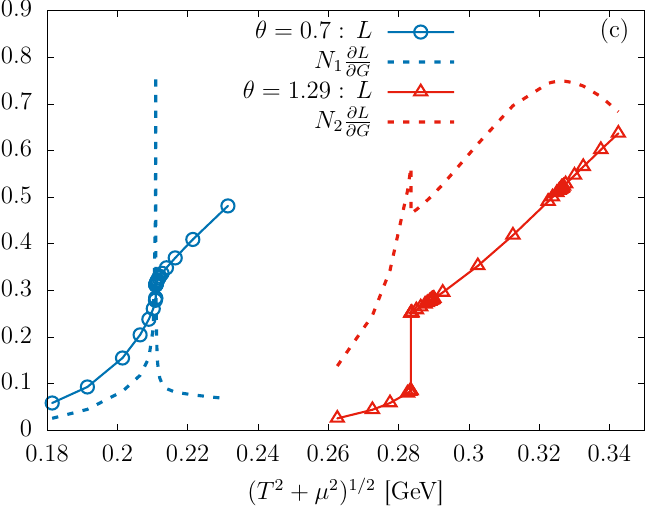} &
\includegraphics[width=0.45\linewidth]{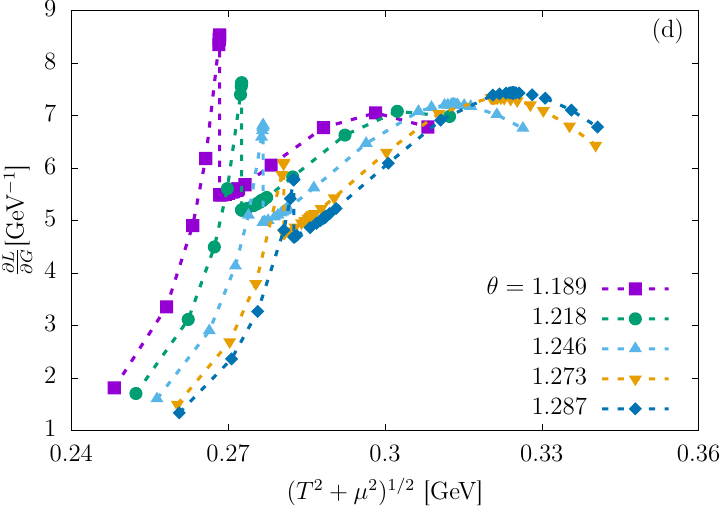}
\end{tabular}
    \caption{Chiral and deconfinement phase diagram with chemical-potential-dependent  $T_0$ in the polynomial Polyakov potential with $b(\mu)$ (a) as in Eq.\eqref{eq:Wambach_bmu}
    and  (b) as in Eq.\eqref{eq:Wambach_bmu_improved}.
    (c) The temperature dependence of the Polyakov loop and its slope for two different values of $\theta$, marked in panel (a), with $N_1 = 0.01$ GeV and $N_2 = 0.1$ GeV. (d) Double-peak structure of the slope $\partial L / \partial G$ of the Polyakov loop with respect to $G = \sqrt{T^2 + \mu^2}$, computed on diagonal trajectories under various angles $\theta$.
    }
    \label{fig:wbcompPD}
\end{figure*}

Lastly, we will consider the chemical-potential-de\-pen\-dent $T_0$. We follow Ref.~\cite{Schaefer:2007pw} and consider only the polynomial Polyakov loop potential. In the presence of quark degrees of freedom, $T_0$ becomes a function of the quark chemical potential $\mu$ and the number of quark flavors $N_f$.
This modification can be obtained by computing the transition temperature via~\cite{Schaefer:2007pw}
\begin{equation}
 T_0 = T_\tau e^{-1 / (\alpha_0 b)},
 \label{eq:Wambach_T0}
\end{equation}
where $\alpha_0 = 0.304$ is the QCD coupling at the $\tau$ scale given by $T_\tau = 1.770$ GeV. The dependence on $\mu$ can be incorporated through the coefficient $b$~\cite{Schaefer:2007pw}. At one-loop level, one can write \cite{Herbst:2010rf,Herbst:2013ail}
\begin{equation}
 b(\mu) = b_0 - b_\mu\frac{\mu^2}{(\hat{\gamma} T_\tau)^2}, \quad
 b_0 = \frac{1}{6\pi} (11 N_c - 2N_f),
 \label{eq:Wambach_bmu}
\end{equation}
where $b_\mu \simeq \frac{16}{\pi} N_f$.
For the $N_f = 2$ model considered in this paper, $T_0$ evaluates to $0.208$ GeV at $\mu = 0$. The parameter $\hat{\gamma}$ is in principle free, however for definiteness we take it as $\hat{\gamma} = 1$.

The model in Eqs.~\eqref{eq:Wambach_T0}--\eqref{eq:Wambach_bmu} has a discontinuity in the critical temperature $T_0$ when $b(\mu) = 0$. As $\mu \to \mu_\infty = T_\tau \sqrt{b_0 / b_\mu}$ from below, $T_0 \rightarrow 0$. When $\mu$ approaches $\mu_\infty$ from above, $T_0 \rightarrow \infty$. This behavior is clearly unphysical and it can be traced to the validity of Eq.~\eqref{eq:Wambach_bmu} only at small $\mu$. However, inducing the phase transition in a bounded system requires a larger $\mu$ and/or $T$ than in the unbounded system, as the boundary generically suppresses fermionic expectation values. Therefore, the study of strongly interacting matter in the presence of a boundary requires a model that gives a regular behavior of $T_0$ for any value of $\mu$. One may extend the model presented in Eq.~\eqref{eq:Wambach_bmu} by replacing $b(\mu)$ by its first-order Pad\'e approximant,
\begin{equation}
 b(\mu) = b_0 \left[1 + \frac{b_\mu}{b_0} \left(\frac{\mu}{\hat{\gamma} T_\tau}\right)^2\right]^{-1},
 \label{eq:Wambach_bmu_improved}
\end{equation}
which has the advantage of giving a finite critical temperature $T_0$ for any value of $\mu$, which tends asymptotically to $0$ as $\mu \rightarrow \infty$.

In the case when $T_0$ depends on the chemical potential, the last relation in Eq.~\eqref{eq:d2FdAdp} is modified to:
\begin{multline}
 \frac{\partial^2 F}{\partial \mu \partial L_\pm} = \frac{\partial^2 V_L}{\partial \mu \partial L_\pm} - 6 N_f \sum_{\varsigma = \pm 1} \varsigma \int \frac{d^3p}{(2\pi)^3} \\
\times f_{\varsigma,\pm} (1 + \delta_{\varsigma,\mp} - 3 f_\varsigma)\,.
\label{eq:d2FdAdp_mu} 
\end{multline}
The second derivative $\partial^2 V_L / \partial \mu \partial L_\pm$ of the Polyakov potential in the logarithmic model \eqref{eq_L_Polyakov} reads:
\begin{multline}
 \frac{\partial^2 V_L}{\partial \mu \partial L_\pm} = - T^4 \frac{\partial T_0}{\partial \mu} \left[\frac{6}{\mathcal{H}} (L_\mp - 2 L_\pm^2 + L_\mp^2 L_\pm) \frac{\partial b}{\partial T_0} \right. \\
 \left. + \frac{L_\mp}{2} \frac{\partial a}{\partial T_0}\right], 
 \label{eq:unb_d2VLdTdL}   
\end{multline}
where $\partial b / \partial T_0 = 3 b_3 T_0^2 / T^3$ and $\partial a / \partial T_0 = (a_1 / T) + (2a_2 T_0 / T^2)$.
In the case of the polynomial potential, we have
\begin{gather}
 \frac{\partial^2 V^{\rm p.}_L}{\partial \mu \partial L_\pm} = - T^4 \frac{\partial T_0}{\partial \mu} \frac{L_\mp}{2} \frac{\partial b_2}{\partial T_0},  \nonumber\\
 \frac{\partial b_2}{\partial T_0} = \frac{a_1}{T} + \frac{2a_2 T_0}{T^2} + \frac{3 a_3 T_0^2}{T^3}.
 \label{eq:unb_poly_d2VLdmudL}
\end{gather}
The derivative of $T_0$ with respect to $\mu$ can be evaluated starting from Eq.~\eqref{eq:Wambach_T0},
\begin{equation}
 \frac{\partial T_0}{\partial \mu} = \frac{T_0}{\alpha_0 b^2(\mu)} \frac{\partial b(\mu)}{\partial \mu}.
\end{equation}
In the case of the potential proposed in Refs.~\cite{Schaefer:2007pw,Herbst:2010rf,Herbst:2013ail}, the derivative of $b(\mu)$ with respect to $\mu$ evaluates to
\begin{equation}
 \frac{\partial b(\mu)}{\partial \mu} = -\frac{2 \mu b_\mu}{(\hat{\gamma} T_\tau)^2},
\end{equation}
while when $b(\mu)$ is represented using the Pad\'e approximant in Eq.~\eqref{eq:Wambach_bmu_improved}, we have
\begin{equation}
 \frac{\partial b(\mu)}{\partial \mu} = -\frac{2\mu b_\mu}{b_0^2 T_\tau^2} b^2(\mu).
\end{equation}

We implemented the above model in our numerical code and used the results reported in Ref.~\cite{Schaefer:2007pw} to validate our implementation. 

In Sec.~\ref{sec:PLSM:phase}, we reported a split between the deconfinement and chiral phase diagrams, for the usual logarithmic Polyakov potential with constant $T_0=0.27$ GeV at higher values of the chemical potential. Figure~\ref{fig:wbcompPD} shows that, with these modifications, the coincidence of the two transitions persists for a longer range in $\mu$.

We would like to point out that the split between the chiral and deconfinement phase transition temperature at high chemical potential can also be observed with the polynomial Polyakov potential used in
Ref.~\cite{Schaefer:2007pw}. In Fig.~\ref{fig:wbcompLsig} we present a comparison between the results we obtained and the results reported in Ref.~\cite{Schaefer:2007pw} for $\sigma/f_\pi$ and $L,L^*$ with the potential and parameters taken from Ref.~\cite{Schaefer:2007pw}. We can see an excellent agreement between the two works. In Fig.~\ref{fig:wbcompLsig}(b), $L$ and $L^*$ are scaled by $\mathcal{N}=1.1$, as also discussed in Ref.~\cite{Schaefer:2007pw}. However, we observe a split in the phase diagram in Fig.~\ref{fig:wbcompPD}(a), between the two transitions, unlike what is reported in Ref.~\cite{Schaefer:2007pw}. Here we have used the algorithm discussed in Sec.~\ref{sec:PLSM:phase} to obtain the phase diagram.

We thus conclude that, in order to incorporate the effects of dynamical quarks on the effective Polyakov potential, a detailed investigation is necessary.

\bibliography{ref}
\end{document}